\pdfoutput=1

\documentclass[11pt,twoside,a4paper,cmspaper,final,collab]{cms-tdr}
\def\svnVersion{91503a8-D}\def\svnDate{2019/05/29}\def\cmsCernNoTag{CERN-EP-2018-335}\def\cmsCernDate{\today}\def\cmsMessage{"Published in the Journal of Instrumentation as \href{http://dx.doi.org/10.1088/1748-0221/14/07/P07004}{\doi{10.1088/1748-0221/14/07/P07004}.}"}

\begin{document}\cmsNoteHeader{JME-17-001}

\hyphenation{had-ron-i-za-tion}
\hyphenation{cal-or-i-me-ter}
\hyphenation{de-vices}
\RCS$HeadURL$
\RCS$Id$

\newcommand{\ptmisstrig}{\ensuremath{p_{\text{T, trig}}^{\text{miss}}}}
\newcommand{\mhttrig}{\ensuremath{H_{\text{T, trig}}^{\text{miss}}}}
\newcommand{\mt}{\ensuremath{M_\mathrm{T}}}
\newcommand{\Zmm}{\ensuremath{\cPZ\to\PGm^+\PGmm}}
\newcommand{\Zee}{\ensuremath{\cPZ\to \Pep\Pem}}
\newcommand{\qt}{\ensuremath{{q}_\mathrm{T}}\xspace}
\newcommand{\Wenu}{\ensuremath{\PW\to \Pe\PGn}}
\newcommand{\Wmn}{\ensuremath{\PW\to \PGm\PGn}}
\newcommand{\vet}{\ensuremath{\vec\varepsilon}}
\newcommand{\like}{\ensuremath{{\cal L}}}
\newcommand{\metsig}{\ensuremath{{\mathcal S}}}
\newcommand{\vqt}{\ensuremath{\vec{q}_\mathrm{T}}\xspace}
\newcommand{\vut}{\ensuremath{\vec{u}_\mathrm{T}}\xspace}
\newcommand{\upar}{\ensuremath{u_\Vert}}
\newcommand{\uperp}{\ensuremath{u_\perp}}
\newcommand{\redupara}{\ensuremath{u_\Vert+\qt}\xspace}
\newcommand{\puppi}{\ensuremath{\text{PUPPI}}\xspace}
\newcommand{\nvtx}{\ensuremath{\text{N}_{\text{vtx}}}}
\providecommand{\et}{\ensuremath{E}_\mathrm{T}}

\cmsNoteHeader{JME-17-001}

\title{Performance of missing transverse momentum reconstruction in  proton-proton collisions at $\sqrt{s} = 13\TeV$ using the CMS detector}

\date{\today}

\abstract{
The performance of missing transverse momentum (\ptvecmiss) reconstruction algorithms for the CMS experiment is presented, using proton-proton collisions at a center-of-mass energy of 13\TeV, collected at the CERN LHC in 2016. The data sample corresponds to an integrated luminosity of 35.9\fbinv. The results include measurements of the scale and resolution of \ptvecmiss, and detailed studies of events identified with anomalous \ptvecmiss. The performance is presented of a \ptvecmiss reconstruction algorithm that mitigates the effects of multiple proton-proton interactions, using the ``pileup per particle identification'' method. The performance is shown of an algorithm used to estimate the compatibility of the reconstructed \ptvecmiss with the hypothesis that it originates from resolution effects.}

\hypersetup{%
pdfauthor={CMS Collaboration},%
pdftitle={Performance of missing transverse momentum in pp collisions at sqrt(s)=13 TeV using the CMS detector},%
pdfsubject={CMS},%
pdfkeywords={CMS, physics, missing transverse momentum, detectors}}

\maketitle

\section{Introduction}
\label{sec:introduction}
Weakly interacting neutral particles produced in proton-proton (pp) collisions at the LHC traverse the collider detectors unobserved.
However, when such particles are produced along with strong or electromagnetically interacting particles, their presence can be inferred through the measured momentum imbalance in the plane
perpendicular to the beam direction, which is referred to as the missing transverse momentum (\ptvecmiss), and its magnitude is \ptmiss.

The precise determination of \ptmiss is critical for standard model (SM) measurements that use final states with neutrinos, such as those containing
leptonic decays of the \PW~boson. In addition, \ptmiss is one of the most important observables in searches for physics beyond the SM that target new weakly interacting particles. The \ptvecmiss\ stemming from weakly interacting particles will be collectively referred to as ``genuine \ptmiss'' in what follows.
However, \ptmiss reconstruction is sensitive to the experimental resolutions, to mismeasurements of reconstructed particles, and to detector artifacts.
The performance of \ptmiss is also affected by additional pp interactions in the same or nearby bunch crossings (pileup). A detailed understanding of all these effects, both in real and simulated data, is important  to achieve optimal \ptmiss performance.

In this paper, we present studies of \ptmiss reconstruction algorithms using Monte Carlo simulation, and data collected in 2016 with the CMS detector~\cite{JINST} at the LHC~\cite{LHC}, corresponding to an integrated luminosity of 35.9\fbinv, and are applicable to the 2015--2018 data--taking period (LHC Run 2). A brief overview of the CMS detector is given in Section~\ref{sec:detector}. Information about event reconstruction is
discussed in Section~\ref{sec:evtreco}, and a description of the different \ptmiss reconstruction algorithms is provided in Section~\ref{sec:reconstruction}. Information about event simulation and selection is provided in Sections~\ref{sec:simulation} and~\ref{sec:simulation_selection}. In Section~\ref{sec:tails}, sources of anomalous \ptmiss measurements from detector and reconstruction artifacts, and methods for identifying and mitigating them, are described. The performance of the \ptmiss reconstruction at the trigger level is discussed in Section~\ref{sec:trigger}. Section~\ref{sec:metperformanceinnomet} details the performance of the \ptmiss algorithms in events with and without genuine \ptmiss. The algorithm that provides an estimate of the \ptmiss significance is described in Section~\ref{sec:metsignificance}. A summary is given in Section~\ref{sec:summary}.

\section{The CMS detector}
\label{sec:detector}
The central feature of the CMS apparatus is a superconducting solenoid of 6\unit{m} internal diameter, providing a magnetic field of 3.8\unit{T}.
Within the solenoid volume are a silicon pixel and strip tracker, a lead tungstate crystal electromagnetic calorimeter (ECAL), and a brass and scintillator hadron calorimeter (HCAL), each composed of a barrel and two endcap sections.
The pseudorapidity ($\eta$) coverage of the ECAL (HCAL) barrel is $\abs{ \eta } < 1.479$ ($\abs{ \eta } < 1.3$) and endcap is $1.479 < \abs{ \eta } < 3.0$ ($1.3 < \abs{ \eta } < 3.0$) respectively.
Forward hadronic calorimeter (HF) extend the $\eta$ coverage up to $\abs{\eta}<5.2$.

In the ECAL and HCAL barrel region, the HCAL cells have widths of 0.087 in $\eta$ and 0.087 radians in azimuth ($\phi$).
In the $\eta$-$\phi$ plane, and for $\abs{\eta} < 1.479$, the HCAL cells map on to 5$\times$5 ECAL crystal arrays (supercrystals) to form calorimeter towers projecting radially outwards from close to the nominal interaction point.
In the ECAL and HCAL endcap regions, the coverage of the towers increases progressively to a maximum of 0.174 in $\Delta \eta$ and $\Delta \phi$. Within each tower, the energy deposits in ECAL and HCAL cells are summed to define the calorimeter tower energies~\cite{JINST}, subsequently used to provide the energies and directions of hadronic jets.

The silicon tracker measures charged particles within the range $\abs{\eta} < 2.5$ (tracker acceptance). It consists of 1440 silicon pixel and 15\,148 silicon strip detector modules.
Tracks with transverse momentum \pt of $\approx$ 100\GeV emitted within $\abs{\eta} < 1.4$ have \pt and impact parameter resolutions of 2.8\% and 10 (20) \mum in the transverse (longitudinal) direction~\cite{TRK-11-001}.

Muons are measured in the range $\abs{\eta} < 2.4$, with detection planes made using three technologies: drift tubes in the barrel, cathode strip chambers (CSC) in the endcaps, and resistive
plate chambers both in the barrel and in the endcaps embedded in the iron flux-return yoke outside the solenoid~\cite{Sirunyan:2018fpa}.

Events of interest are selected using a two-tiered trigger system~\cite{Khachatryan:2016bia}. The first level (L1), composed of custom hardware processors, uses information from the calorimeters and muon detectors to select events at a rate of around 100\unit{kHz}. The second level, known as the high-level trigger (HLT), consists of a farm of processors running a version of the full event reconstruction software optimized for fast processing, and reduces the event rate to an average of 1\unit{kHz} before data storage.

A more detailed description of the CMS detector, together with a definition of the coordinate system used and the relevant kinematic variables,
can be found in Ref.~\cite{JINST}.

\section{Event reconstruction}
\label{sec:evtreco}
The CMS particle-flow (PF) algorithm~\cite{Sirunyan:2017ulk} aims to reconstruct and identify each individual particle with an optimized combination of information from the various components of the detector. Particles are identified as a mutually exclusive list of PF candidates:  charged or neutral hadrons, photons, electrons, or muons.
The PF candidates are then used to build higher-level objects, such as jets and \ptmiss.

Events are required to have at least one reconstructed vertex. When multiple vertices are reconstructed due to pileup, the vertex with the largest value of summed physics-object $\pt^2$ is the primary pp interaction vertex (PV).

Photon candidates are reconstructed from energy deposits in the ECAL using algorithms that check the compatibility of the clusters to the size and shape expected from a photon~\cite{CMS:EGM-14-001}.
The identification of the candidates is based on shower-shape and isolation variables~\cite{Khachatryan:2015hwa}. For a photon to be considered isolated, the scalar \pt sum of PF candidates originating from the PV, within a cone of $\DR \equiv \sqrt{\smash[b]{(\Delta\eta)^{2}+(\Delta\phi)^{2}}}< 0.3$ around the photon candidate, is required to be smaller than a given threshold.
Only PF candidates that do not overlap with the electromagnetic shower of the candidate photon are included in the isolation sums. The exclusion of PF candidates associated with the photon in the isolation sum, also known as ``footprint removal'', is significantly improved for the LHC Run 2.

The analyses described in this paper use two sets of photon
identification criteria: ``loose'' and ``tight''.
The loose photon candidates are required to be reconstructed within $\abs{\eta} < 2.5$, whereas tight photon candidates
are required to be reconstructed in the ECAL barrel ($\abs{\eta} < 1.44$).
Tight photon candidates, used in the performance measurements discussed in
Section~\ref{sec:metperformanceinnomet},
are also required to pass identification and isolation criteria that ensure an efficiency of 80\% for the  selection of prompt photons and a sample purity of 95\%.
In the barrel section of the ECAL, an energy resolution of about 1\% is achieved for unconverted or late-converting photons in the tens of \GeV energy range. The remaining barrel photons have a resolution of about 1.3\% up to $\abs{\eta} = 1$, rising to about 2.5\% at $\abs{\eta} = 1.4$. In the endcaps, the resolution of unconverted or late-converting photons is about 2.5\%, whereas the remaining endcap photons have a resolution between 3 and 4\%~\cite{CMS:EGM-14-001}.

Electrons within the geometrical acceptance $\abs{\eta} < 2.5$ are reconstructed by associating tracks reconstructed in the silicon detector with clusters of energy in the ECAL.
Electron candidates are required to satisfy
identification criteria~\cite{Khachatryan:2015hwa} based on the shower shape of the energy deposit in the ECAL and the consistency of the electron track with the PV. Electron candidates that are identified as coming from photon conversions in the detector material are removed. The isolation requirement
is based on the energy sum of the PF candidates originating from the PV within a cone of $\DR < 0.3$ around the electron direction, excluding PF candidates associated to the electron or identified as muons.
The mean energy deposited in the isolation cone of the electron
from  pileup is estimated following the method described in Ref.~\cite{Khachatryan:2015hwa} and is subtracted from the isolation sum.
Two types of electron identification selection requirements are also used: ``loose'' and ``tight''.
The loose electrons are selected with an average efficiency of 95\% and up to 5\% misidentification rate.
The loose identification requirements are used in some of the analyses presented in this paper as part of selection requirements designed to remove backgrounds containing electrons \eg \Zee events.
The tight electrons are selected with an average efficiency of 70\% and an average misidentification rate of 1\%, and are used to select events used in the performance measurements (Section~\ref{sec:metperformanceinnomet}).

Muons within the geometrical acceptance $\abs{\eta} < 2.4$ are reconstructed by combining information from the silicon tracker and the muon system~\cite{Sirunyan:2018fpa}.
They are required to pass a set of quality criteria based on the number of spatial points measured in the tracker and in the muon system, the fit quality of the muon track and its consistency with the PV.
The isolation requirements for muons are based on the energy sum of the PF candidates originating from the PV within a cone of $\DR < 0.3$ around the muon direction, excluding PF candidates identified as electrons or muons.
The muon isolation variable is corrected for pileup effects from neutral particles by subtracting half of the \pt sum of the charged particles that are inside the isolation cone and not associated with the PV.
Two types of muon identification selection requirements are used: ``tight'' and ``loose''.
The tight muons are selected with an average efficiency of 95\% and are used to select the events analyzed in the performance measurement (Sections~\ref{sec:metperformanceinnomet} and~\ref{sec:metsignificance}), whereas the loose muons are selected with an average efficiency of 98\% and are used when appropriate to veto background events with additional muons. The \pt resolution for muons with
$20 <\pt < 100\GeV$ is 1\% in the barrel and
better than 3\% in the endcaps. The \pt resolution in the
barrel is better than 10\% for
muons with \pt up to 1\TeV~\cite{Sirunyan:2018fpa}.

Hadronically decaying \PGt lepton candidates detected within $\abs{\eta} < 2.3$ are required to pass identification criteria using the hadron-plus-strips algorithm~\cite{Khachatryan:2015dfa}.
The algorithm identifies a jet as a hadronically decaying \PGt lepton candidate if a subset of the particles assigned to the jet is consistent with the decay products of a \PGt candidate.
In addition, \PGt candidates are required to be isolated from the surrounding activity in the event.
The isolation requirement is computed by summing the \pt of the PF charged and PF photon candidates within an isolation cone of $\DR = 0.5$, around the \PGt candidate direction.
A more detailed description of the isolation requirement can be found in Ref.~\cite{Khachatryan:2015dfa}.

Jets are reconstructed by clustering PF candidates using the infrared- and collinear-safe anti-\kt algorithm~\cite{Cacciari:2008gp} with a distance parameter of 0.4.
To reduce the effect of pileup collisions, charged PF candidates that originate from pileup vertices are removed~\cite{Khachatryan:2016kdb} before the jet clustering.
The jet momentum is determined as the vector sum of all particle momenta in the jet, and is found from simulation to be within 5 to 10\% of the true momentum over the full \pt spectrum and detector acceptance.
An energy correction is applied to jet energies to subtract the contribution from pileup.
Jet energy corrections, are derived from simulation to adjust the measured jets based on a ratio of the average measured jets to the simulated average jets.
Measurements done in situ of the momentum balance in dijet, quantum chromodynamics (QCD) multijet, $\PGg+$jet, and leptonic $\PZ+$jet events are used to correct for any residual differences in jet energy scale (JES) in data and simulation~\cite{Khachatryan:2016kdb}.

Jets originating from the hadronization of bottom (\cPqb) quarks are identified (``tagged``) via a combined secondary vertex algorithm~\cite{BTV-16-002}.
The working point of this algorithm provides an average efficiency of $\sim$80\% for the identification of jets originating from \cPqb\ quarks whereas the misidentification rate for light quarks or gluons
is $\sim$10\%, and $\sim$40\% for charm quarks.

\section {Reconstruction and calibration of \texorpdfstring{\ptmiss}{pT miss}} \label{sec:reconstruction}
At hadron colliders, the reconstructed \ptmiss is a useful quantity because the net momentum in the plane transverse to the beam is known to be nearly zero from the initial conditions. Therefore, the total \pt of weakly interacting final-state particles can be inferred from the negative vector~\ptvec sum of all visible final-state particles. CMS event reconstruction employs two distinct \ptmiss\ reconstruction algorithms, described in the following, both based on PF candidates.

\subsection{The \texorpdfstring{\ptmiss}{pT miss} reconstruction algorithms} \label{sec:reconstruction_sub}
The first \ptmiss reconstruction algorithm, referred to as  PF \ptmiss in this paper, defines \ptvecmiss~as the negative vector~\pt sum of all the PF candidates in the event~\cite{2010,Khachatryan:2014gga}. The PF \ptmiss is used in the majority of CMS analyses, since it provides a simple, robust, yet very performant estimate of the \ptmiss reconstruction. A second algorithm has been developed to further reduce the dependence on pileup. This algorithm relies on the ``pileup per particle identification'' (\puppi) method~\cite{Bertolini:2014bba}, and uses local shape information around each PF candidate in the event, event pileup properties, and tracking information to reduce the pileup dependence of jet and \ptmiss observables.

The \puppi \ptmiss method employs a local shape variable $\alpha$,  which is sensitive to differences  between the collinear configuration of particles produced by the hadronization of quarks and gluons produced via QCD mechanisms and the soft diffuse radiation coming from pileup. The $\alpha$ variable is computed for each neutral particle, using the surrounding charged particles compatible with the PV within the tracker acceptance ($\abs{\eta} < 2.5$), and using both charged and neutral particles in the region outside of the tracker coverage. The momenta of the neutral particles are then rescaled according to the probability that they originate from the PV deduced from the local shape variable~\cite{Bertolini:2014bba}, superseding the need for jet-based pileup corrections~\cite{CMS-PAS-JME-16-003}.

In CMS, the \puppi algorithm is implemented using PF candidates. A different $\alpha$ definition is adopted for PF candidates within and outside the tracker acceptance. For a given PF candidate $i$, the $\alpha$ variable is defined as:
\begin{linenomath}
\begin{equation}
\label{eq:puppi}
 \alpha_i =    \log \sum_{j \neq i, \Delta R_{ij}<0.4 } \left(\frac{p_{\mathrm{T} j}}{\Delta R_{ij}}\right)^{2}
 \begin{cases}
   \text{for } |\eta _i | < 2.5, & j \text{ are charged PF candidates from PV} \\
   \text{for } |\eta _i | > 2.5, & j \text{ are all kinds of reconstructed PF candidates} \\
 \end{cases},
\end{equation}
\end{linenomath}
where $j$ refers to neighboring charged PF candidates originating from the PV within a cone of radius $R$ in $\eta$-$\phi$ space around $i$, and $\Delta R_{ij}$ is the distance in $\eta$-$\phi$ space between the $i$ and $j$ PF candidates. In addition, charged PF candidates not associated with the PV are used in the calculation if they satisfy $d_\text{z} < 0.3$\unit{cm}, where $d_\text{z}$ is the distance in z between the track and the PV. In the absence of tracking coverage, the $j$ in Eq.~(\ref{eq:puppi}) extends to all PF candidates within a cone of radius 0.4.

A $\chi^{2}$ approximation
\begin{linenomath}\begin{equation}
\label{eq:puppi_chi2}
\chi^{2}_{i} = \frac{(\alpha_i -  \overline{\alpha}_{\mathrm{PU}})^{2}}{\mathrm{RMS}_{\mathrm{PU}}^{2}},
\end{equation}\end{linenomath}
is used to determine the likelihood that a PF candidate came from pileup. In this equation, $\overline{\alpha}_{\mathrm{PU}}$ is the median value of the $\alpha_i$ distribution for pileup particles in the event (pileup PF candidates) in the event, and $\rm{RMS}_{\mathrm{PU}}$ is the corresponding root-mean-square ($\rm{RMS}$) of the $\alpha_{i}$ distribution. Within the tracker acceptance ($|\eta|<2.5$), the values of $\overline{\alpha}_{\mathrm{PU}}$ and $\rm{RMS}_{\mathrm{PU}}$ are calculated using all charged pileup PF candidates, and are $\sim3.5$. Outside the tracker acceptance, the $\overline{\alpha}_{\mathrm{PU}}$ and $\rm{RMS}_{\mathrm{PU}}$ are first estimated in the $|\eta|<2.5$ region and then, with the aid of simulation, are extrapolated in the forward region by means of transfer factors. We define two forward regions: $2.5 < \abs{\eta} < 3$ and $\abs{\eta} > 3$. The typical values of $\overline{\alpha}_{\mathrm{PU}}$ and $\rm{RMS}_{\mathrm{PU}}$ in the $2.5 < \abs{\eta} < 3$ region, are $\sim5.5$ and $\sim2.5$, respectively, whereas in the $\abs{\eta} > 3$ region, are $\sim4.5$ and $\sim2$, respectively.  The $\chi^{2}$ variable in Eq.~(\ref{eq:puppi_chi2}) is transformed to a weight using:
\begin{linenomath}\begin{equation}
\label{eq:puppi_wgt}
w_{i}=F_{\chi ^ 2,\text{NDF}=1}(\chi_{i}^{2}),
\end{equation}\end{linenomath}
where $F_{\chi ^{2},\text{NDF}=1}$ is the cumulative distribution function, which approximates the $\chi^{2}$ distribution  with one degree of freedom of all PF candidates in the event. The weights range from zero, for PF candidates originating from a pileup vertex, to close to one, for PF candidates originating from the PV. Charged PF candidates associated with the PV take the value of one. Once a weight per PF candidate is determined, the \ptmiss can be computed using the sum of PF candidate four-vectors weighted by their $w_i$. In addition, the \puppi-weighted PF candidates can be used as inputs to the jet clustering algorithm. No additional pileup corrections are applied to jets clustered from these weighted inputs. The results presented in this paper are based on jets without \puppi\ corrections applied.

The $w_{i}$ are required to be larger than 0.01 and the minimum scaled $\pt$ of neutral PF candidates is required to be $w_i \, p_\text{T ,i} > (A+B\, {N}_{\text{vtx}})$, where ${N}_{\text{vtx}}$ is the reconstructed vertex multiplicity. In this equation, A and B are adjustable parameters that depend on $\eta$.
An optimization of the tunable parameters to achieve the best jet $\pt$ and \ptmiss resolutions is performed separately for jets in the regions $\abs{\eta} < 2.5$,  $2.5 < \abs{\eta} < 3$, and $\abs{\eta} > 3$. The resulting algorithm parameters are similar to those recommended in Ref.~\cite{Bertolini:2014bba}, ranging from 0.2--2.0 and 0.015--0.8, for A and B, respectively.

\subsection{Calibration of \texorpdfstring{\ptmiss}{pT miss}} \label{sec:metcorrections}
Examples of sources that can lead to an inaccurate estimation of \ptmiss are the nonlinearity in the calorimeter response to hadrons, the minimum energy thresholds in the calorimeters, and the minimum \pt\ thresholds and inefficiencies in track reconstruction.
The estimation of \ptmiss is improved by propagating the correction of the \pt of the jets, $\vec{p}_\text{T, jet}^\text{corr}$, described in Ref.~\cite{Khachatryan:2016kdb}
to \ptmiss in the following way:
\begin{linenomath}\begin{equation}
 \ptvecmiss
=\vec{p}_\text{T}^\text{ miss, raw} - \sum_\text{jets} (\vec{p}_\text{T, jet}^\text{corr}-\vec{p}_\text{T, jet}),
\label{eq:Type1MET}
\end{equation}\end{linenomath}
where $\vec{p}_\text{T}^\text{~miss, raw}$ is the uncorrected \ptmiss. The sum is over jets with $\pt>15\GeV$.
The results in Section~\ref{sec:metperformanceinnomet} show that this choice for the jet \pt\ threshold reduces the contribution from jets from pileup interactions and gives a \ptmiss\ response close to unity.

The corresponding threshold for LHC Run 1, with lower pileup, was 10\GeV~\cite{2010,Khachatryan:2014gga}.
To remove the overlap of jets with electrons and photons, jets with more than 90\% of their energy associated to the ECAL are not included in the sum.
In addition, if a muon reconstructed using the outer tracking system overlaps with a jet, its four momentum is subtracted from the four momentum of the jet, and the JES correction~\cite{Khachatryan:2016kdb} appropriate for the modified jet momentum is used in the \ptmiss calculation.

The \ptmiss relies on the accurate measurement of the reconstructed physics objects, namely muons, electrons, photons, hadronically decaying taus, jets, and unclustered energy ($E_{U}$).
The $E_{U}$ is the contribution from the PF candidates not associated with any of the previous physics objects.
Uncertainties related to the \ptmiss measurement depend strongly on the event topology.
To estimate the uncertainty in \ptmiss, the uncertainty in the momenta of all reconstructed objects is propagated to  \ptmiss\ by varying the estimate of each PF candidate flavor within its uncertainty and recomputing \ptmiss.

The JES uncertainties are less than 3\% for jets within the tracker acceptance and 1--12\% for those outside. The jet energy resolution (JER) uncertainties typically range between 5--20\%.
The muon energy scale uncertainty is 0.2\%, and the electron and photon energy scale uncertainties are 0.6\% in the barrel and 1.5\% in the endcap.
For hadronically decaying \PGt leptons the energy scale uncertainty is 1.2\%. The uncertainties related to the leptons are small, compared to those from the JES and JER uncertainties, and are not considered in the results presented in this paper.

The uncertainty in the $E_{U}$ for LHC Run 1 was assessed as a uniform 10\%, and it accounted for the differences observed between the data and the simulation~\cite{Khachatryan:2014gga}. The method is improved for LHC Run 2.
The $E_{U}$ uncertainty is evaluated based on the momentum
resolution of each PF candidate, which depends on the type of the candidate.
A detailed description of the PF candidate calibration can be found in Refs.~\cite{Sirunyan:2017ulk,TRK-11-001,CMS:EGM-14-001}.  The \pt measurement for PF charged hadrons is dominated by the tracker resolution.
For PF neutral hadrons, the \pt resolution is dominated by the
resolution of the HCAL. The ECAL resolution dominates the PF photon \pt measurement, whereas HF intrinsic resolution dominates that for the PF particles in the HF. The largest contributions to the $E_{U}$ uncertainty are due to the PF neutral hadrons and PF candidates in the HF. Table~\ref{tab:unclust_unc} lists the functional forms of the resolutions
of the PF candidate classes contributing to the $E_{U}$.
\begin{table}[!ht]
\centering
\renewcommand{\arraystretch}{1.2}
\topcaption{\label{tab:unclust_unc} Functional forms of the resolutions in the \pt measurement for each PF candidate flavor contributing to the $E_{U}$~\cite{Sirunyan:2017ulk,TRK-11-001,CMS:EGM-14-001}. The mathematical symbol $\oplus$ indicates that the quantities are added in quadrature.}
\begin{tabular}{cc}  \hline
        PF candidate flavor & Resolution functions\\ \hline
        \setlength\extrarowheight{6pt}
        Charged hadron & \begin{math}  (0.00009 \, \pt)^{2} + (0.0085/ \sqrt{\sin  (2  \arctan(\re^{-\eta}) )} )^{2} \end{math}\\
        \setlength\extrarowheight{3pt}
        Neutral hadron ($\abs{\eta}<1.3$)      & $\min(0.25, (0.8/\pt) \oplus 0.05)$ \\
        Neutral hadron ($\abs{\eta} \geq 1.3$) & $\min(0.30, (1/\pt) \oplus 0.04)$ \\
                photon         & $(0.03/\pt) \oplus 0.001$ \\
        HF    & $(1./\pt)  \oplus 0.05$ \\ \hline
\end{tabular}
\end{table}
\renewcommand{\arraystretch}{1.}

\section{Simulated events}\label{sec:simulation}
For comparison with data, simulated Monte Carlo (MC) events are produced for $\PGg+$jet and QCD multijet processes at leading order (LO) using the \MGvATNLO 2.2.2~\cite{Alwall:2014hca}
generator with up to four additional partons in the matrix element calculations.
Samples for the $\PZ+$jets and $\PW+$jets processes are also produced at next-to-leading order (NLO) using the \MGvATNLO generator with up to two additional partons in the matrix element calculations.
The \ttbar and single top quark background processes are simulated at NLO using \POWHEG 2.0 and 1.0, respectively~\cite{Oleari:2010nx,Alioli:2009je}.
The diboson samples (\PW\PW, \PW\PZ , and $\PZ\PZ$) are simulated at NLO using \MGvATNLO and \POWHEG.
A set of triboson samples ($\PW\PW\PW$, $\PW\PW\PZ$, $\PW\PZ\PZ$, $\PZ\PZ\PZ$) is simulated at NLO using \MGvATNLO.
Lastly, the $\PZ\PGg$ and $\PW\PGg$ processes, collectively referred to as $\text{V}\PGg$ in the following, are simulated at LO with \MGvATNLO.

The MC samples produced using \MGvATNLO and \POWHEG generators are interfaced with
\PYTHIA 8.2~\cite{Sjostrand:2014zea} using the CUETP8M1 tune~\cite{Khachatryan:2015pea} for the fragmentation, hadronization, and underlying event description.
For the \MGvATNLO
samples, jets from the matrix element calculations are matched
to the parton shower following the MLM~\cite{Mangano:2006rw} (FxFx~\cite{Frederix:2012ps}) prescription
for LO (NLO) samples. The NNPDF3.0~\cite{Ball:2014uwa} parton distribution functions (PDFs) are used for all samples, with the order matching the matrix element calculations.
The simulation of the interactions
of all final-state particles with the CMS detector
is done
with \GEANTfour~\cite{Agostinelli:2002hh}. The simulated events are reconstructed using the same algorithms used for the data. The simulated events include the effects of pileup, with the number of additional pp interactions matching that observed in data. The average number of pileup interactions per proton bunch crossing is 23 for the data sample used in this analysis~\cite{CMS:2017sdi}.

\section{Event selection}\label{sec:simulation_selection}
In this paper, several final states are used to evaluate the performance of \ptmiss\ reconstruction algorithms.
Monojet and dijet samples are primarily used to study the performance of the algorithms developed to reject spurious events with anomalous \ptmiss, and are discussed in Section~\ref{sec:tails}.
Dilepton and single-photon samples are used to study the \ptmiss scale and resolution.
A single-lepton sample, which contains events with a genuine \ptmiss\ originating from a neutrino escaping without detection, is used to study the performance of the \ptmiss\ reconstruction algorithm.
Finally, the single-lepton and dilepton samples are also used to study the performance of the \ptmiss significance. The selection criteria used for each sample are discussed below.

\subsection{Monojet and dijet event samples}
\label{sec:dijetselection}
The events in the monojet sample are selected using triggers with requirements on both {\ptmisstrig} and \mhttrig, where {\ptmisstrig} is the magnitude of the vector \ptvec sum of all PF candidates reconstructed at the trigger level, and {\mhttrig} is  the magnitude of the vector \ptvec sum of jets with $\pt > 20\GeV$ and $\abs{\eta} < 5.0$ reconstructed at the trigger level.
Candidate events are required to have $\ptmiss > 250\GeV$, and the highest $p_\mathrm{T}$ (leading) jet in the event is required to have \pt$> 100\GeV$ and $\abs{\eta} < 2.4$.
The background from processes including $\PW$ bosons decaying leptonically is suppressed by imposing a veto on events containing one or more loose muons or electrons with \pt$>10\GeV$, or \PGt leptons with $\pt>18\GeV$.
Events that contain a loose, isolated photon with $\pt>15\GeV$ and $\abs{\eta} < 2.5$ are also vetoed. This helps suppress electroweak (EW) backgrounds with a photon radiated from an initial state parton.
To reduce the contamination from top quark backgrounds, events are rejected if they contain a b-tagged jet with $\pt > 20\GeV$ and $\abs{\eta} < 2.4$.

The QCD multijet background with \ptmiss arising from mismeasurements of jet momenta is suppressed by requiring the angle between the \ptvecmiss direction and each of the first four leading jets with $\pt >30\GeV$ is at least 0.5 radians.
This selection facilitates the study of sources that could lead to artificially large (``spurious'') \ptmiss due a malfunctioning detector (Section~\ref{sec:tails}).

The events in the dijet sample are also selected using
the {\ptmisstrig} and {\mhttrig} triggers. Candidate events are
required to have \ptmiss greater than 250\GeV and the leading (subleading) jet in the event is required to have \pt$> 500~(200)\GeV$. As for the monojet sample, events with an identified loose lepton, photon, or a b-tagged jet are rejected.

\subsection{Dilepton event samples}
\label{sec:zselection}
\begin{figure}[!htb]
  \centering
  \includegraphics[width=0.42\textwidth]{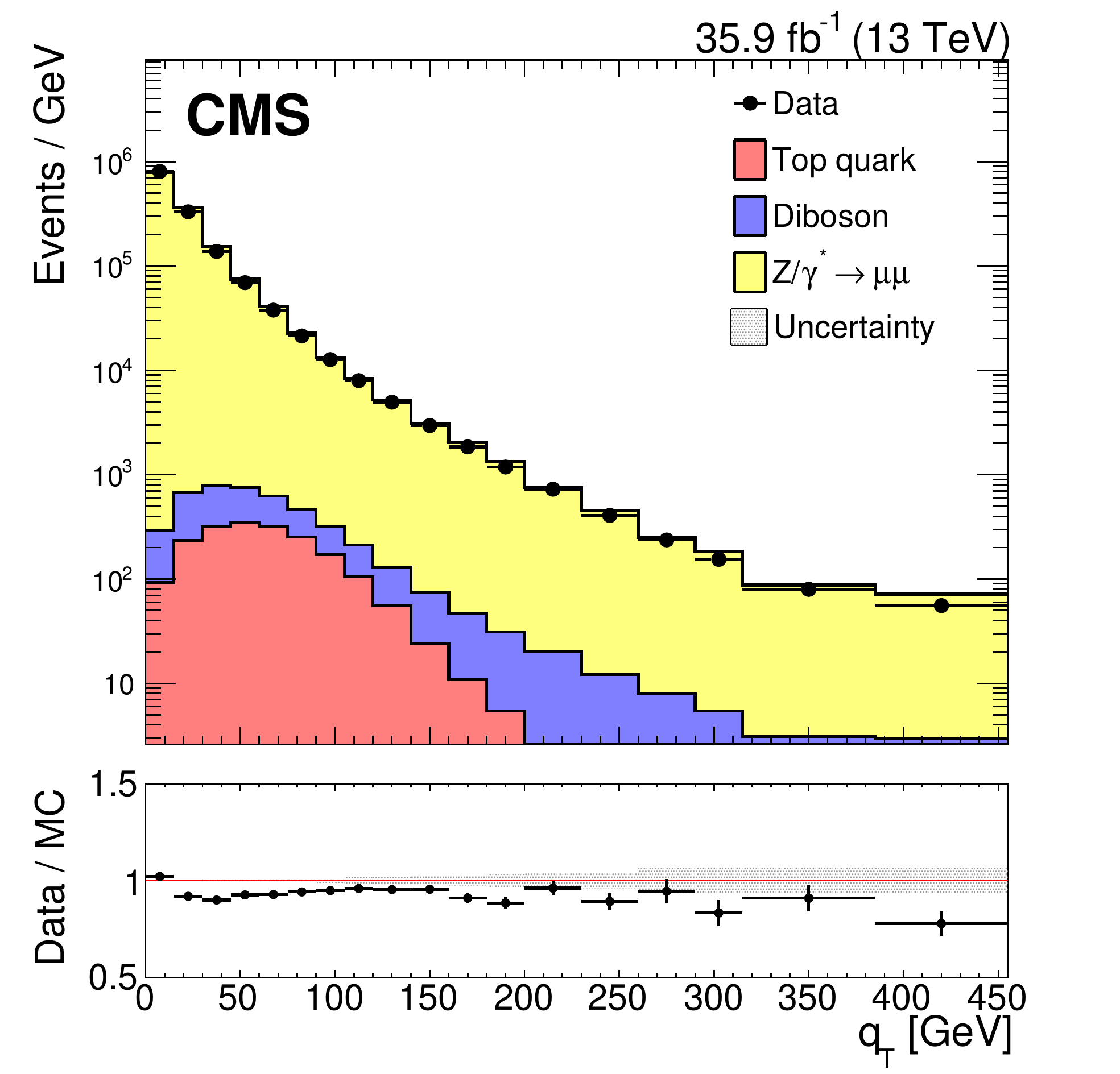}
  \includegraphics[width=0.42\textwidth]{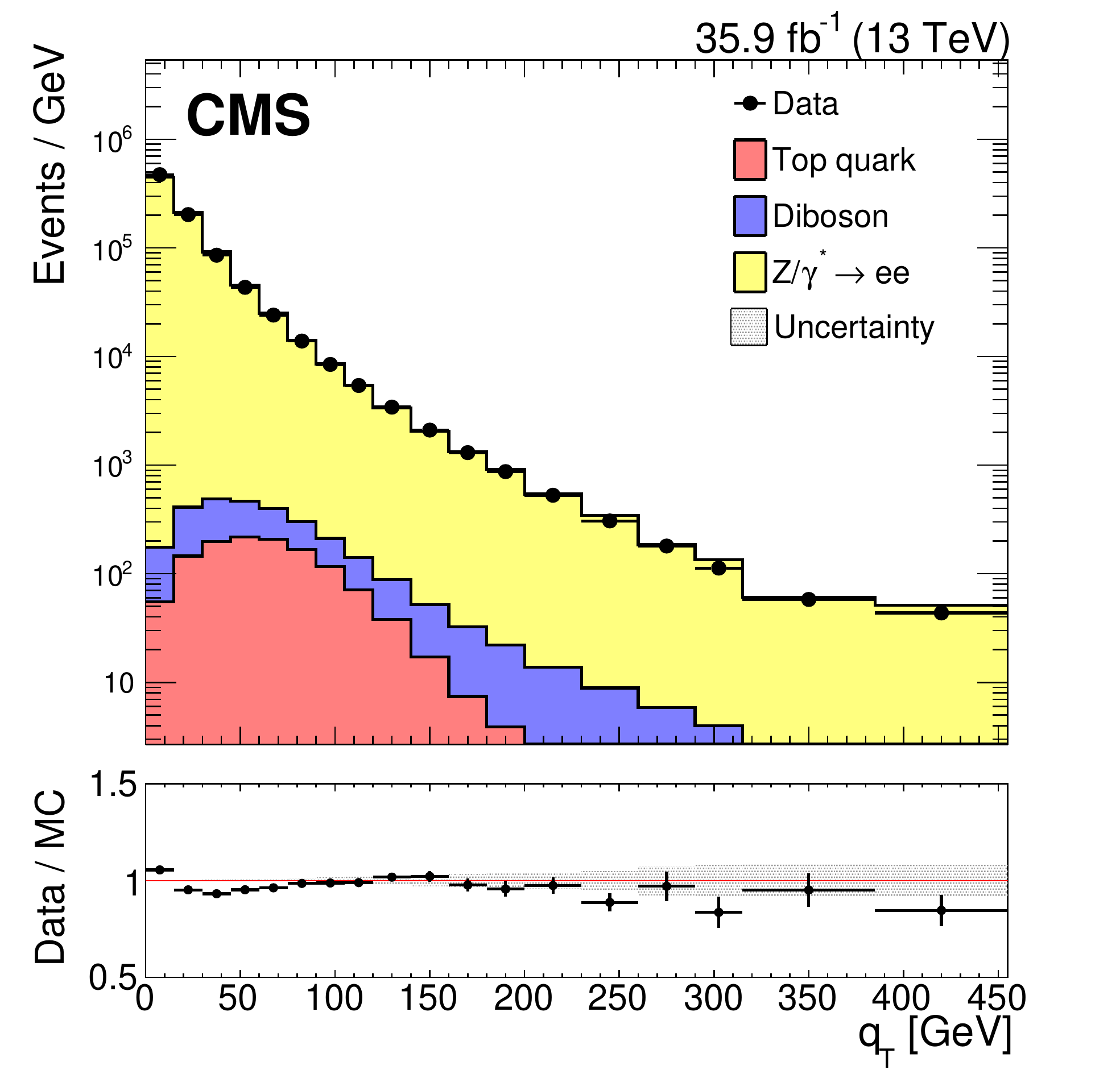}
  \caption{Upper panels: Distributions of \PZ boson \qt in \Zmm\ (left) and \Zee\ (right) samples.  The diboson contribution corresponds to processes with two electroweak bosons produced in the final state. The top quark contribution corresponds to the top pair and single top production processes. The last bin includes all events with $\qt>385\GeV$. Lower panel: Data to simulation ratio. The band corresponds to the statistical uncertainty in simulated samples. }
  \label{fig:zbosonpt}
\end{figure}

The dilepton samples are subdivided into two categories based on the flavor of the lepton, namely \Zmm\ and \Zee. The events for the \Zmm\ sample are recorded using dimuon triggers that select events where the \pt of each of the
two leading muons is above an asymmetric threshold.
Candidate events are required to have both the leading (subleading) muon \pt greater than 25 (20)\GeV and an invariant mass in the range of 80 to 100\GeV, compatible with the mass of the
Z boson~\cite{PhysRevD.98.030001}.
Events are vetoed if there is an additional muon or electron with $\pt > 20\GeV$. The events in the \Zee\ samples are recorded using dielectron triggers that have asymmetric selection requirements on
the \pt of the two leading electrons. Candidate events are required to have the leading (subleading) electron \pt greater than 25 (20)\GeV. As in the dimuon case, the invariant mass of the dielectron system is required to be in the range of 80 to 100\GeV. Events are vetoed if there is an additional muon or electron with $\pt > 20\GeV$. The spectrum of the \PZ boson transverse momentum, \qt, is shown in Fig.~\ref{fig:zbosonpt} where only the statistical uncertainty in the simulated samples is considered because the dilepton energy resolution is very good.

\subsection{Single-photon event sample}
\label{sec:photonselection}
The events in the single-photon sample are selected using a set of isolated single-photon triggers with varying thresholds. The \pt thresholds of the triggers
are 30, 50, 75, 90, 120, and 165\GeV. The first five of these triggers used different, luminosity dependent, L1 accept rates (prescales) during the data--taking periods. Candidate events are weighted based on the prescale values of the triggers.

Candidate events are required to
have a tight photon with $\pt > 50\GeV$. To match the trigger conditions, the leading photon is further required to have
the ratio of the energy deposited in a $3\times3$ crystal region of the ECAL, which is centered around the crystal containing an energy deposit greater than all of its immediate neighbors, to the energy of the entire deposit of the photon greater than 0.9.

\begin{figure}[!htb]
  \centering
  \includegraphics[width=0.42\textwidth]{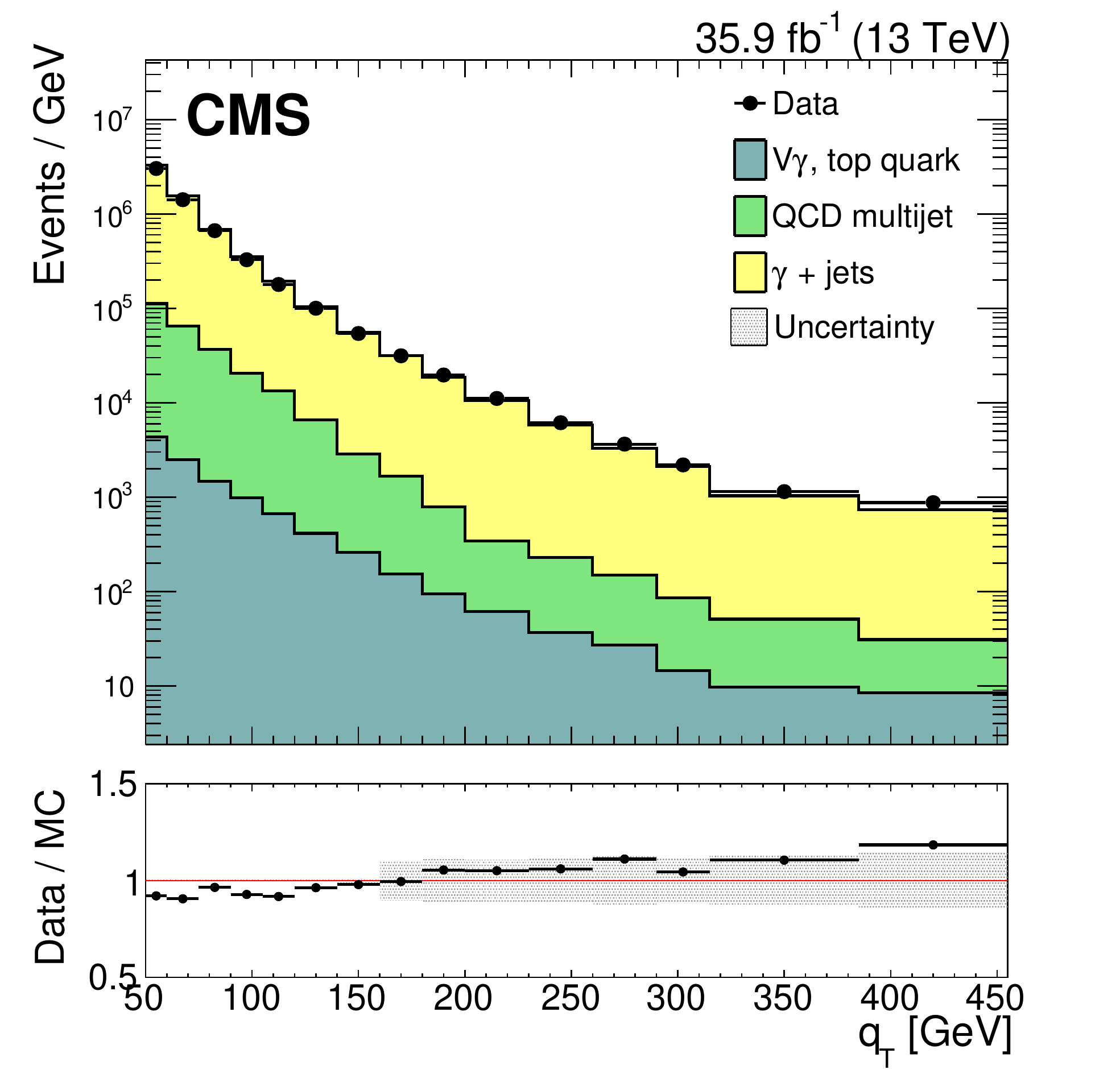}
  \caption{Upper panel: Distribution of the photon \qt in the single-photon sample. The V$\PGg$, top quark contribution corresponds to the $\PZ\PGg$, $\PW\PGg$, top pair and single top production processes. The last bin includes all events with $\qt>385\GeV$. Lower panel: Data to simulation ratio. The band corresponds to the statistical uncertainty in the simulated samples.  }
  \label{fig:gbosonpt}
\end{figure}

The single-photon sample events are also required to have at least one jet with \pt greater than 40\GeV, and events with leptons with \pt greater than 20\GeV are vetoed. The  photon \qt spectrum is shown in Fig.~\ref{fig:gbosonpt}. As in Fig.~\ref{fig:zbosonpt}, only the statistical uncertainty in the simulated samples is considered because the photon energy resolution is very good.

\subsection{Single-lepton event samples}
\label{sec:onelepselection}

The single-lepton samples are subdivided into two categories based on the flavor of the lepton.
These events in the single-muon (single-electron) sample are selected using triggers based on the \pt and the isolation of the muon (electron).
Candidate events are required to have a tight muon (electron) with \pt greater than 25 (26)\GeV. Events with an additional lepton with \pt greater than 10\GeV, or with a b-tagged jet, are rejected.

These single-lepton samples consist mainly of $\PW+$jets events. One source of background stems from QCD multijet events containing a jet misidentified as a lepton.
The simulation indicates that the magnitude of this background is small. However, since the uncertainties in simulating this background can be significant, we use a data control region to estimate it. The data control sample is selected by inverting the requirement on the relative isolation of the lepton and is dominated by QCD multijet events. The normalization of this background is then corrected by comparing the observed and expected number of events in the data control sample. Other processes are estimated from simulation.

\begin{figure}[!htb]
  \centering
  \includegraphics[width=0.42\textwidth]{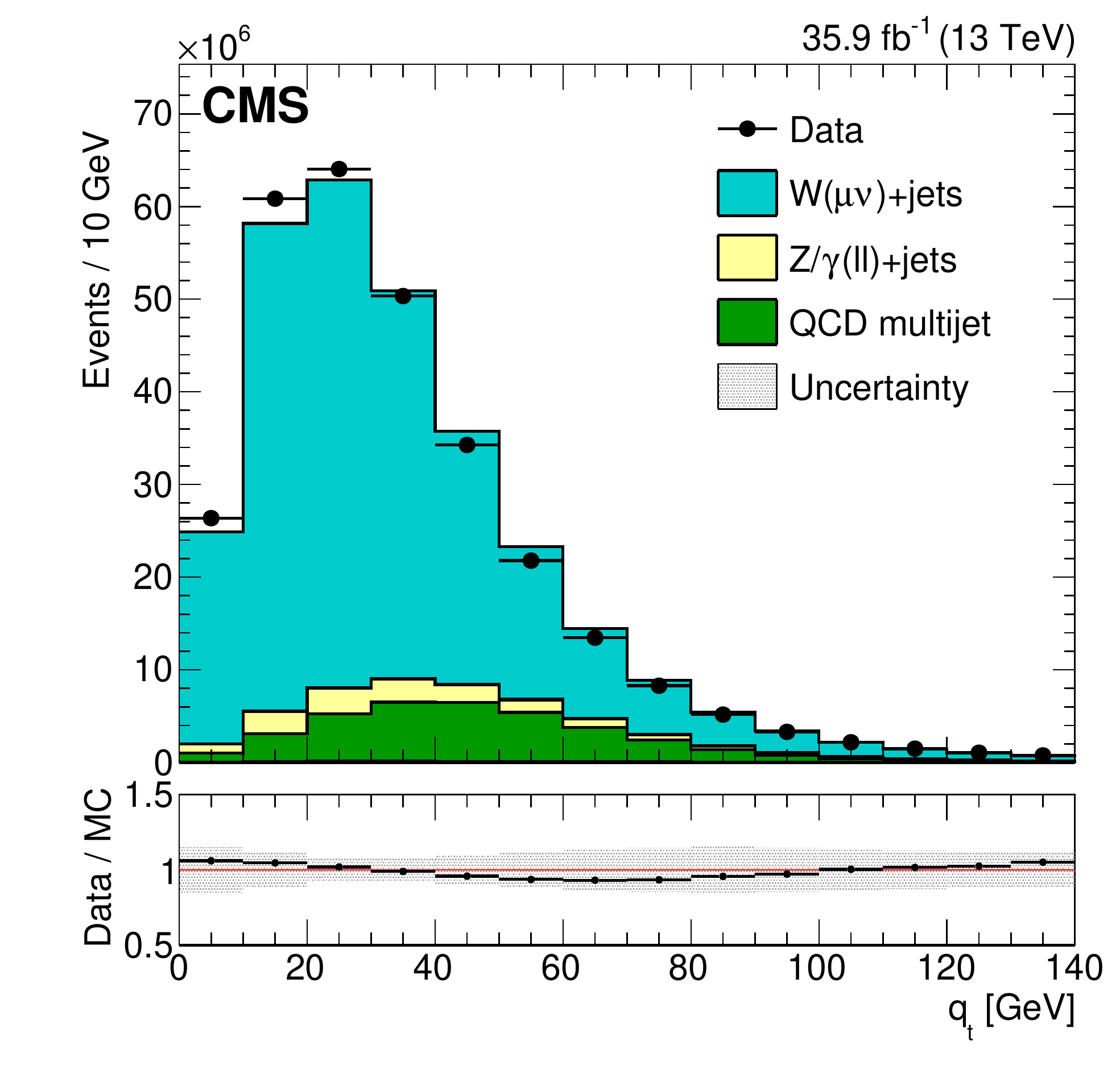}
  \includegraphics[width=0.42\textwidth]{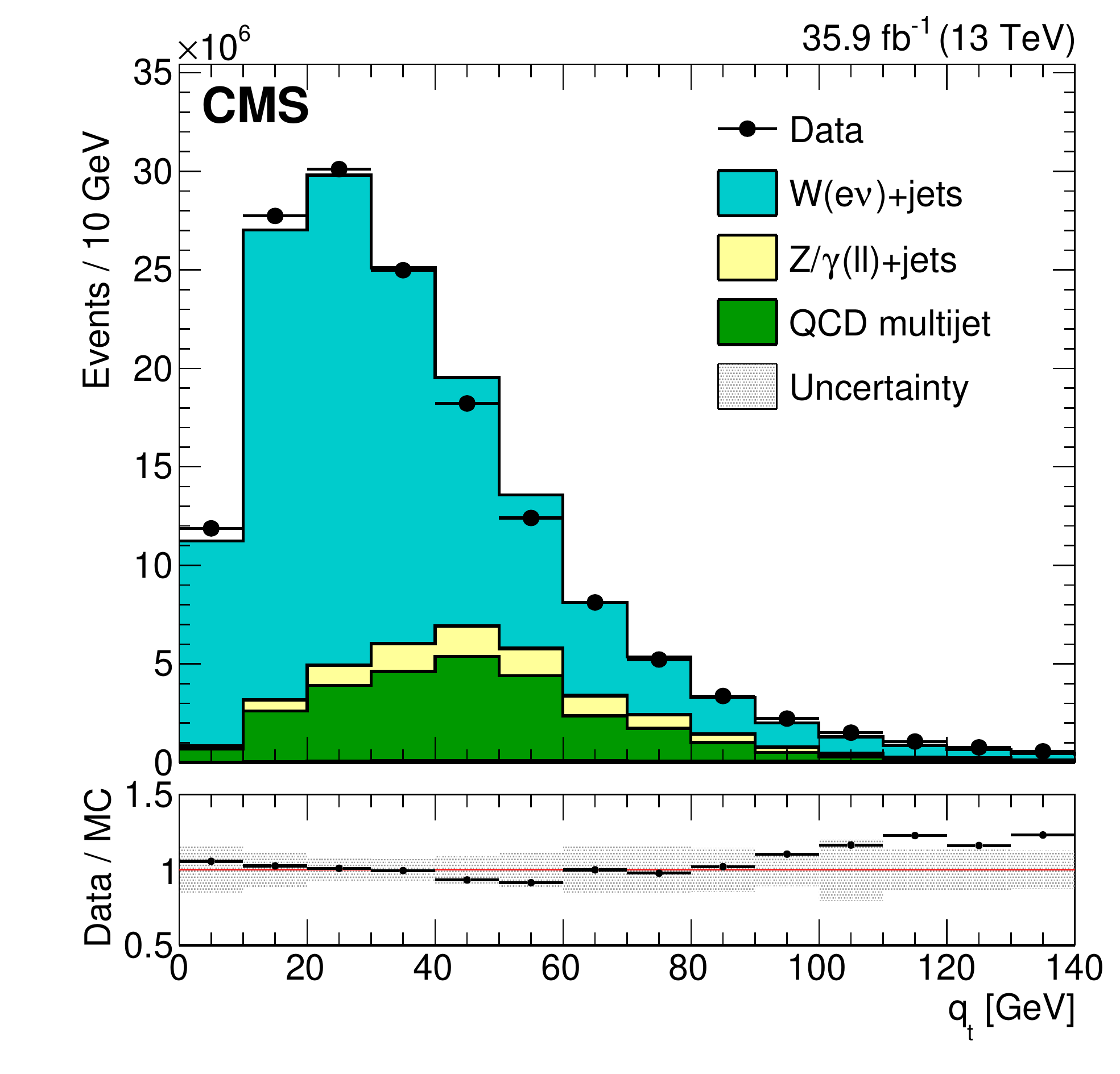}
  \caption{Upper panels: Distributions of $\PW$ boson \qt in single-muon (left) and single-electron (right) samples. The last bin includes all events with $\qt>130\GeV$.
Lower panels: Data to simulation ratio. The systematic uncertainties due to the JES, the JER, and variations in the $E_{U}$ are added in quadrature and displayed with a band. }
  \label{fig:wbosonpt}
\end{figure}

The spectrum of the $\PW$ boson transverse momentum \qt is shown in Fig.~\ref{fig:wbosonpt}.
In contrast to Figs.~\ref{fig:zbosonpt} and~\ref{fig:gbosonpt}, the effects of the systematic uncertainties from the JES, JER, and $E_{U}$ are sizable and are included in addition to the systematic uncertainty from the limited statistics in the simulated samples.

\section{Anomalous \texorpdfstring{\ptmiss}{pT miss} events}
\label{sec:tails}
Anomalous high-\ptmiss events can arise because of a variety of reconstruction failures or malfunctioning detectors.
In the ECAL, spurious deposits may appear due to noisy sensors in the ECAL  photodetectors, or from genuine showers with noncollision origins, such as those caused by the production of muons when beam protons undergo collisions upstream of the detector (beam halo).
An additional source of artificial \ptmiss is the presence of dead cells, leading to underestimation of the energy.
In the HCAL, spurious energy can arise from  noise in the hybrid photodiode (HPD) and in the readout box (RBX) electronics, as well as from direct particle interactions with  the light guides and photomultiplier tubes of the HF.
These sources have been studied extensively in the data collected in LHC Run 1~\cite{2010,Khachatryan:2014gga}.
Algorithms (filters) developed during LHC Run 1 to identify and suppress events with anomalously high \ptmiss are also used for this data (LHC Run 2) with the necessary modifications  for the upgraded detector~\cite{Collaboration:1355706} and the different data--taking conditions. An additional set of filters was also developed during this run to identify new sources of artificial \ptmiss. Details of the various filters are given below.
\begin{itemize}
\item {\bf HCAL filters}

The geometrical patterns of HPD or RBX channels as well as the pulse shape  and timing information are used by various HCAL barrel and endcap (HBHE) algorithms to identify and eliminate noise. These filter algorithms operate both in ``noise filtering'' and ``event filtering'' modes.
In the noise filtering mode,  the anomalous energy deposits are removed
from the event reconstruction;
in the filtering mode, the event is removed from the data set.
In addition, there is
an isolation-based noise filter that utilizes a
topological algorithm, where energy deposits in HCAL and ECAL are combined
and compared with measurements from the tracker to identify isolated
anomalous activity in HB/HE.
An additional noise filter based on pulse shapes uses information at the cluster reconstruction level and searches for uncharacteristic noise signals in the HB/HE HPD channels.  It relies on the known pulse shapes of HPDs, and is similar the RBX pulse shape filters~\cite{Chatrchyan:2009hy}, but explicitly corrects for the presence of in-time and out-of-time pileup when testing for anomalous pulse shapes.

\item {\bf ECAL filters}

For the ECAL, much of the electronics noise and spurious signals from particle interactions with the photodetectors is removed
during reconstruction using  the
topological and timing information.
The remaining effects that lead to high-\ptmiss signatures, such as
anomalously high energy deposits in supercrystals, and the lack of information for channels
that have nonfunctioning readout electronics,
are removed through dedicated noise filters.

During this data--taking run (LHC Run 2), five ECAL endcap supercrystals
produced large, anomalous pulses,
leading to spurious \ptmiss. These crystals are
removed from the readout, and their energies are not considered.
Furthermore, in about 0.7\% of ECAL towers (i.e. 5$\times$5 ECAL crystals),
the crystal-by-crystal information is not available.
The trigger primitive (TP)~\cite{Khachatryan:2016bia} information,
however, is still available, and is used to estimate the energy.
The TP information saturates above 127.5\GeV.
Events with a TP close to saturation in any of these ECAL towers are removed.

\item {\bf Beam halo filter}

Machine-induced backgrounds, especially beam halo, can cause
anomalously large \ptmiss.
Beam halo particles travel nearly parallel to the collision axis
and can sometimes interact in the calorimeters, leaving
energy deposits along a line with constant $\phi$.
In addition, interactions in the CSC,
a subdetector with good reconstruction performance for
both collision and noncollision muons, will often be in
line with the calorimeter deposits.
The beam halo filter was redesigned for LHC Run 2.
In LHC Run 1 the filter was based solely on information from the CSC.
However, the LHC Run 2 filter exploits information
from both the CSC and the calorimeters, resulting in a
significant improvement in performance.

\item {\bf Reconstruction filters}

An additional source of anomalous high-\ptmiss events during LHC Run 2
was poor reconstruction  of muons
during the muon-tracking iteration step~\cite{Sirunyan:2018fpa}.
If a high-\pt track has a low quality reconstruction,
it could contribute to \ptmiss
either as a poorly reconstructed PF muon,
or as a poorly reconstructed PF charged hadron.
The poorly reconstructed muons and charged hadrons are identified based on the ratio of the relative \pt uncertainty of the track \pt, determined by the Tune-P algorithm~\cite{Sirunyan:2018fpa}, or the inner track \pt. Once a poorly reconstructed muon or a charged hadron is identified, dedicated filters are designed to reject these events.
\end{itemize}

\begin{figure}[!htp]
  \centering
  \includegraphics[width=0.42\textwidth]{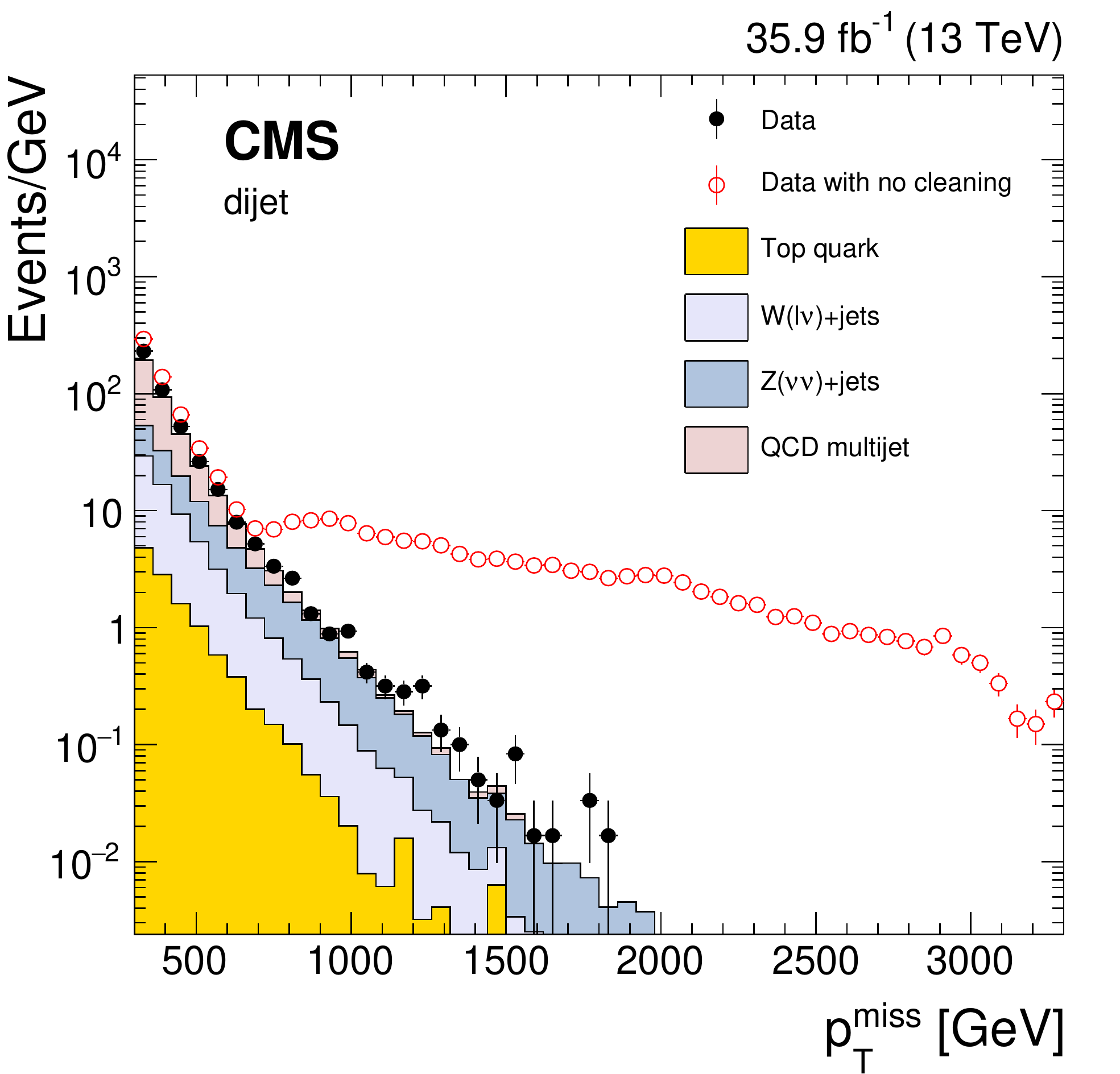}
  \includegraphics[width=0.42\textwidth]{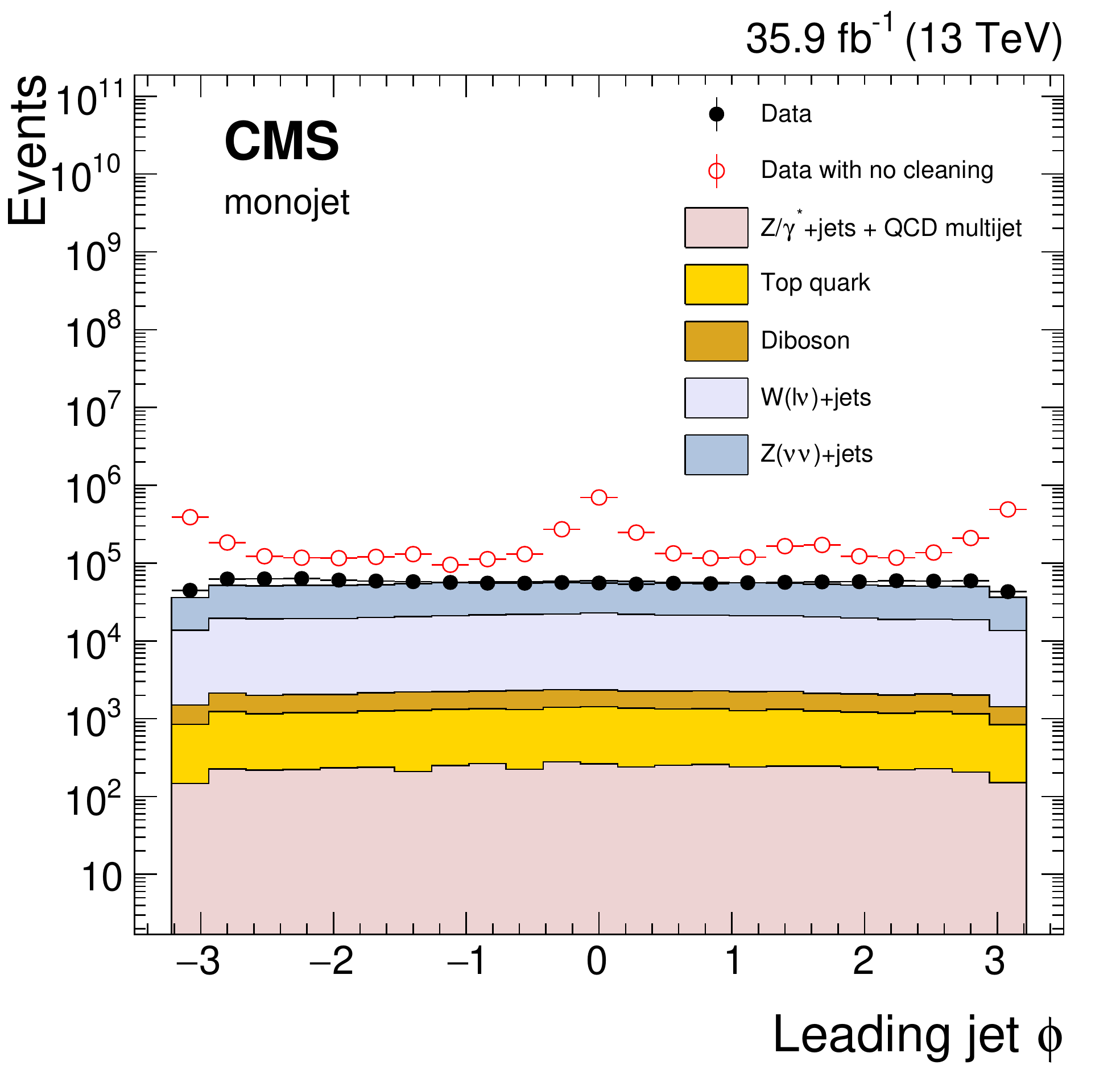}
  \caption{The \ptmiss (left) and jet $\phi$ (right) distributions for events passing the dijet (left) and monojet (right) selection with the event filtering algorithms applied, including that based on jet
identification requirements (filled markers), without the     event filtering algorithms applied (open markers), and from simulation (solid histograms). }
  \label{fig:dijet}
\end{figure}

Figure~\ref{fig:dijet} shows a comparison of the \ptmiss (left) and jet $\phi$ (right) distributions before and after the application of the event filters for the dijet and monojet samples, respectively. The anomalous events with large \ptmiss in the dijet sample are mostly due to electronic noise in the calorimeters.
The jet $\phi$ distribution in the monojet sample is used to validate the performance of the beam halo filter. The angular distribution of beam halo events is dictated by the shape of the LHC tunnel and the beamline elements~\cite{Tosi:2017nzv} and results in an excess of events with jet $\phi\approx 0$ or $\phi\approx\pi$. These events are removed  by the beam halo filter.
In both samples, the simulated \ptmiss and jet $\phi$ distributions are  in good agreement with data after the application of all the filters.
The event filters are designed to identify more than 85--90\% of the spurious high-\ptmiss events with a mistag rate of less than 0.1\%. In addition to the event filtering algorithms, a jet identification selection is imposed, which requires the neutral hadron energy fraction of a jet be less than 0.9. This selection rejects more than 99\% of the noise jets, independent of jet
\pt, with a negligible mistag rate.

\section{Performance of \texorpdfstring{\ptmiss}{pT miss} reconstruction at the trigger level}
\label{sec:trigger}
At L1, \ptmiss is computed at the global calorimeter trigger (GCT) level~\cite{Khachatryan:2016bia}, which is the last stage of the L1 calorimeter trigger chain.
The trigger-level quantities computed by the GCT use data from the regional calorimeter trigger (RCT)~\cite{Khachatryan:2016bia}, which receives the transverse energies, \et, and quality flags from ECAL and HCAL.
At GCT level, the \ptmiss is calculated by summing the regional transverse energy values and rotating the resulting vector by 180$^{\circ}$.
A more detailed description can be found in~\cite{Khachatryan:2016bia}.
Although the RCT coverage could be extended to $\abs{\eta}$ of 5.0, the \ptmiss algorithm at L1 only uses information from trigger towers within $\abs{\eta}<3.0$, due to the bandwidth restrictions of the trigger system.

Two reconstruction algorithms are used at the HLT.  A \ptmiss variable using only information from the calorimeters (Calo \ptmiss) is used as a prefilter to a more complex, PF-based \ptmiss reconstruction.
The Calo \ptmiss is computed by taking the negative vector~ \et\ sum of all calorimeter towers, whereas PF \ptmiss is based on the negative vector \pt sum of all reconstructed PF jets without a \pt requirement, as in  the case of the offline reconstruction algorithms.

To maintain the lowest possible thresholds for the \ptmiss triggers, event filtering algorithms are applied at the trigger level.
In contrast to the offline case, at the trigger level the calorimeter energy deposits flagged as being consistent either with HB/HE noise or beam halo are removed from the energy sum, and \ptmiss is recomputed.
The noise filtering algorithms used at the HLT are fully efficient with respect to the offline filtering algorithms, and reduce the rate of \ptmiss triggers by up to a factor of 2.5, depending on the \ptmiss\ threshold.

\begin{figure}[th!]
  \centering
  \includegraphics[width=0.42\textwidth]{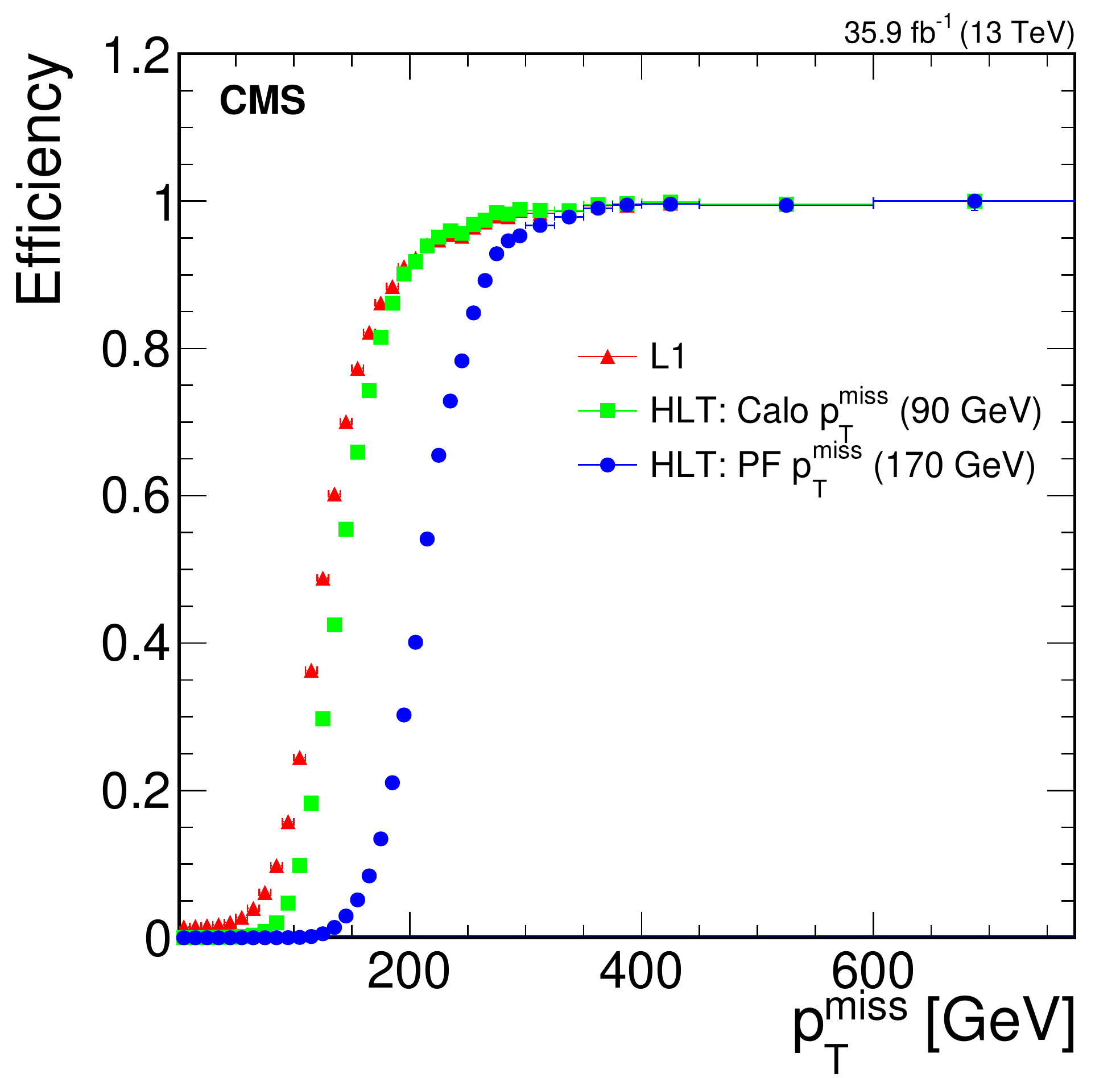}
  \caption{The \ptmiss trigger efficiency as a function of offline \ptmiss, measured using a single-electron sample. The efficiency of each reconstruction algorithm, namely the L1, the calorimeter, and the PF-based \ptmiss algorithms, is shown separately. The numbers in parentheses correspond to the HLT \ptmiss thresholds. The logical OR of the L1 \ptmiss triggers with requirements on \ptmiss\ greater than 50, 60, 70, 80, 90, 100 and 120\GeV are used. }
  \label{fig:trigger_eff}
\end{figure}

As with the offline reconstruction, HLT PF \ptmiss is calibrated by correcting the \pt of the jets using the jet energy corrections.
In contrast to the offline calibration, the corrections for the jets are only propagated to the \ptmiss if the jet \pt is above 35\GeV.
The performance of the \ptmiss triggers is measured in single-electron samples.
The efficiency for each trigger-level \ptmiss\ object type is shown in Fig.~\ref{fig:trigger_eff}.
The calibrated \ptmiss at the HLT level yields an improved efficiency at lower \pt.
As a result, online trigger thresholds are set to higher values, typically $\gtrsim$170\GeV, yielding the same performance offline, for up to 10\% rate reduction depending on the \ptmiss\ threshold.

\section{Performance of \texorpdfstring{\ptmiss}{pT miss} algorithms}
\label{sec:metperformanceinnomet}
\begin{figure}[!htb]
\vskip3ex
\centerline{
\includegraphics[width=0.4\textwidth]{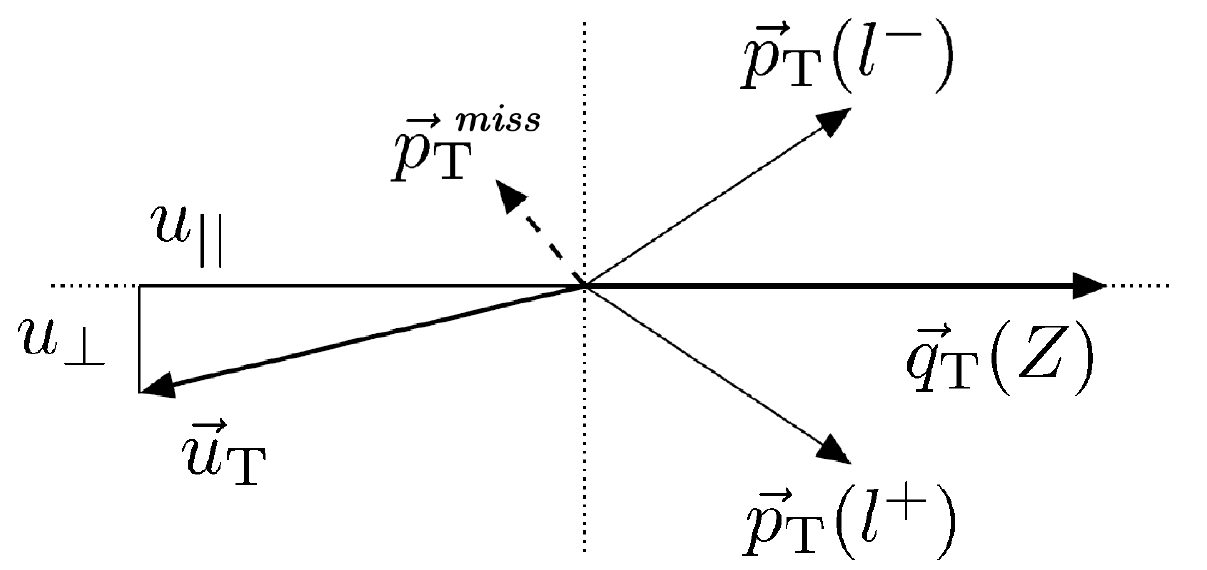}
~~~~~~~~~
\includegraphics[width=0.4\textwidth]{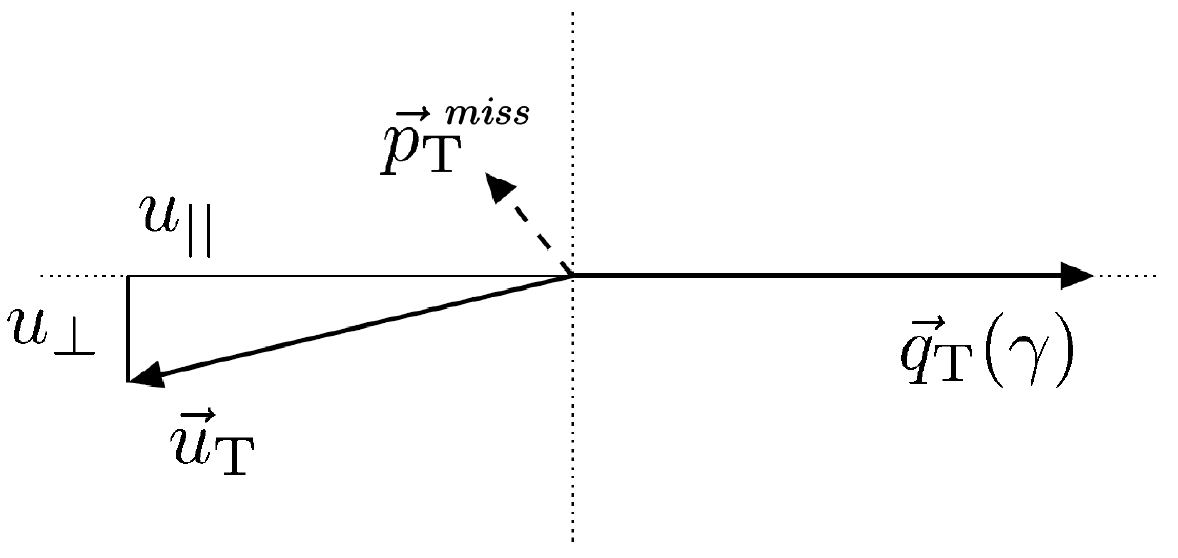}}
\setlength{\abovecaptionskip}{18pt}
\caption{Illustration of the \PZ boson (left) and photon (right) event kinematics in the transverse plane.
The vector \vut denotes the vectorial sum of all particles reconstructed in the event
except for the two leptons from the \PZ decay (left) or the photon (right).}
\label{fig:recoilDef}
\end{figure}

A well-measured \PZ/{\cPgg} boson provides a unique event axis and a precise momentum scale.
To this end, the response and resolution of \ptmiss is studied in samples with an identified \PZ boson decaying to a pair of electrons or muons, or with an isolated photon.
Such events should have little or no genuine \ptmiss, and the performance is measured by comparing the momenta of the vector boson to that of the hadronic recoil system.
The hadronic recoil system is defined as the vector \pt sum of all PF candidates except for the vector boson (or its decay products in the case of the \PZ boson decay).
In Fig.~\ref{fig:recoilDef} the kinematic representations of the transverse momenta of the vector boson and the hadronic recoil, \vqt and \vut, are shown.
Momentum conservation in the transverse plane imposes $\vqt+\vut+\ptvecmiss = 0$.

The components of the hadronic recoil parallel and perpendicular to the boson axis are denoted by \upar\ and \uperp, respectively.
These are used to study the \ptmiss response and resolution.
Specifically, the mean of the distribution of the magnitude of $\vec{u}_{||}+\vec{q}_{\perp}$, denoted as \redupara, is used to estimate the \ptmiss\ response, whereas the $\rm{RMS}$ of the \redupara\ and \uperp\ distributions are used to estimate the resolution of \upar\ and \uperp, denoted by $\sigma(\upar)$ and $\sigma(\uperp)$, respectively. The response of \ptmiss is defined as $-\langle \upar \rangle /\langle \qt \rangle$ where $\langle\, \rangle$ indicates the mean of the distributions.

An alternative method insensitive to tails in the distributions is also used.
The \redupara and \uperp\ are parametrized using a Voigtian function, defined as the convolution of a Breit--Wigner and a Gaussian distribution.
The results obtained with the alternative method agree within 2\% with those obtained using the primary method (\ie, mean/$\rm{RMS}$), indicating that the effect of the nonGaussian tails on the \ptmiss\ performance is small. In the following sections, the performance of the PF and \puppi \ptmiss algorithms is shown using the primary method.

\subsection{Performance of the PF \texorpdfstring{\ptmiss}{pT miss} algorithm}

\begin{figure}[!htbp]
  \centering
  \includegraphics[width=0.42\textwidth]{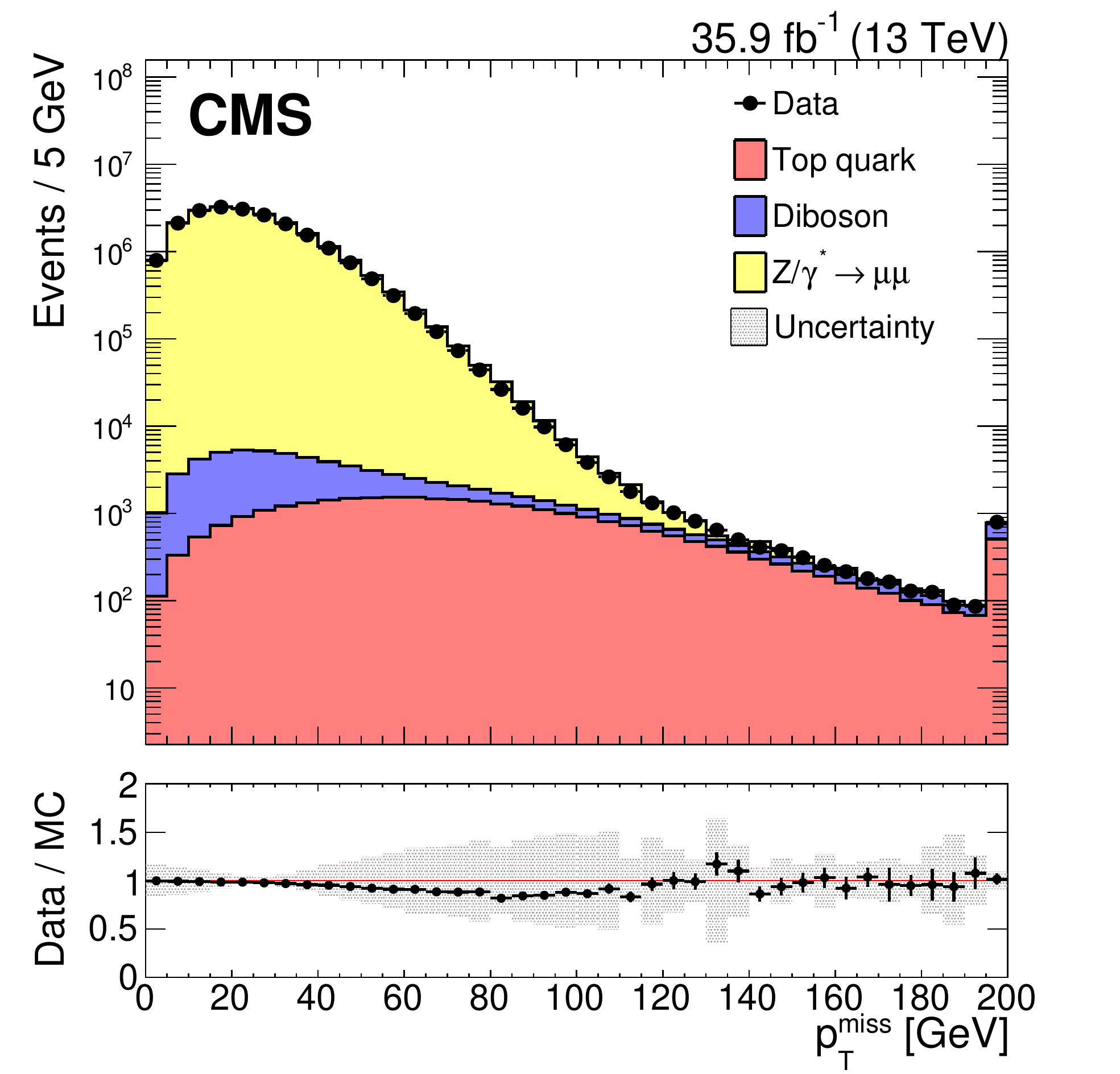}
  \includegraphics[width=0.42\textwidth]{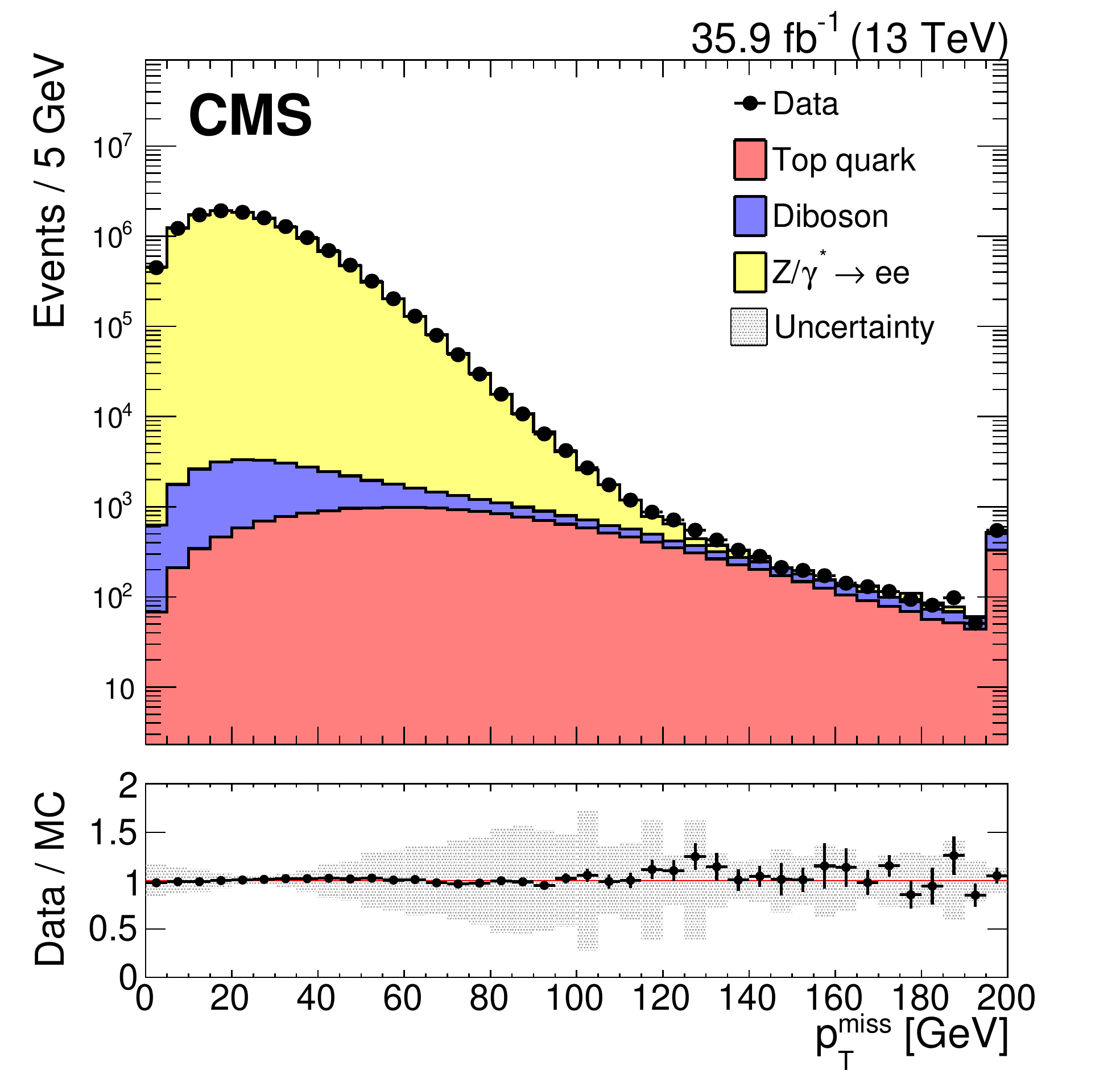}\\
  \includegraphics[width=0.42\textwidth]{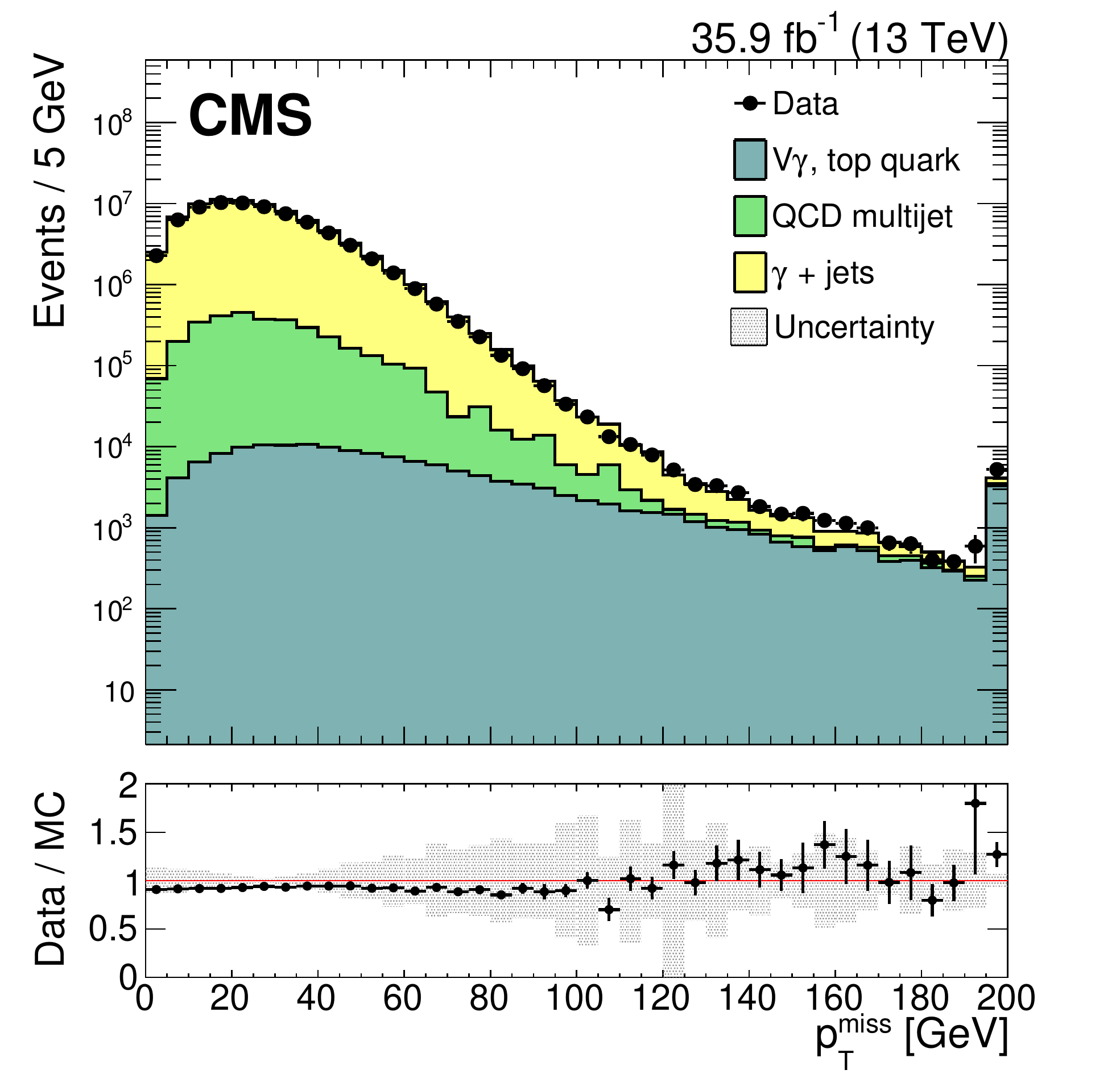}
  \caption{Upper panel: Distributions of \ptmiss in \Zmm\ (top left), \Zee\ (top right), and $\PGg+$jets events (lower middle) in data and simulation. The last bin includes all events with $\ptmiss > 195\GeV$.
Lower panel: Data to simulation ratio. The systematic uncertainties due to the JES, the JER, and variations in the $E_{U}$ are added in quadrature and represented by the shaded band.}
  \label{fig:pfmet}
\end{figure}

The PF \ptmiss distributions in dilepton and photon samples are shown in Fig.~\ref{fig:pfmet}. The data distributions are modeled well by the simulation.

The \ptmiss resolution in these events is dominated by the resolution of the hadronic activity, since the momentum resolution for leptons  and photons is $\sigma_{\pt}/\pt \lesssim 1.5\%$~\cite{CMS:EGM-14-001,Sirunyan:2018fpa}, compared to  5--20\% for the jet momentum resolution~\cite{Khachatryan:2016kdb}. The uncertainty shown in the figures includes uncertainties in the JES, the JER,
and the energy scale of unclustered particles, added in quadrature. The increase in the uncertainty band around 40\GeV is related to the JES and the JER sources in events with at least one jet and no genuine \ptmiss. For higher values of \ptmiss, where processes with genuine \ptmiss, \eg, top quark background, are present, the uncertainty is somewhat smaller.

\begin{figure}[!phtb]
  \centering
  \includegraphics[width=0.42\textwidth]{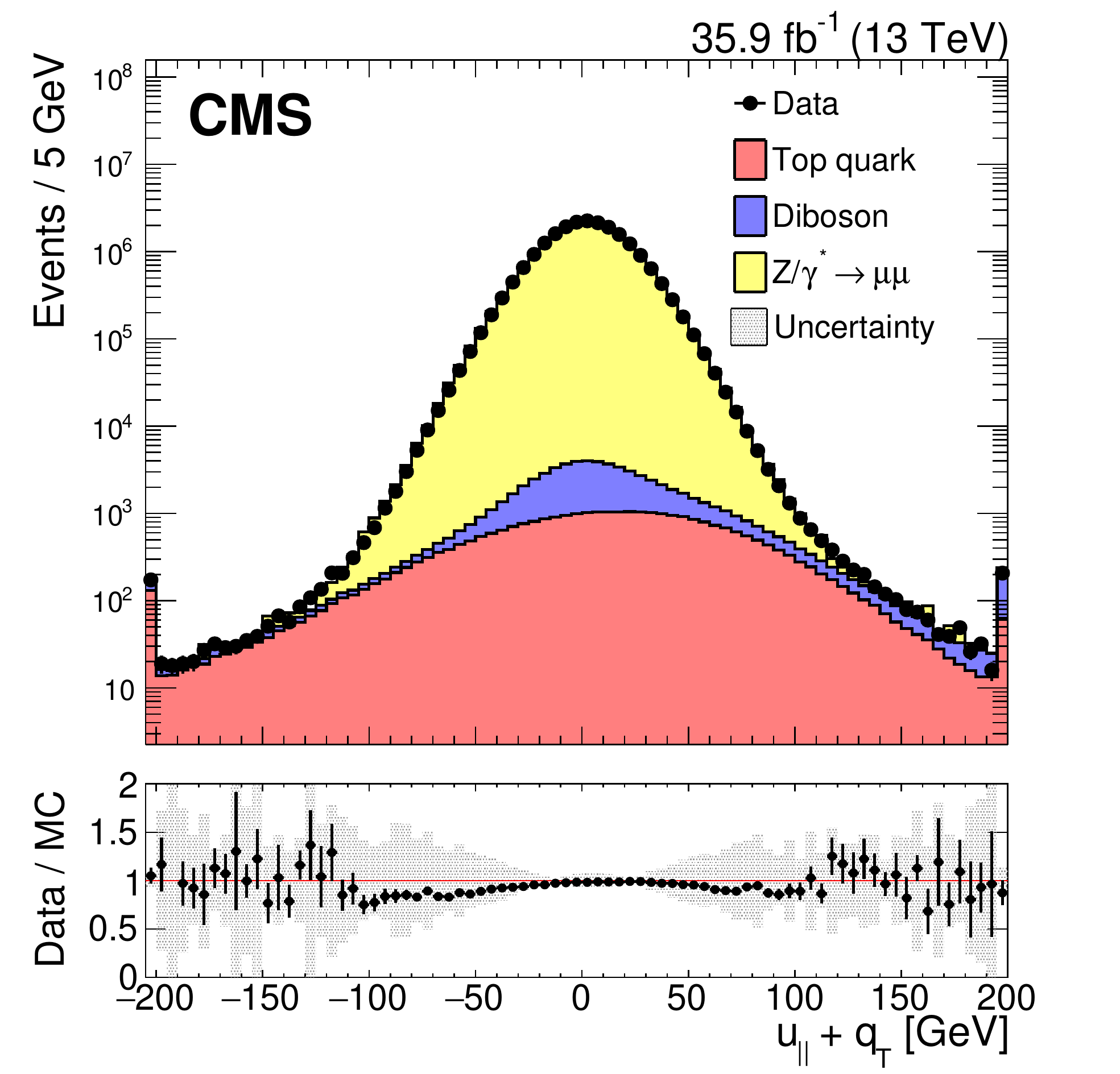}
  \includegraphics[width=0.42\textwidth]{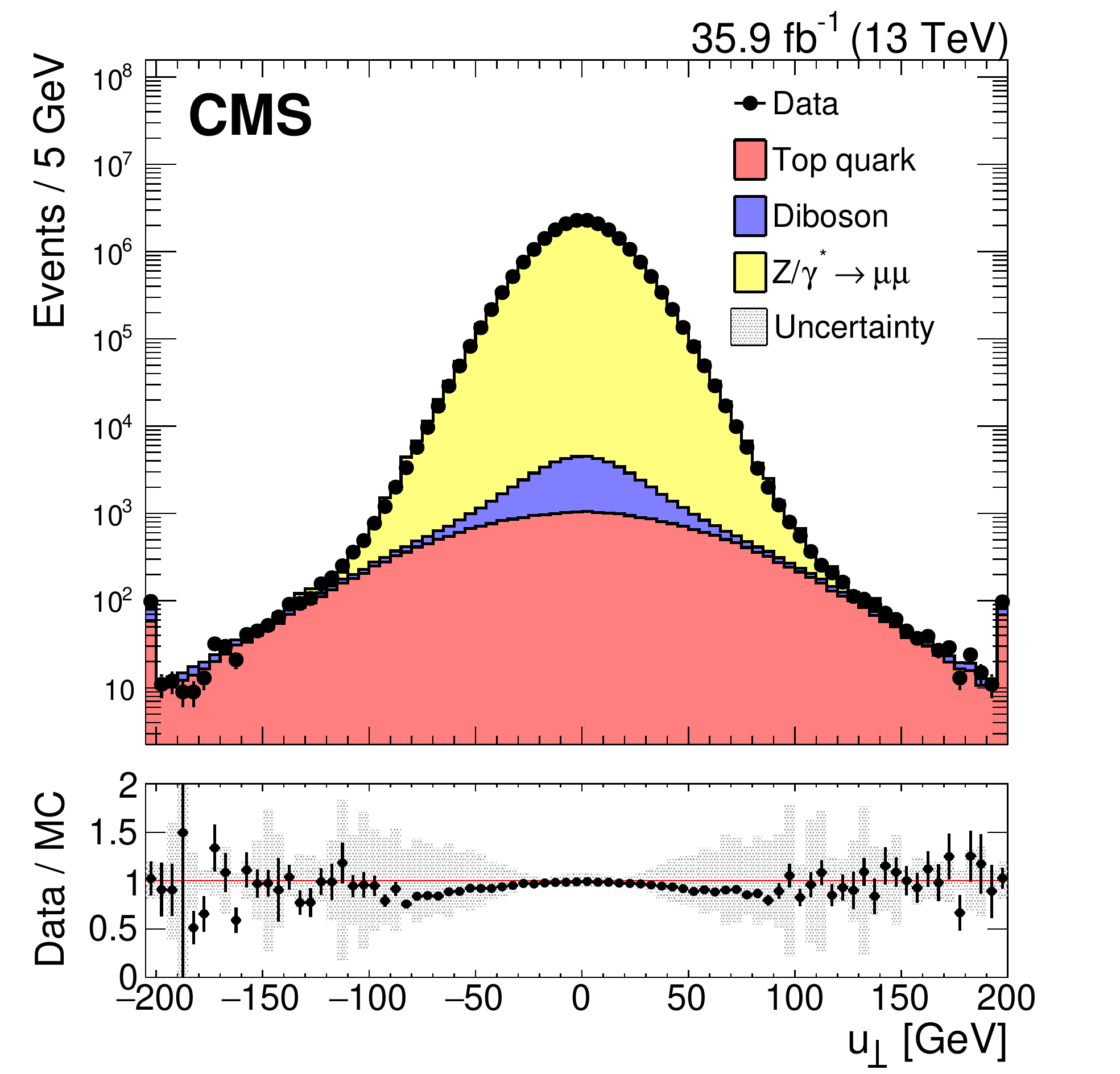}
  \includegraphics[width=0.42\textwidth]{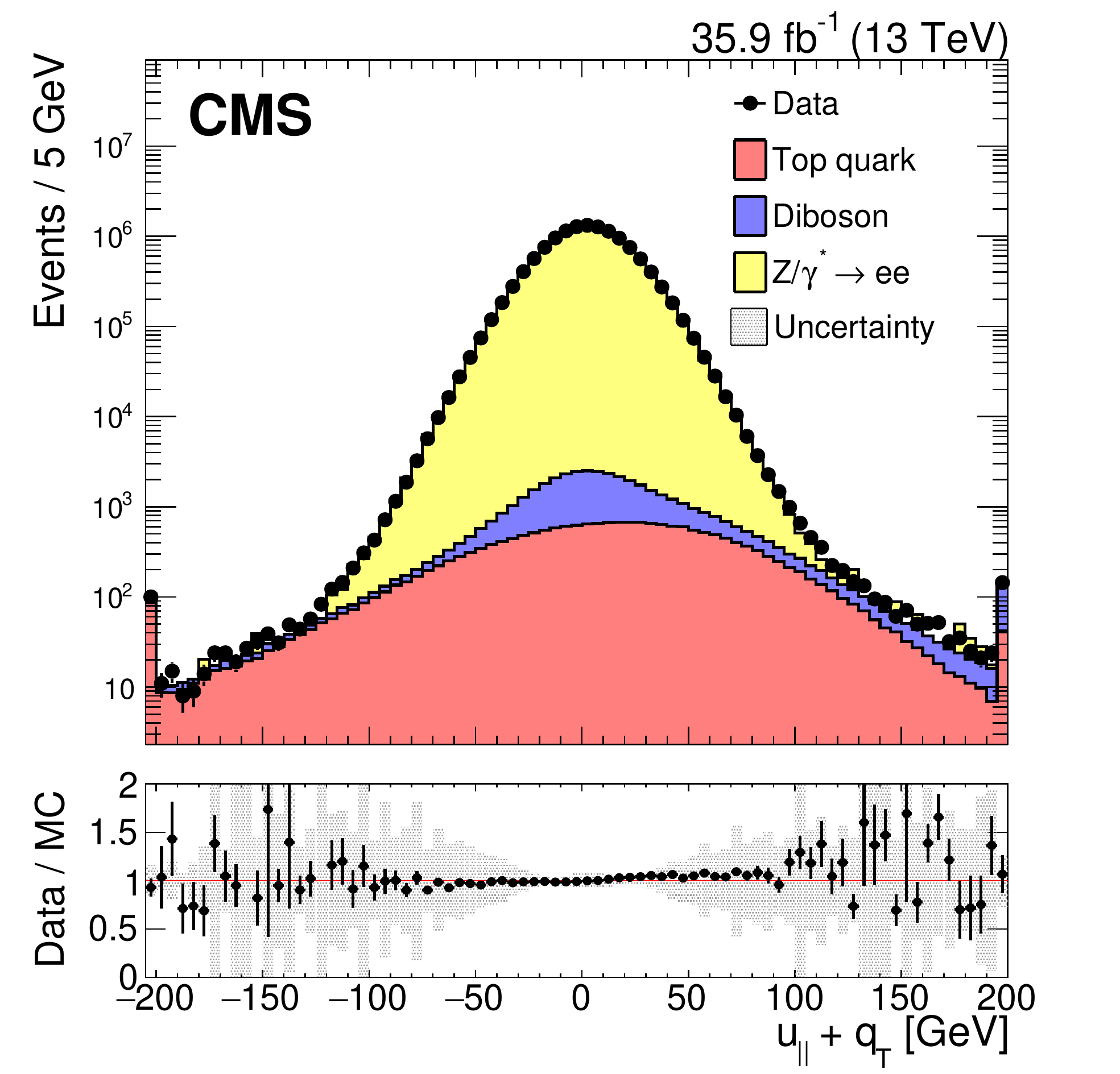}
  \includegraphics[width=0.42\textwidth]{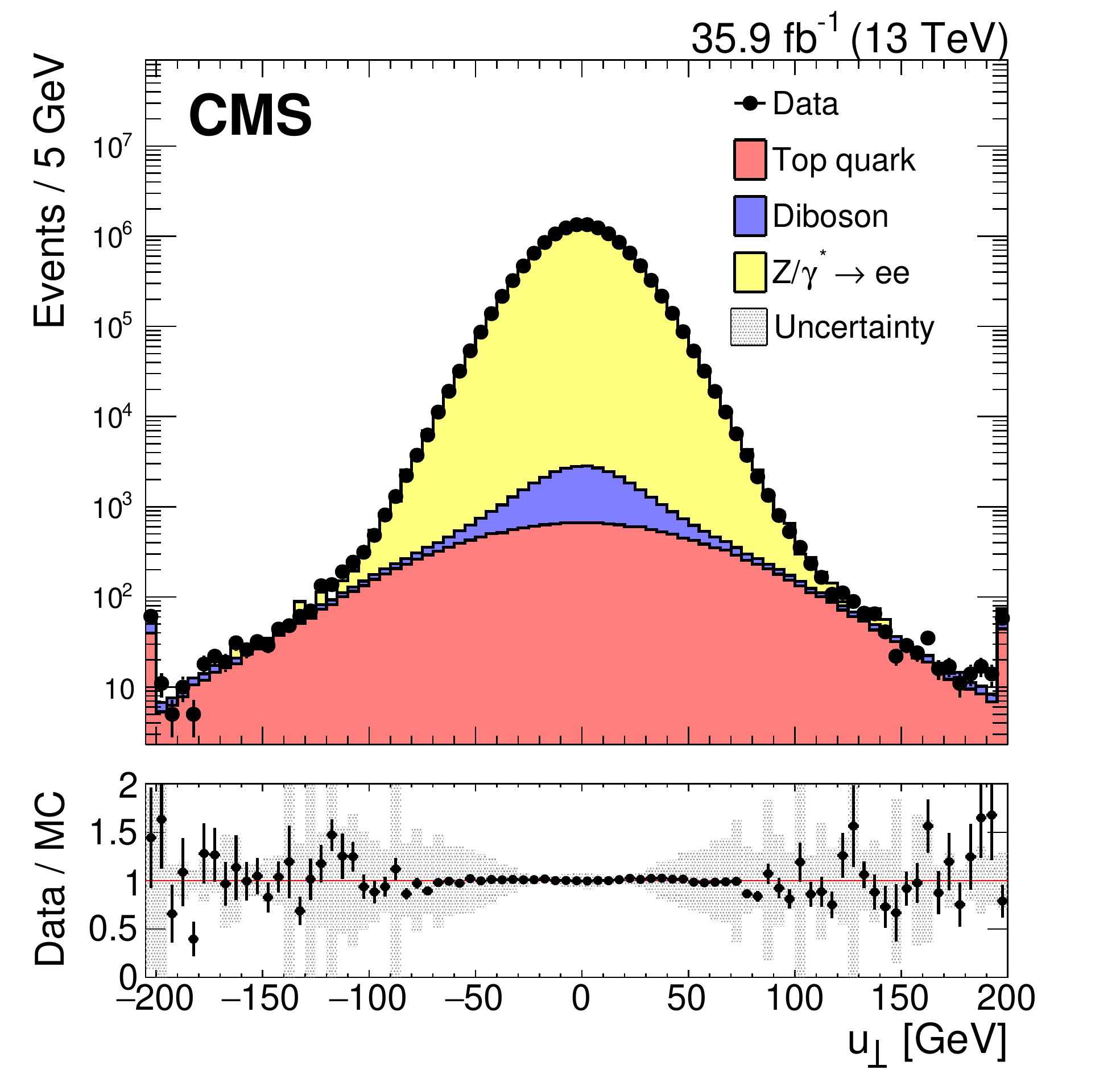}\\
  \includegraphics[width=0.42\textwidth]{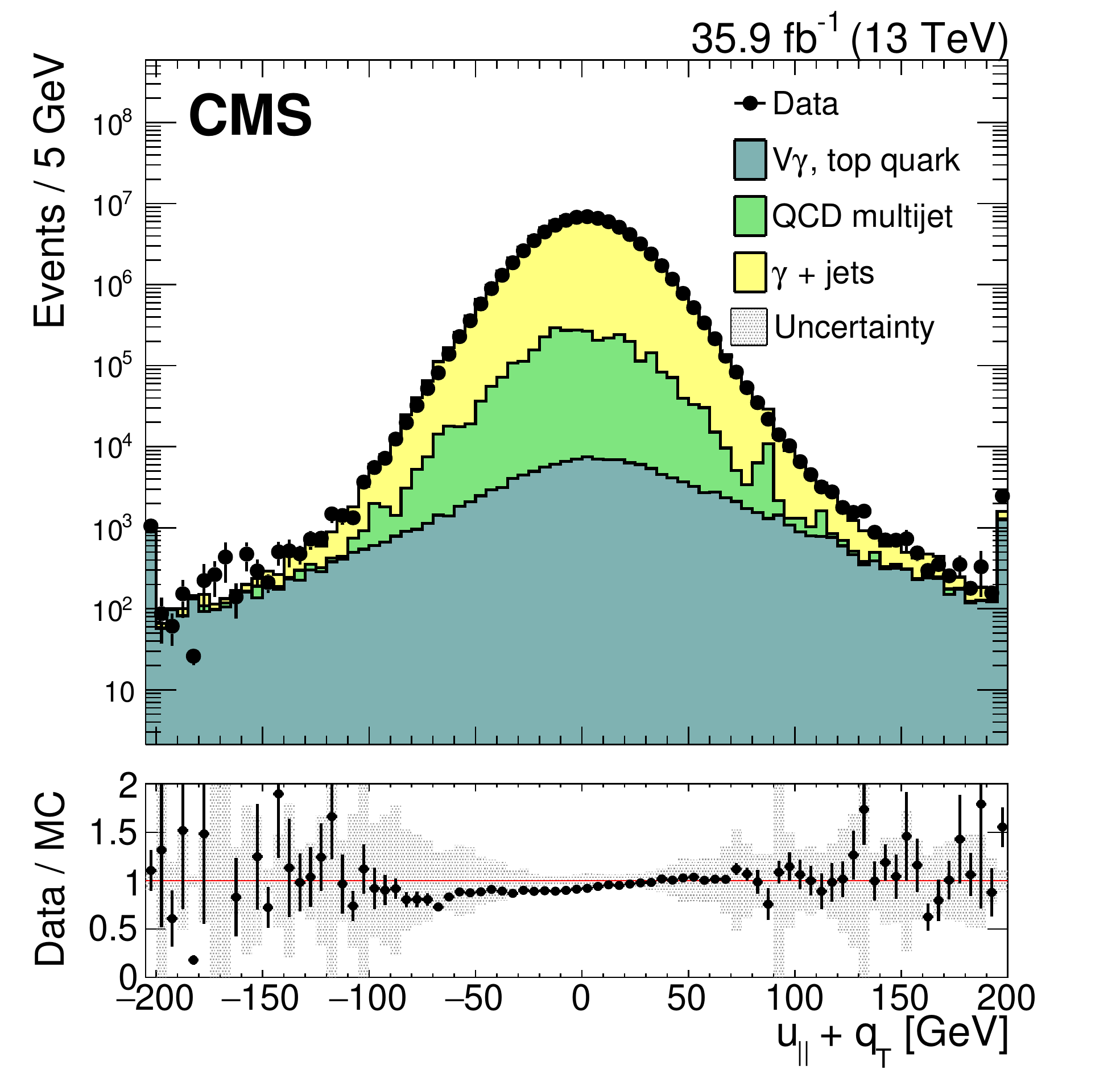}
  \includegraphics[width=0.42\textwidth]{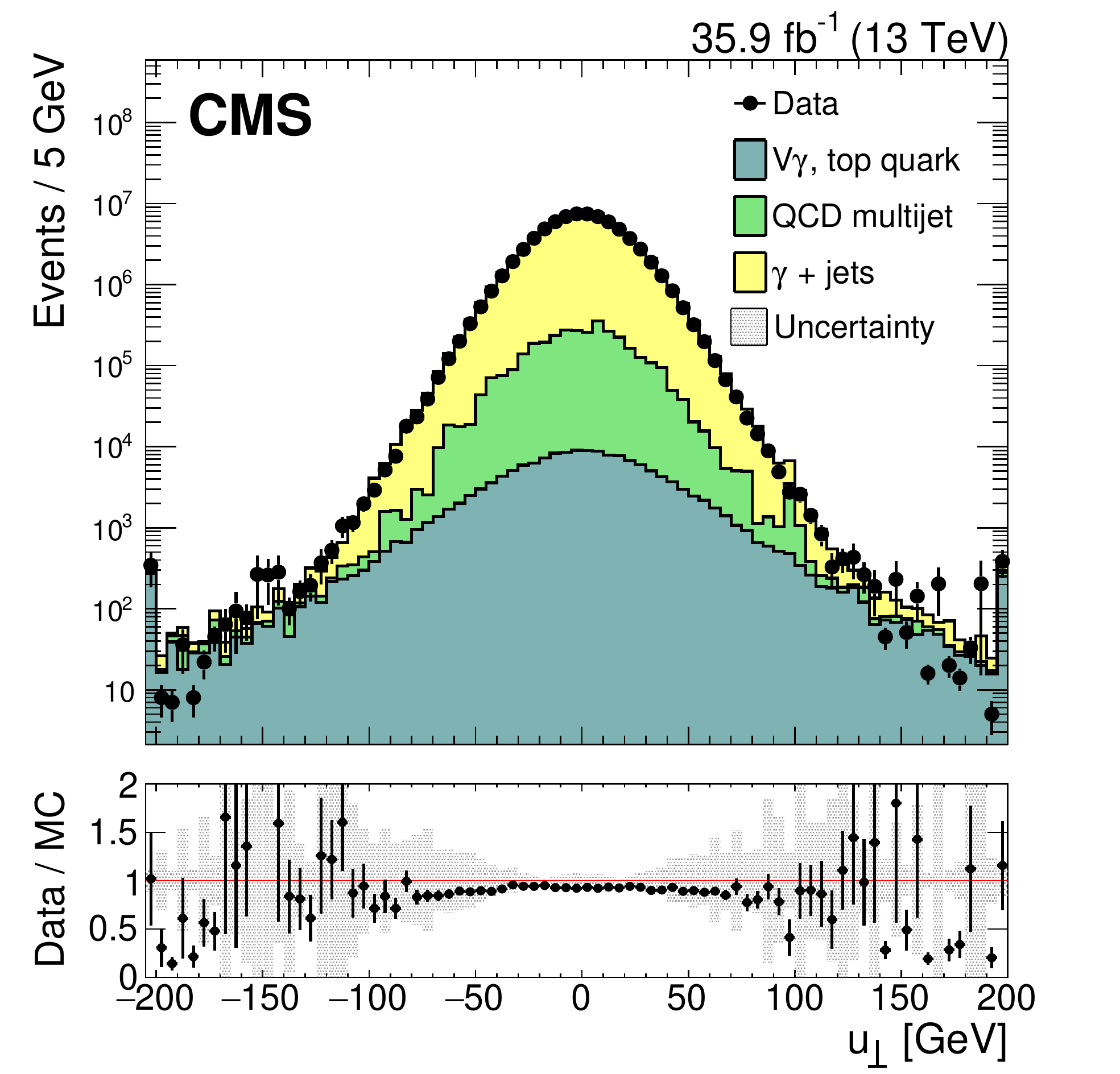}
  \caption{Distribution of \upar+\qt and \uperp\ components of the hadronic recoil, in data (filled markers) and simulation (solid histograms), in the  \Zmm\ (upper), \Zee\ (middle), and $\PGg+$jets (lower) samples. The first and the last bins include all events below -195 and above +195, respectively.
The points in the lower panel of each plot show the data to simulation ratio. The systematic uncertainties due to the JES, the JER, and variations in the $E_{U}$ are added in quadrature and represented by the shaded band.}
  \label{fig:uparuperp}
\end{figure}

Distributions of $\upar+\qt$ and \uperp\ in \Zmm, \Zee\, and $\PGg+$jets events are shown in Fig.~\ref{fig:uparuperp}. The kinematic definition of \upar\ dictates that for processes with no genuine \ptmiss, \upar\ is balanced with the boson \qt. Therefore, the vectorial sum of \upar\ and \qt results in a symmetric distribution, centered at zero; any deviations from this behavior imply imperfect calibration of \ptmiss. Events with genuine \ptmiss\ due to the presence of neutrinos, \upar\ and \qt are not balanced, leading to an asymmetric distribution. The \uperp\ distribution is symmetric with a mean value of zero. This symmetry is due to the assumed isotropic nature of the energy fluctuations of the detector noise and underlying event. Good agreement is observed between data and simulation for all the distributions.

\begin{figure}[!htp]
  \centering
   \includegraphics[width=0.55\textwidth]{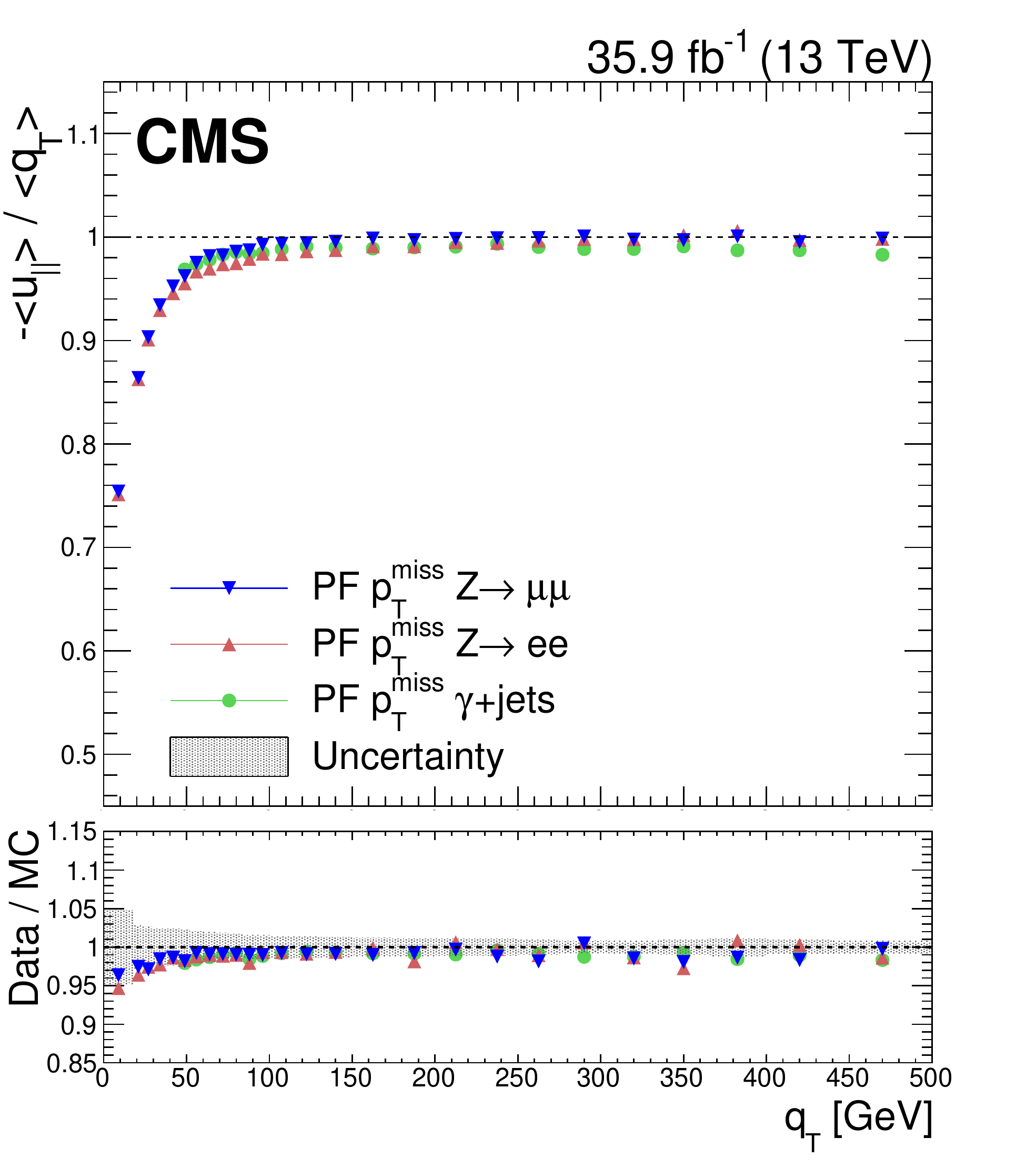}
  \caption{Upper panel: Response of \ptmiss, defined as $-\langle \upar \rangle /\langle \qt \rangle$,  in data in \Zmm\ (blue), \Zee\ (red), and $\PGg+$jets (green) events.  Lower panel: Ratio of the \ptmiss response in data and simulation. The band corresponds to the systematic uncertainties due to the JES, the JER, and variations in the $E_{U}$ added in quadrature, estimated from the \Zee\ sample.}
  \label{fig:response}
\end{figure}

Figure \ref{fig:response} shows the \ptmiss response as a function of \qt, in data and simulation, in \Zmm, \Zee, and photon events.
The response reaches unity for boson $\pt > 100\GeV$. Deviations from unity indicate imperfect calibration of the hadronic energy scale.
The underestimation of the hadronic response observed at smaller $\qt \lesssim 100\GeV$ is due to the significant contribution of the uncalibrated component of \ptmiss, which mainly consists of jets with $\pt < 15$\GeV and unclustered particles.
There is no dedicated response correction for the $E_{U}$.
The response of \ptmiss agrees for all three samples within 2\%; a significant improvement with respect to the results from the LHC Run 1~\cite{2010, Khachatryan:2014gga}.
The ``footprint removal'' discussed in Section~\ref{sec:evtreco} plays an important role in this improvement.
The residual response difference among the samples stems from the different mechanism used to differentiate muons, electrons, and photons from jets used in the correction of the \ptmiss, as discussed in Section~\ref{sec:metcorrections}.
Simulation studies have shown that in the case of electrons and photons, a small fraction ($\lesssim$10\%) of jets survive the differentiation criteria yet overlap with prompt electrons and photons.
As a result, these jets wrongly contribute to the \ptmiss calibration, leading to a 1--2\% lower response in the electron and photon channels. Future studies will aim at further improving the electron/photon and jet differentiation mechanism. Overall, we observe good agreement between data and simulation.

\begin{figure}[h!p]
  \centering
   \includegraphics[width=0.4\textwidth]{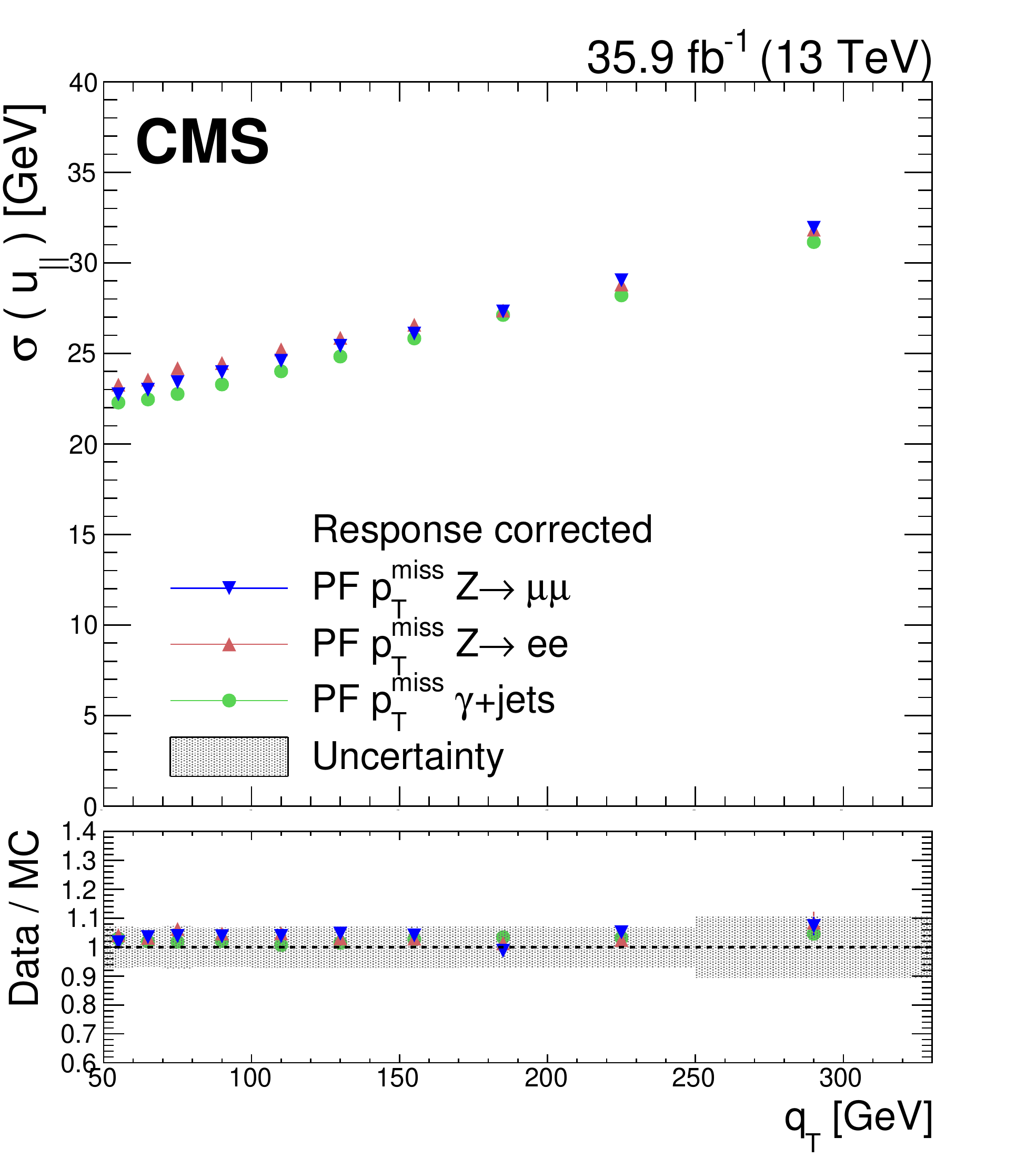}
   \includegraphics[width=0.4\textwidth]{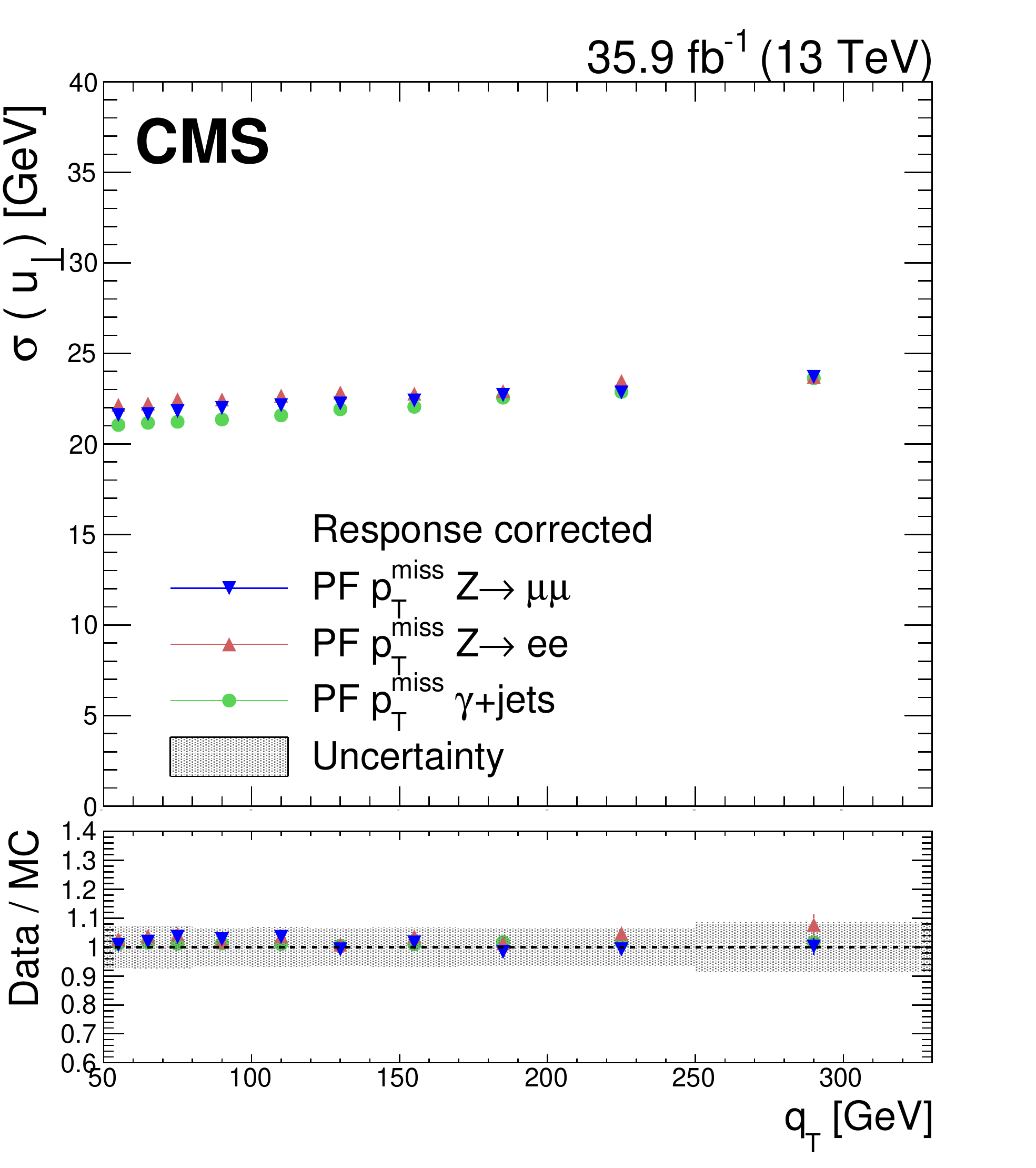}
   \includegraphics[width=0.4\textwidth]{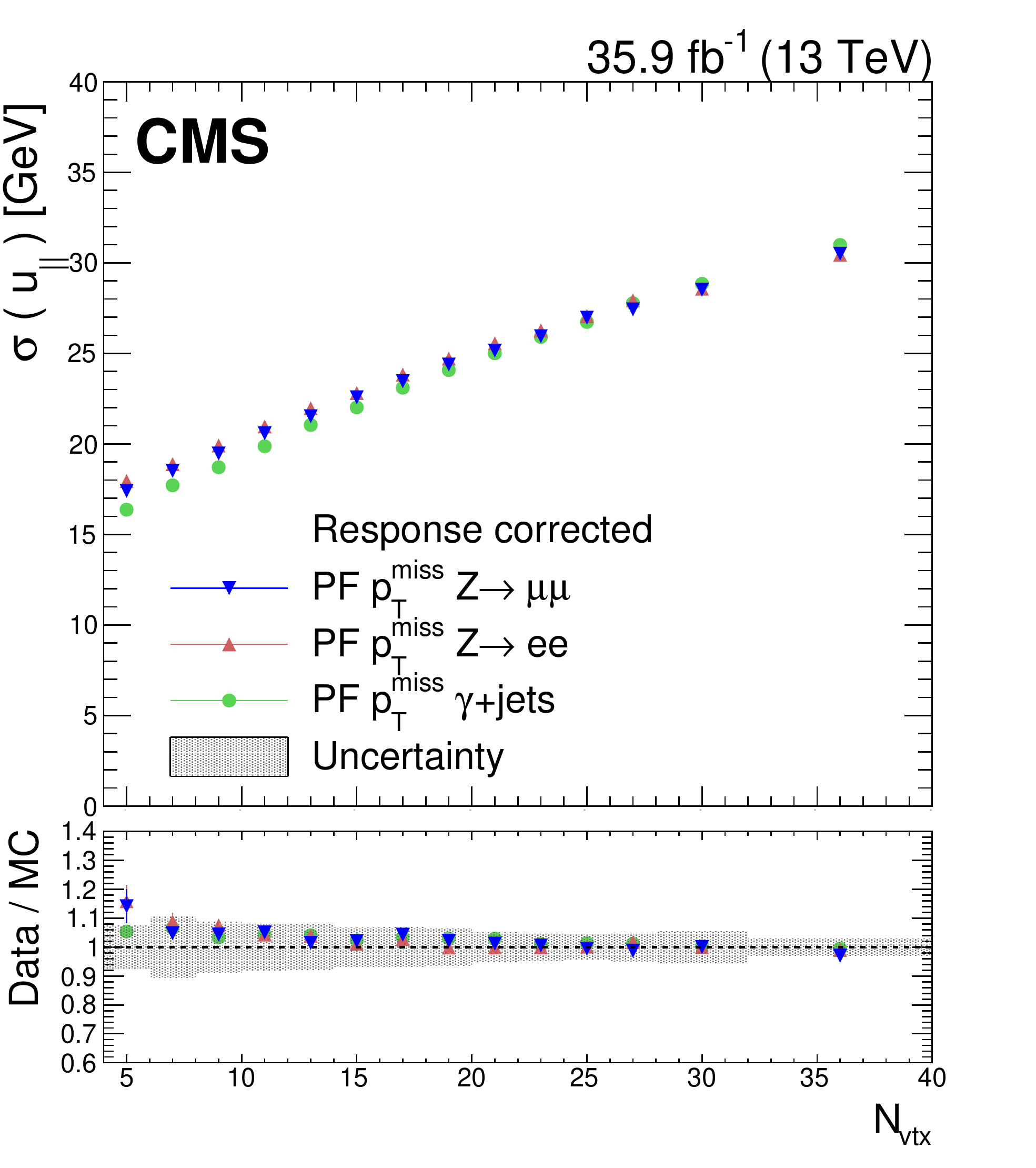}
   \includegraphics[width=0.4\textwidth]{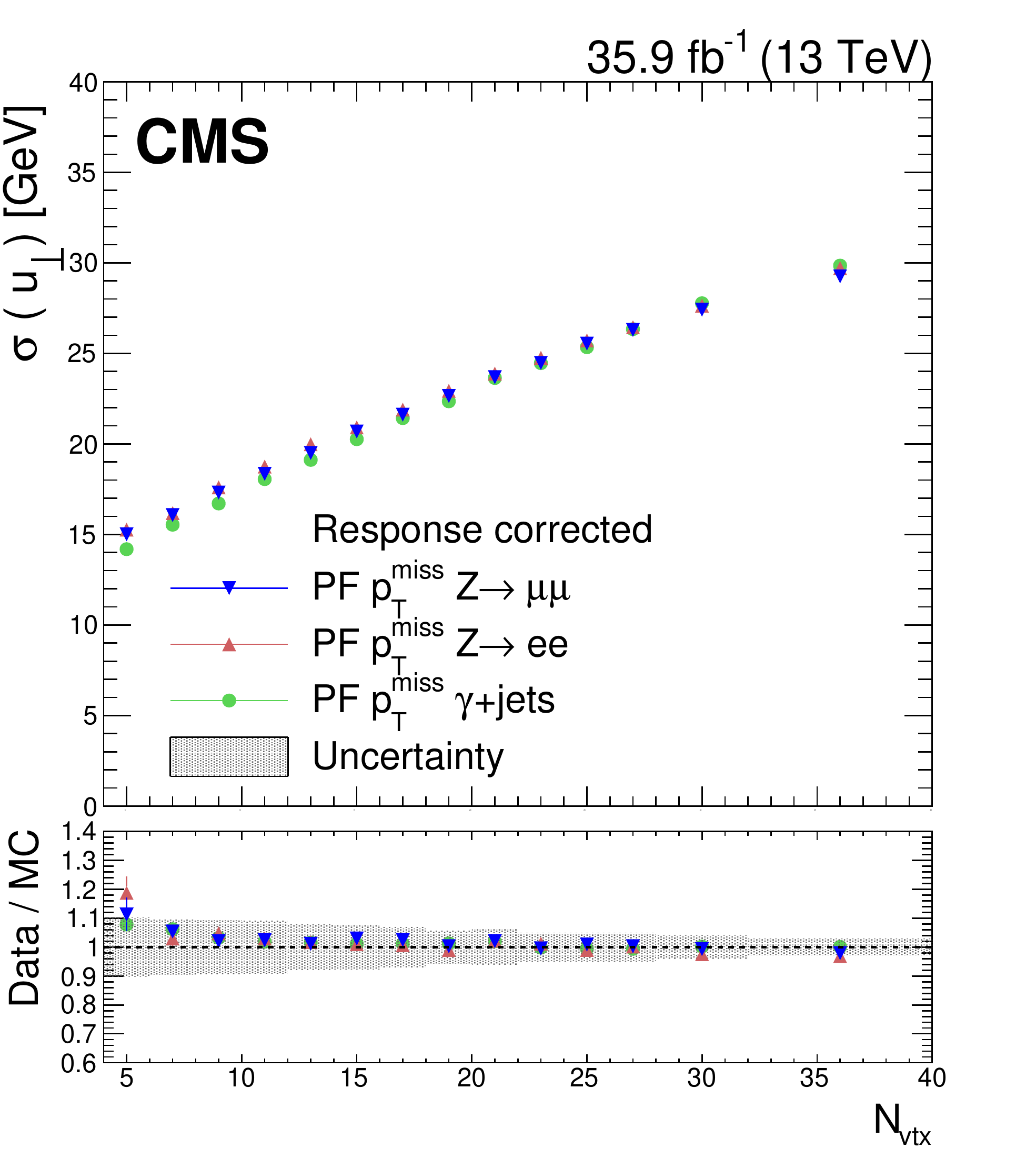}
   \includegraphics[width=0.4\textwidth]{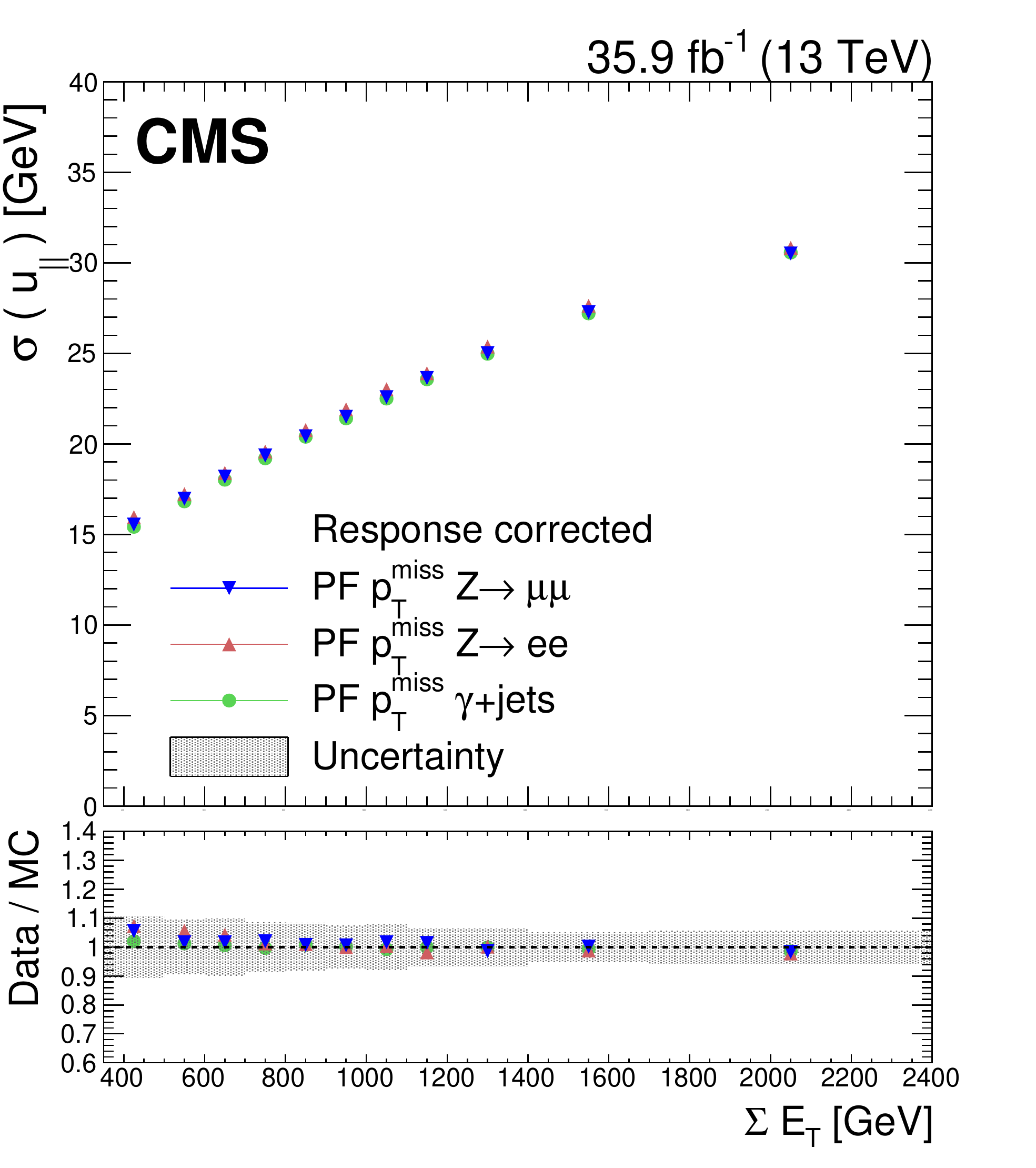}
   \includegraphics[width=0.4\textwidth]{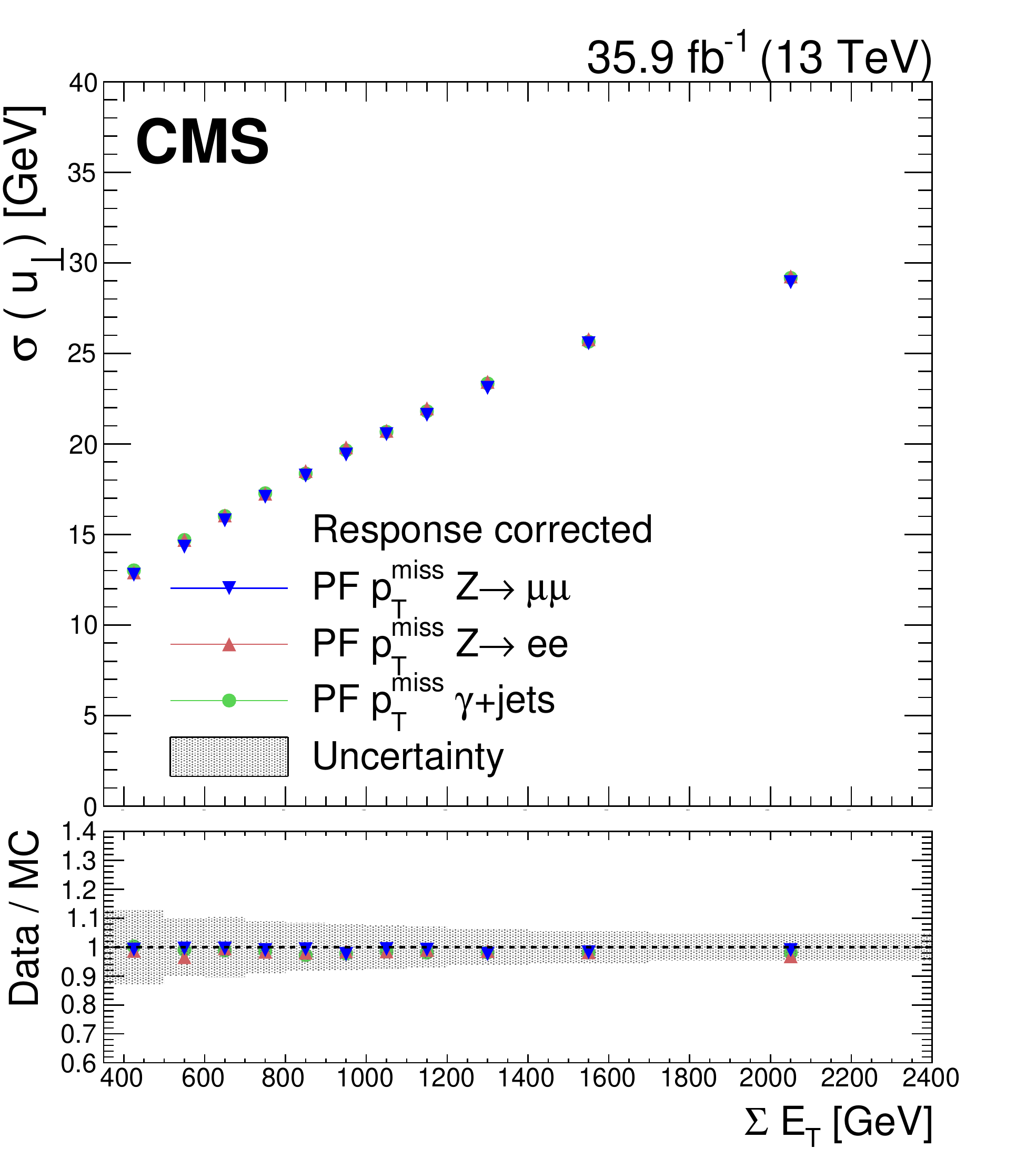}
  \caption{Resolution of the \upar\ and \uperp\ components of the hadronic recoil as a function of \qt (upper row), the reconstructed vertices (middle row), and the scalar \pt sum of all PF candidates (lower row), in \Zmm, \Zee, and $\PGg+$jets events. In each plot, the upper panel shows the resolution in data, whereas the lower panel shows the ratio of data to simulation. The band corresponds to the systematic uncertainties due to the JES, the JER, and variations in the $E_{U}$ added in quadrature, estimated from the \Zee\ sample. }
  \label{fig:resolution_sumet}
\end{figure}

The resolution of \ptmiss for the \upar\ and \uperp\ components of the hadronic recoil as a function of \qt is shown in Fig.~\ref{fig:resolution_sumet} (upper row).
To compare the resolution of \ptmiss consistently across the samples, the resolution in each sample is corrected for the differences observed in the response.
The correction has a negligible impact on the results. The resolutions measured in different samples are  in good agreement.
The relative resolution, both in \upar\ and \uperp, improves as a function of \qt because of the improved energy resolution in the calorimeters.
Furthermore, due to the isotropic nature of energy fluctuations stemming from detector noise and the underlying event, the dependence of the resolution of \uperp\ on \qt is smaller than for \upar.
For $\qt>200\GeV$, the \ptmiss resolution is $\approx$13\% and $\approx$9\%, for \upar\ and \uperp, respectively.

The resolution of the \upar\ and \uperp\ components of the hadronic recoil as a function of \nvtx, are shown in Fig.~\ref{fig:resolution_sumet} (middle row).
The resolutions measured in different samples, and in data and simulation, are  in good agreement. However, the resolution shows strong dependence on \nvtx,
since pileup mitigation techniques are employed only for the PF jets, but not for the PF \ptmiss algorithm.

The resolution is parametrized as a function of \nvtx:
\begin{linenomath}
\begin{equation} \label{eq:npv}
f(\nvtx) = \sqrt{\sigma_{\mathrm{c}}^2 + \frac{\nvtx}{0.70}\sigma_{\mathrm{PU}^2}},
\end{equation}
\end{linenomath}
where $\sigma_{\mathrm{c}}$ is the resolution term induced by the hard scattering interaction and $\sigma_{\mathrm{\mathrm{PU}}}$ is the average contribution to the resolution from each additional pileup interaction.
The factor 0.70 accounts for the vertex reconstruction efficiency~\cite{CMS-DP-2017-015}. Results of the parametrization for the \upar\ and \uperp\ components
are given in Table~\ref{tab:tab4lcontrol_par}. Good agreement is observed between data and simulation and no additional corrections are used for the \ptmiss\ calibration. Every additional pileup vertex degrades the resolution of each component by 3.8--4.0\GeV.

Lastly, Fig.~\ref{fig:resolution_sumet} (lower row) shows an alternative parametrization of the resolution of \upar\ and \uperp\, as a function of the scalar \pt sum of all PF candidates ($\sum \et$). The resolutions measured in different samples, and in data and simulation, are  in good agreement. The relative \ptmiss resolution improves with increasing $\sum \et$, driven by the amount of the activity in the calorimeters. The resolution in different samples is parametrized as:
\begin{linenomath}\begin{equation}
f(\sum{\et}) = \sigma_{0} + \sigma_{\text{s}}\sqrt{\sum{\et}},
\end{equation}\end{linenomath}
where $\sigma_{\mathrm{0}}$ is the resolution term induced by intrinsic detector noise and $\sigma_{\mathrm{s}}$ is the stochastic resolution term. Results of the parametrization for the \upar\ and \uperp\ components are given in Table~\ref{tab:tab4lcontrol_par_sumet}. The results are found to be consistent between data and simulation and no additional corrections are used for the \ptmiss\ calibration.

\begin{table}[h!t]

\centering
\bgroup
\def\arraystretch{1.2}
\topcaption{Parametrization results of the resolution curves for the \upar\ and \uperp\ components as a function of \nvtx. The parameter values for $\sigma_{\mathrm{c}}$ are obtained from data and simulation, and the values for $\sigma_{\mathrm{PU}}$ are obtained from data, along with a ratio $R_{\mathrm{PU}}$ of data and simulation. The uncertainties displayed for both components are obtained from the fit, and for simulation the JES, the JER, and $E_{U}$ uncertainties are added in quadrature.
}
\label{tab:tab4lcontrol_par}
\begin{tabular}{l c c c c}
\hline
Process & $\sigma_{c}(\text{data}) [\GeVns{}]$ & $\sigma_{c}(\mathrm{MC}) [\GeVns{}]$ & $\sigma_{\mathrm{PU}}(\text{data}) [\GeVns{}]$  & $R_{\mathrm{PU}}=\sigma_{\mathrm{PU}}(\text{data})/\sigma_{\mathrm{PU}}(\mathrm{MC})$\\ \hline
\multicolumn{5}{c}{\upar\ component} \\
$\Zmm$          & 13.9 $\pm$ 0.07 & 11.9 $\pm$ 1.53 & 3.82 $\pm$ 0.01 & 0.95 $\pm$ 0.04\\
$\Zee$          & 14.6 $\pm$ 0.09 & 12.0 $\pm$ 1.09 & 3.80 $\pm$ 0.02 & 0.95 $\pm$ 0.03\\
$\PGg+$jets       & 12.2 $\pm$ 0.10 & 10.2 $\pm$ 1.98 & 3.97 $\pm$ 0.02 & 0.97 $\pm$ 0.05\\

\multicolumn{5}{c}{\uperp\ component} \\
$\Zmm$          & 10.3 $\pm$ 0.08 & 8.58 $\pm$ 2.20 & 3.87 $\pm$ 0.01 & 0.97 $\pm$ 0.04\\
$\Zee$          & 10.7 $\pm$ 0.10 & 8.71 $\pm$ 1.76 & 3.89 $\pm$ 0.01 & 0.96 $\pm$ 0.03\\
$\PGg+$jets       & 9.04 $\pm$ 0.11 & 6.93 $\pm$ 2.70 & 3.94 $\pm$ 0.01 & 0.97 $\pm$ 0.04\\
\hline
\end{tabular}
\egroup

\end{table}

\begin{table}[h!b]

\centering
\def\arraystretch{1.2}
\topcaption{Parametrization results of the resolution curves for \upar\ and \uperp\ components as a function of the scalar \pt sum of all PF candidates. The parameter values for $\sigma_{\mathrm{0}}$ are obtained from data and simulation, whereas the $\sigma_{s}$ are obtained from data along with the ratio $R_{\mathrm{s}}$, the ratio of data and simulation. The uncertainties displayed for both components are obtained from the fit, and for simulation the JES, the JER, and $E_{U}$ uncertainties are added in quadrature.
}
\label{tab:tab4lcontrol_par_sumet}
\begin{tabular}{l c c  c c}
\hline
Process        & $\sigma_{0}(\text{data}) [\GeVns{}]$ & $\sigma_{0}(\mathrm{MC}) [\GeVns{}]$ & $\sigma_{s}[\mathrm{GeV^{1/2}}]$ & $R_{\mathrm{s}}=\sigma_{s}(\text{data})/\sigma_{s}(\mathrm{MC})$\\ \hline
\multicolumn{5}{c}{\upar\ component} \\
$\Zmm$              & 1.98 $\pm$ 0.07 & 0.85 $\pm$ 2.45 & 0.64 $\pm$ 0.01 & 0.95 $\pm$ 0.11\\
$\Zee$              & 2.18 $\pm$ 0.09 & 0.19 $\pm$ 2.90 & 0.64 $\pm$ 0.01 & 0.92 $\pm$ 0.11\\
$\PGg+$jets           & 1.85 $\pm$ 0.09 & 0.94 $\pm$ 2.52 & 0.64 $\pm$ 0.01 & 0.96 $\pm$ 0.11\\

\multicolumn{5}{c}{\uperp\ component} \\
$\Zmm$              & -1.63 $\pm$ 0.06 & -1.72 $\pm$ 2.53 & 0.68 $\pm$ 0.01 & 0.99 $\pm$ 0.11\\
$\Zee$              & -1.42 $\pm$ 0.08 & -1.98 $\pm$ 2.95 & 0.69 $\pm$ 0.01 & 0.96 $\pm$ 0.12\\
$\PGg+$jets           & -1.16 $\pm$ 0.08 & -1.31 $\pm$ 2.53 & 0.68 $\pm$ 0.01 & 0.98 $\pm$ 0.11\\
\hline
\end{tabular}
\end{table}

\subsection{Performance of the  \texorpdfstring{\puppi \ptmiss}{PUPPI pT miss} algorithm}

\begin{figure}[!htp]
  \centering
  \includegraphics[width=0.42\textwidth]{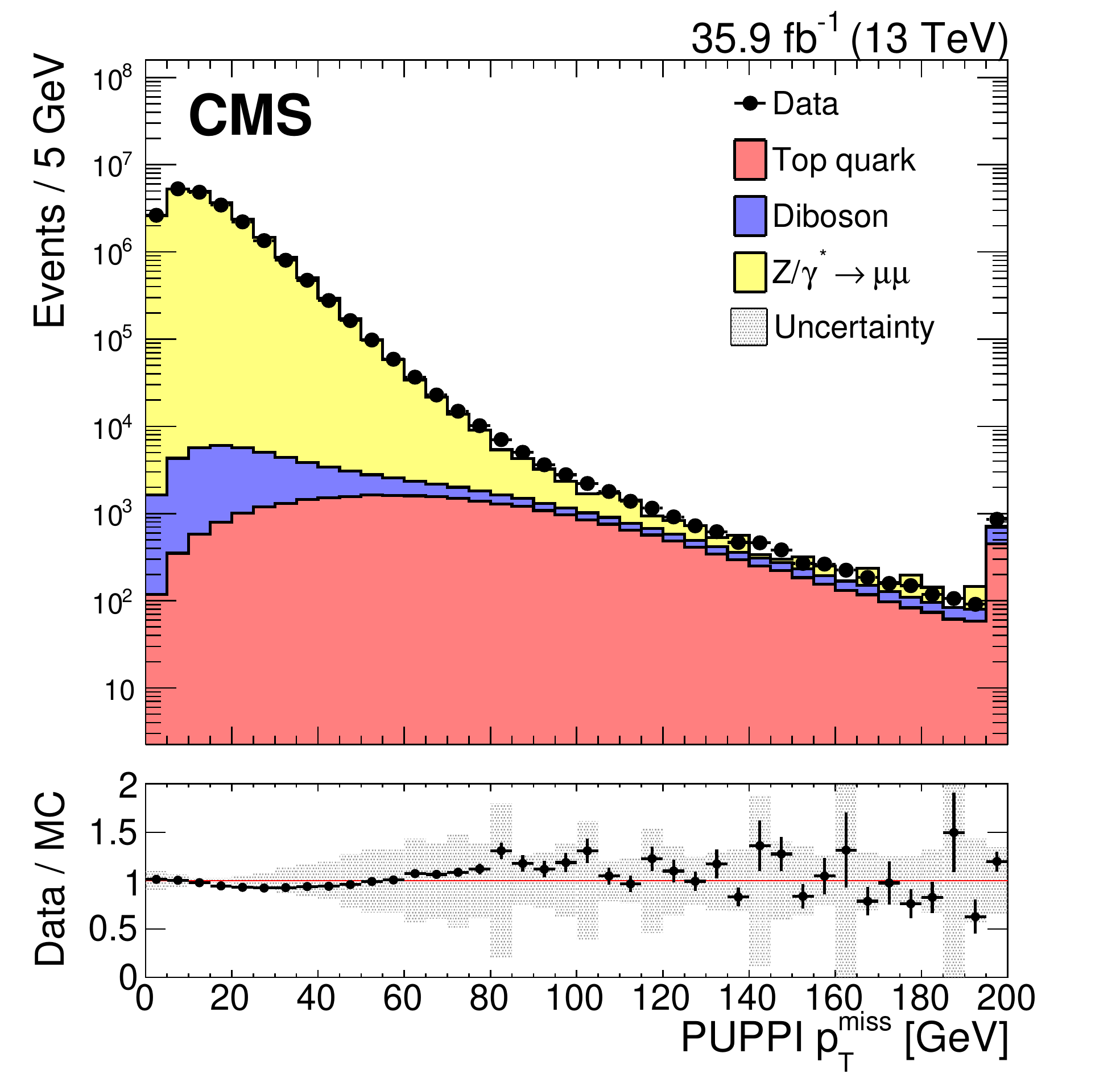}
  \includegraphics[width=0.42\textwidth]{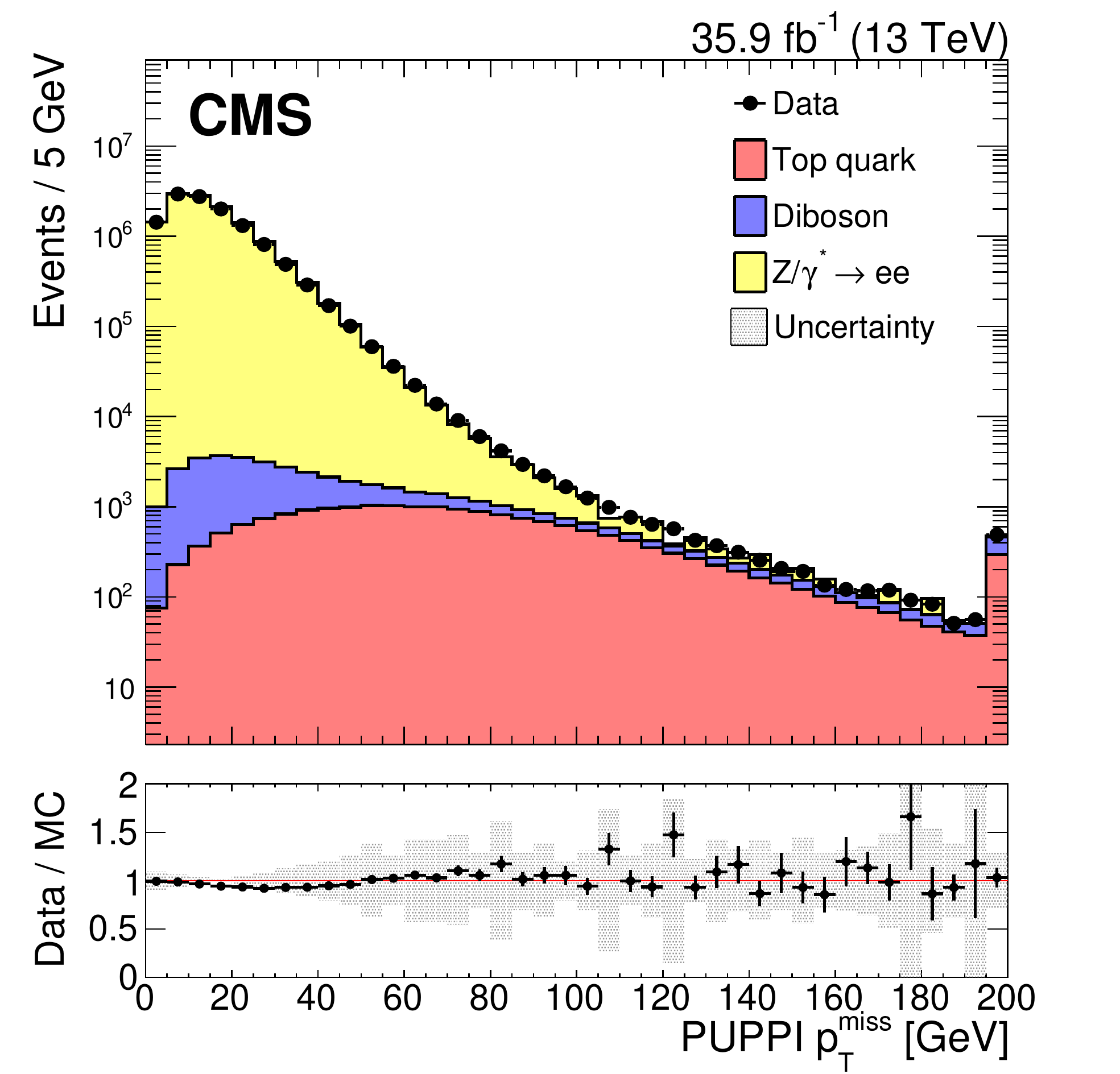}

  \caption{Upper panels: Distributions of \puppi \ptmiss in \Zmm\ (left) and \Zee\ (right) events. The last bin includes all events with $\ptmiss>195\GeV$.
Lower panels: Data-to-simulation ratio. The band corresponds to the systematic uncertainties due to the JES, the JER, and variations in the $E_{U}$ added in quadrature, estimated from the \Zee\ sample.}
  \label{fig:puppimet}
\end{figure}

The \puppi \ptmiss distributions in the dilepton samples are shown in Fig.~\ref{fig:puppimet}. The data distributions are modeled well by the simulation,
 in both the muon and the electron channels.
As in the case of PF \ptmiss, the \ptmiss resolution in these events is dominated by the resolution of the hadronic activity,
but the \puppi-weighted PF candidates yield improved resolution for jets compared to the PF case. This is also reflected in the uncertainty shown in the figures, which includes the uncertainties due to JES and JER, and the energy scale of the unclustered particles.

\begin{figure}[!htp]
  \centering
  \includegraphics[width=0.42\textwidth]{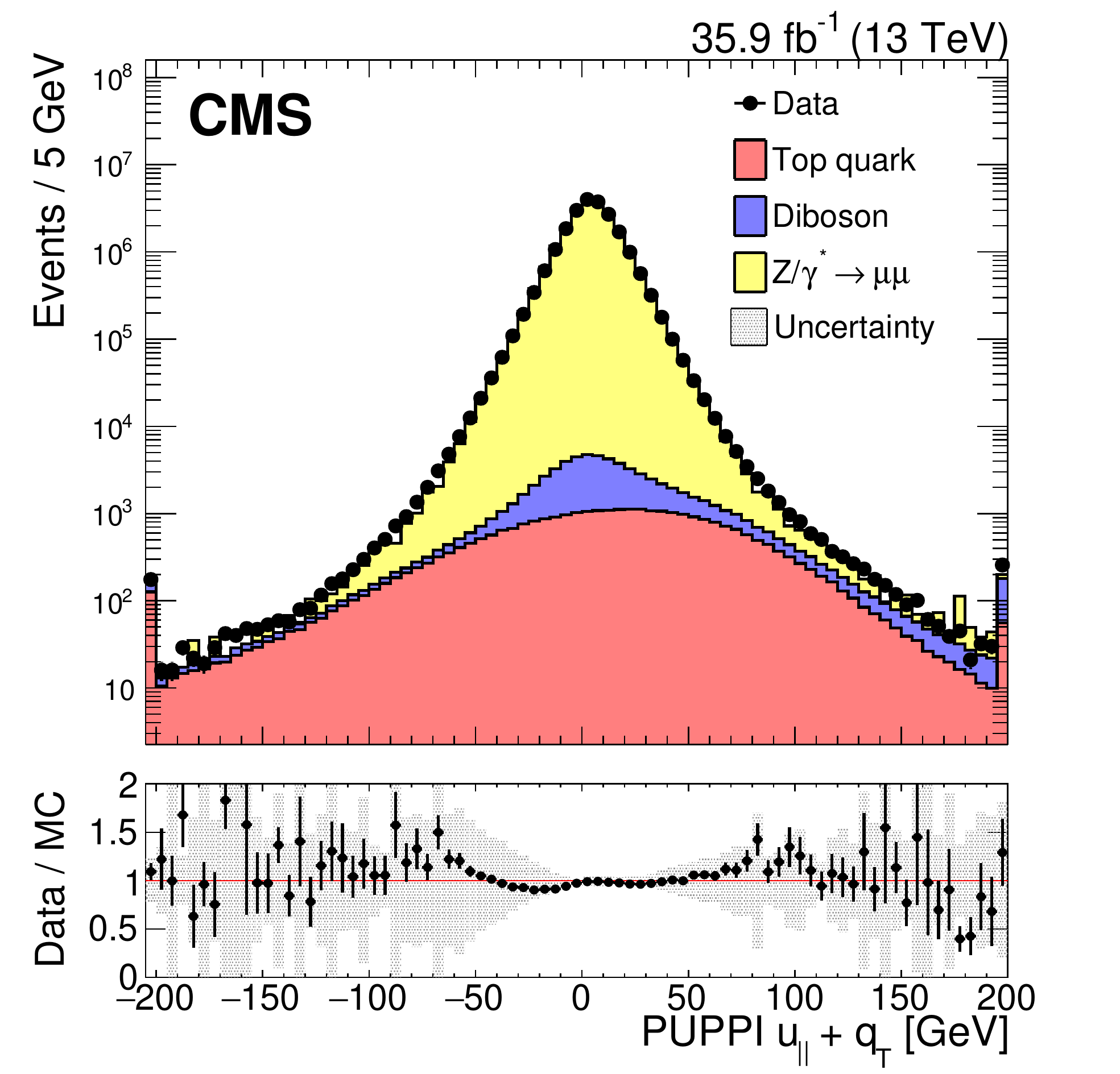}
  \includegraphics[width=0.42\textwidth]{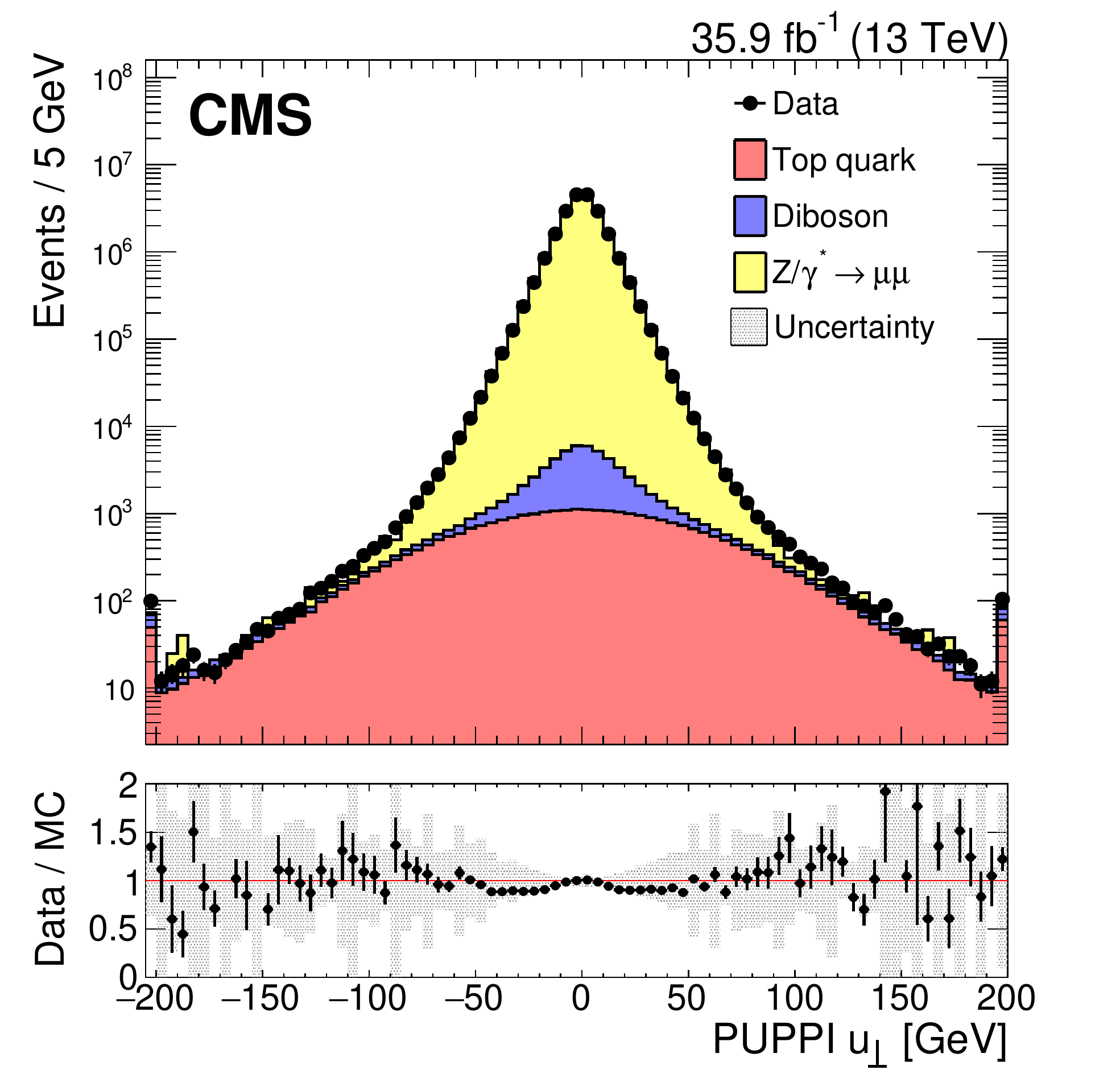}\\
  \includegraphics[width=0.42\textwidth]{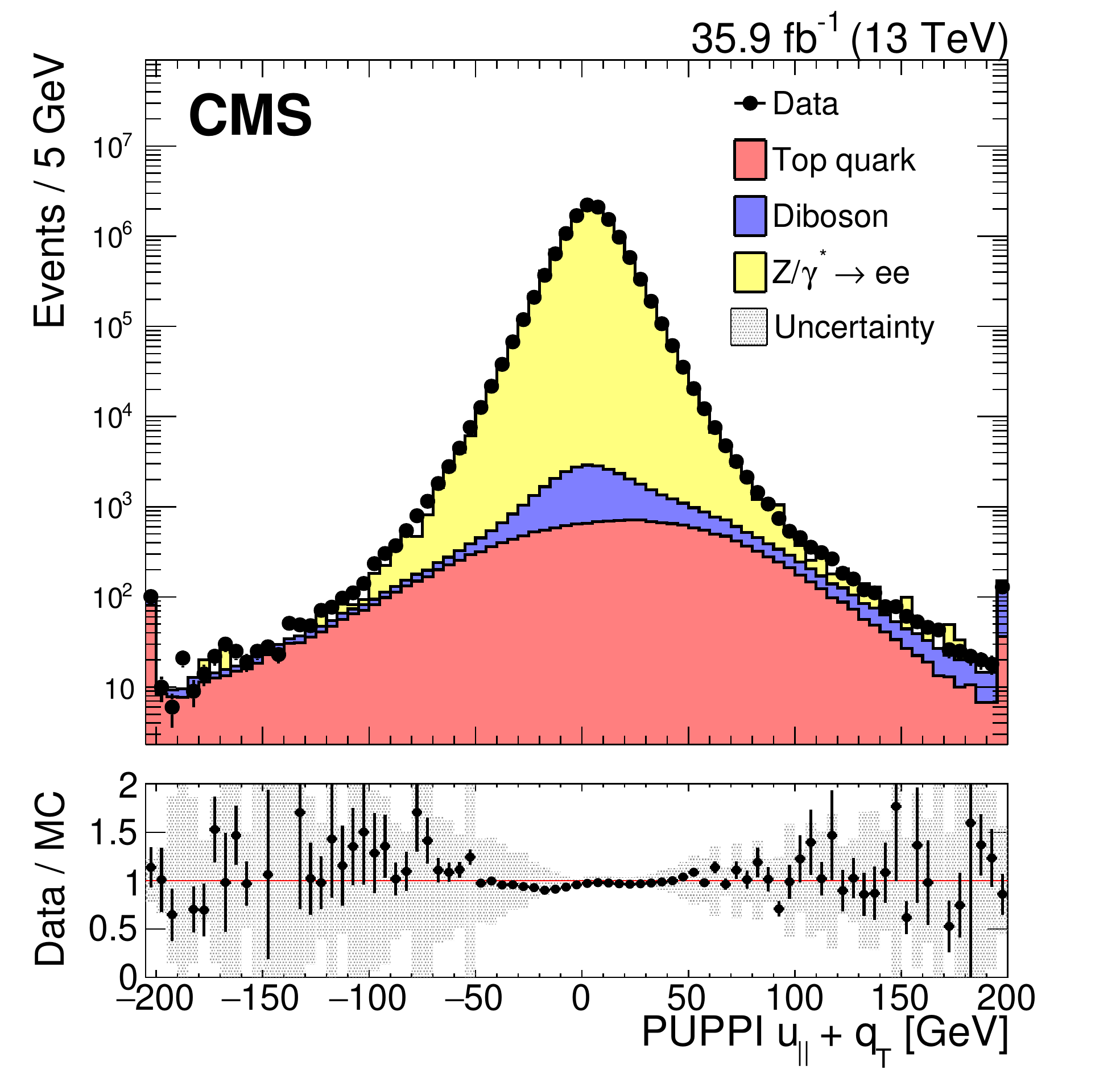}
  \includegraphics[width=0.42\textwidth]{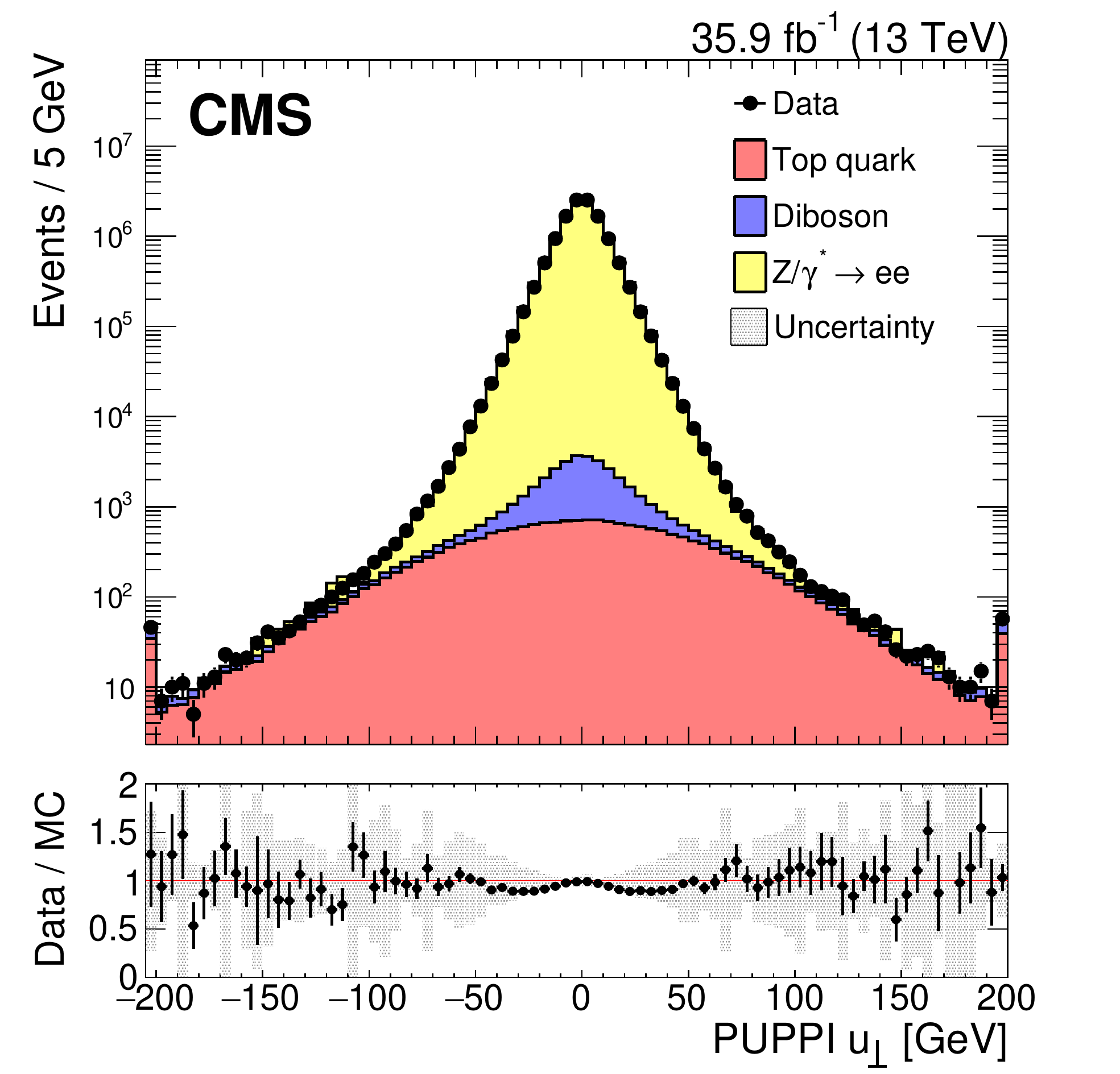}
  \caption{Upper panels: Distributions of the \upar+\qt and \uperp\ components of the hadronic recoil, in data (filled markers) and simulation (solid histograms), for the \Zmm\ (upper) and \Zee\ (lower) events. The first and the last bins include all events below -195 and above +195, respectively.
Lower panel: Data-to-simulation ratio. The band corresponds to the systematic uncertainties due to the JES, the JER, and variations in the $E_{U}$ added in quadrature, estimated from the \Zee\ sample.}
  \label{fig:uparuperp_puppi}
\end{figure}

The distributions in \Zmm\ and \Zee\ events
of the vectorial sum \upar + \qt and of \uperp\, using \puppi \ptmiss ,  are shown in Fig.~\ref{fig:uparuperp_puppi}. Following the same arguments as in the PF \ptmiss case, in events with no genuine \ptmiss\ the vectorial sum of \upar\ and \qt is symmetric around zero, whereas for processes with genuine \ptmiss an asymmetric behavior is observed. The distribution of \uperp\ is symmetric around zero. Simulation describes data well for all distributions.

\begin{figure}[!htp]
  \centering
   \includegraphics[width=0.42\textwidth]{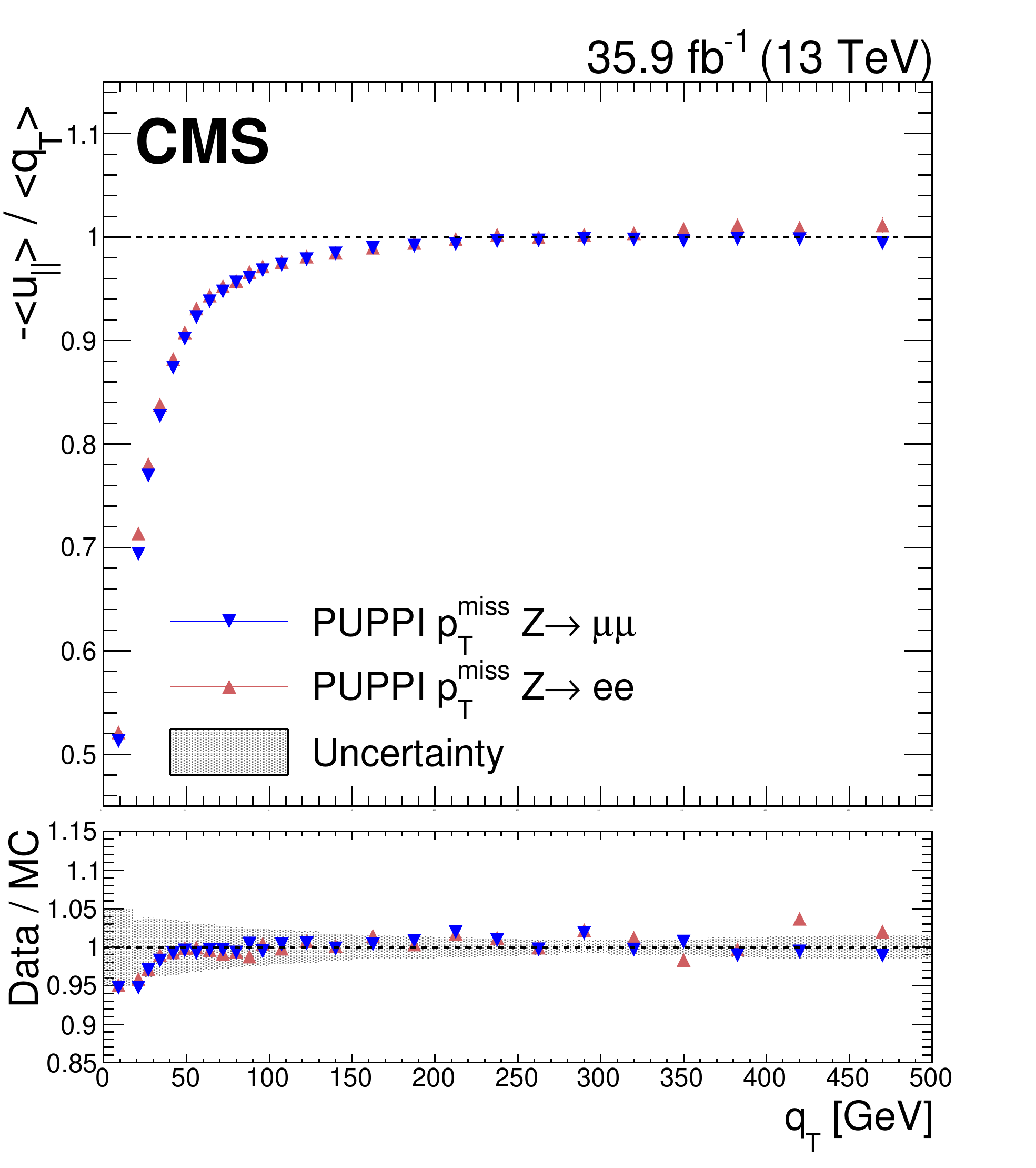}
  \caption{Upper panel: Response of \puppi \ptmiss, defined as $-\langle \upar \rangle /\langle \qt \rangle$, in data in \Zmm\ and \Zee\ events. Lower panel: ratio of the \puppi \ptmiss response in data and simulation. The band corresponds to the systematic uncertainties due to the JES, the JER, and variations in the $E_{U}$ added in quadrature, estimated from the \Zee\ sample.}
  \label{fig:response_puppi}
\end{figure}

Figure \ref{fig:response_puppi} shows the \puppi \ptmiss response as a function of \qt\ for data and simulation in \Zmm\ and \Zee\ events. The response rises to unity for \Zmm\ events at a $\PZ$ boson \pt of 150\GeV,
whereas for PF~\ptmiss the reaches unity at 100\GeV.
The slower rise of the response to unity is due to the removal of PF
candidates that are wrongly associated with pileup interactions
by the \puppi algorithm. As in PF~\ptmiss, there is no response correction for the $E_{U}$ in the  \puppi~\ptmiss,
which results in an underestimated response at low \qt. The response of \ptmiss agrees for the different samples  within 2\%.

\begin{figure}[!htp]
  \centering
   \includegraphics[width=0.42\textwidth]{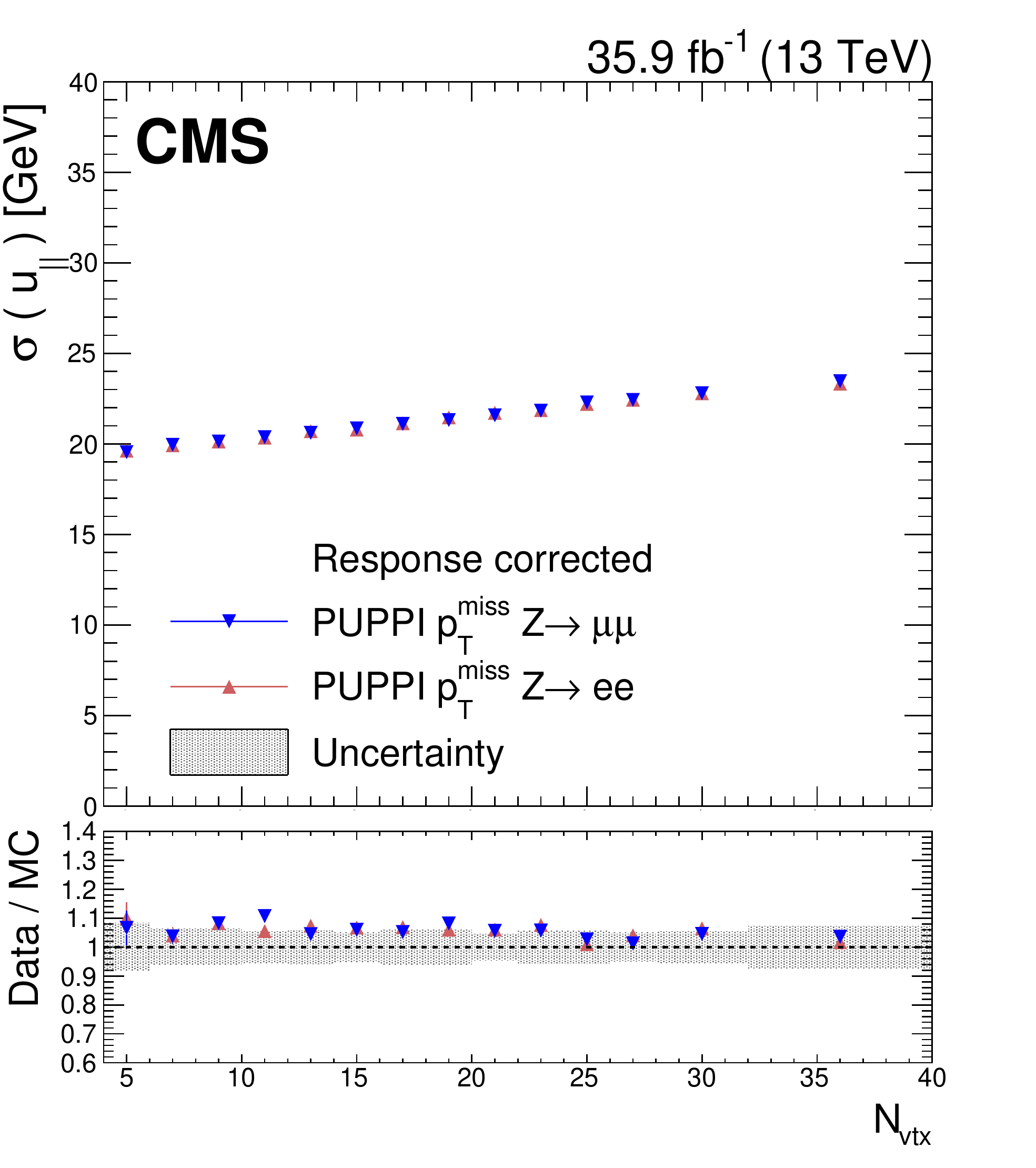}
   \includegraphics[width=0.42\textwidth]{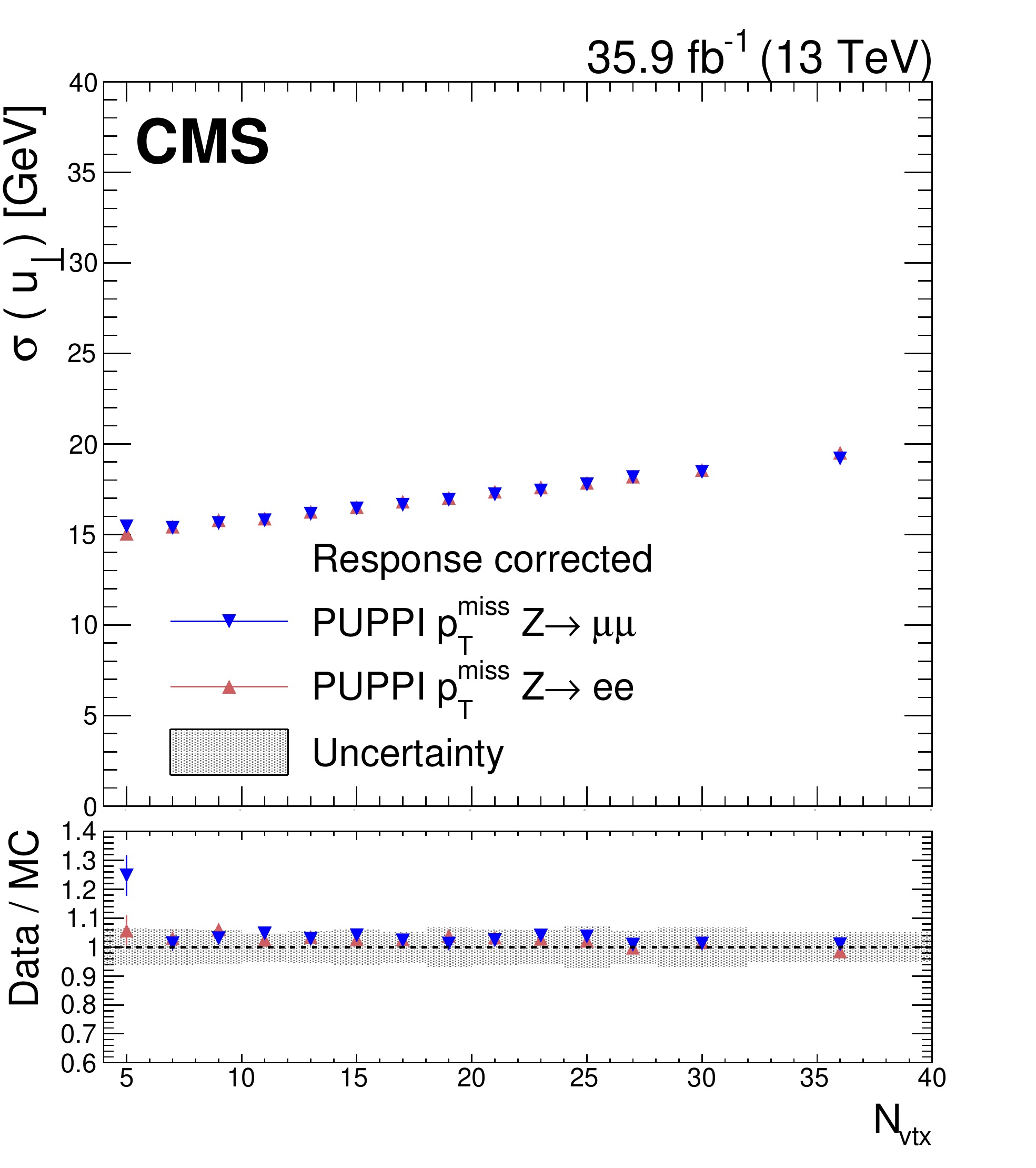}
  \caption{\puppi \ptmiss resolution of the \upar\ (left) and \uperp\ (right) components of the hadronic recoil as a function of \nvtx, in \Zmm\ and \Zee\ events. In each plot, the upper panel shows the resolution in data, whereas the lower panel shows the ratio of data to simulation. The band corresponds to the systematic uncertainties due to the JES, the JER, and variations in the $E_{U}$ added in quadrature, estimated from the \Zee\ sample. }
  \label{fig:puppiresolution_npv}
\end{figure}

\begin{figure}[!htp]
  \centering
   \includegraphics[width=0.42\textwidth]{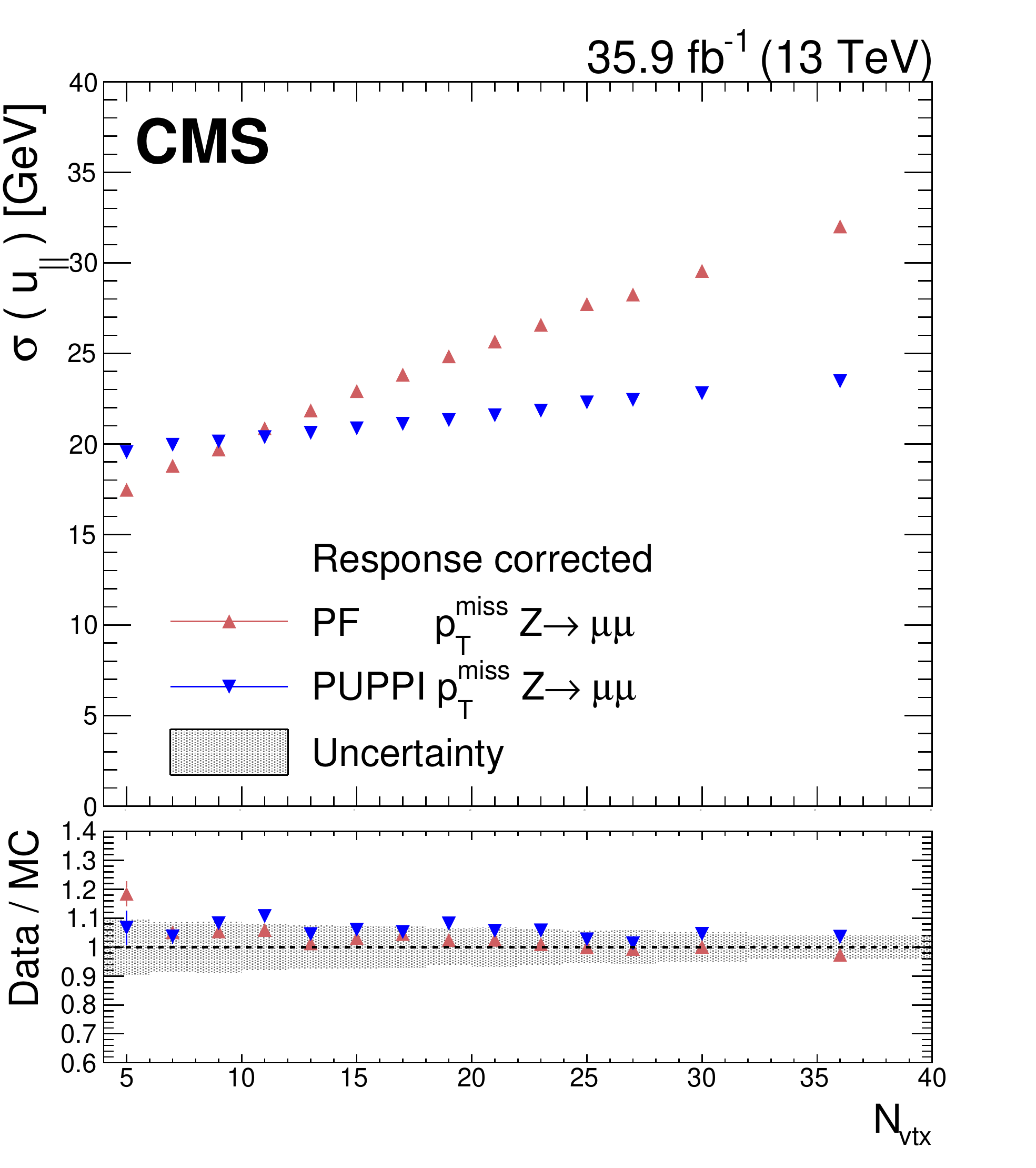}
   \includegraphics[width=0.42\textwidth]{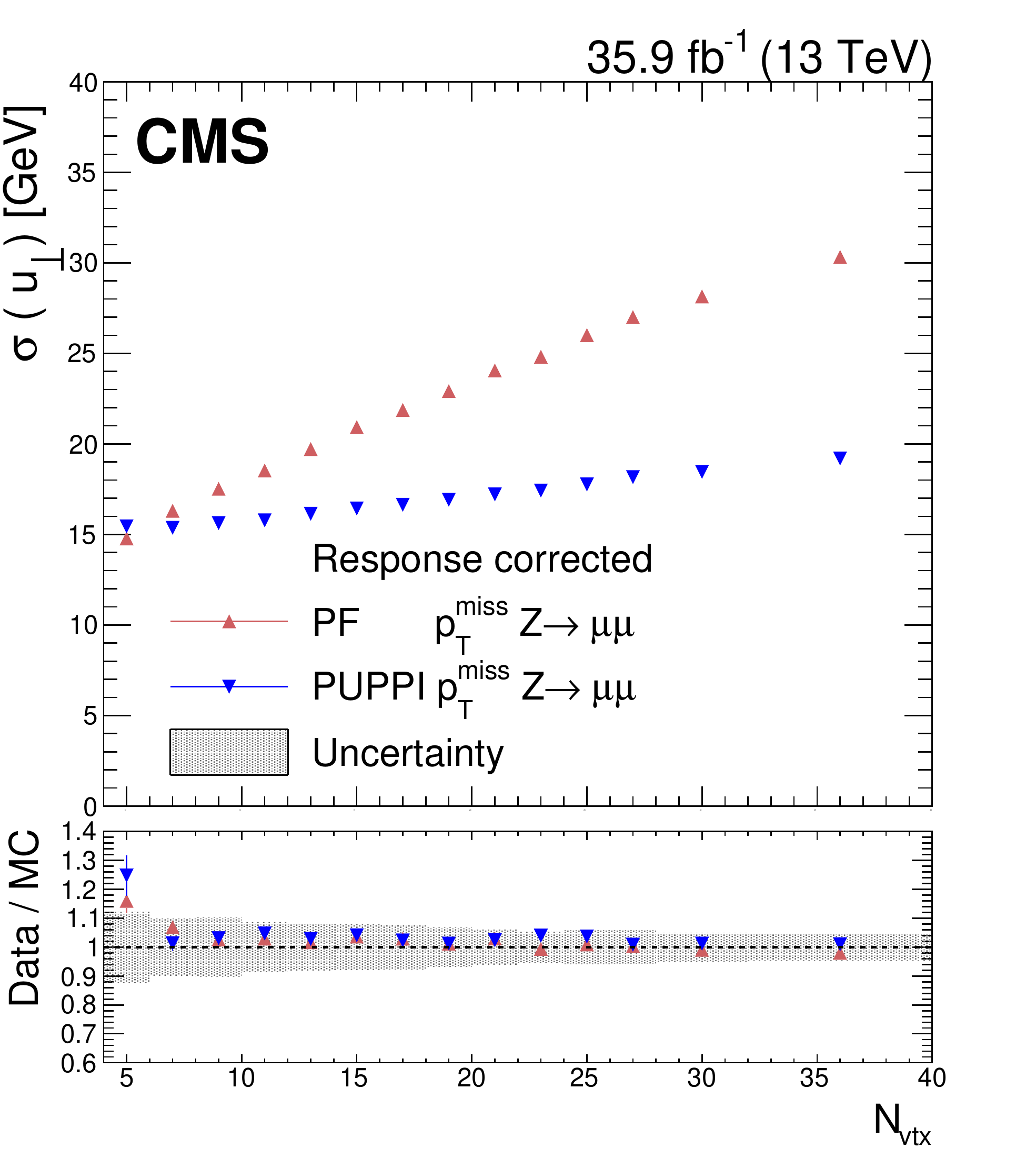}
   \caption{Upper panels: \puppi and PF \ptmiss resolution of \upar\ (left) and \uperp\ (right) components of the hadronic recoil as a function of \nvtx, in \Zmm\ events. Lower panels: Data-to-simulation ratio. The systematic uncertainties due to the JES, the JER, and variations in the $E_{U}$ are added in quadrature and represented by the shaded band.}
   \label{fig:Res_vs_PileUpHighPU}
\end{figure}

The resolution of the \puppi \ptmiss for the \upar\ and \uperp\ components of the hadronic recoil as a function \nvtx\ is shown in Fig.~\ref{fig:puppiresolution_npv}. To compare the resolution of \ptmiss consistently across the samples, the resolution in each sample is corrected for the differences observed in the scale. The resolutions measured in different samples are in good agreement. In Fig.~\ref{fig:Res_vs_PileUpHighPU}, the results obtained for the case of \puppi \ptmiss\ are overlaid with the ones obtained using PF~\ptmiss. Compared to the case of PF~\ptmiss, the resolutions show a much reduced dependence on the number of pileup interactions.

The resolutions in different samples are parametrized using Eq.~(\ref{eq:npv}), and the results of the parameterization are given in Table~\ref{tab:tab4lcontrol_par_puppi}. Good agreement is observed between data and simulation and no additional corrections are used in the \ptmiss\ calibration. Each additional pileup interaction degrades the resolution of each component by up to 2\GeV. This degradation in resolution corresponds to half of that observed in the case of PF \ptmiss.

\begin{table}[hbtp]
\centering
\def\arraystretch{1.2}
\topcaption{Parameterization results of the resolution curves for \puppi \upar\ and \uperp\ components as a function of \nvtx. The parameter values for $\sigma_{\mathrm{c}}$ are obtained from data and simulation, and the values for $\sigma_{\mathrm{\mathrm{PU}}}$ are obtained from data, along with the ratio $R_{\mathrm{PU}}$ of data and simulation. The uncertainties displayed for both the components are obtained from the fit, and for simulation the JES, the JER, and $E_{U}$ uncertainties are added in quadrature.
}
\label{tab:tab4lcontrol_par_puppi}
\begin{tabular}{l c c c c}
\hline
Process & $\sigma_{c}(\text{data}) [\GeVns{}]$ & $\sigma_{c}(\mathrm{MC}) [\GeVns{}]$ & $\sigma_{\mathrm{PU}}(\text{data}) [\GeVns{}]$  & $R_{\mathrm{\mathrm{PU}}}=\sigma_{\mathrm{PU}}(\text{data})/\sigma_{\mathrm{PU}}(\mathrm{MC})$\\ \hline
\multicolumn{5}{c}{\upar\ component} \\
$\Zmm$        & 18.9 $\pm$ 0.05 & 17.5 $\pm$ 0.74 & 1.93 $\pm$ 0.02 & 0.97 $\pm$ 0.11\\
$\Zee$        & 18.9 $\pm$ 0.06 & 17.4 $\pm$ 0.80 & 1.94 $\pm$ 0.03 & 0.98 $\pm$ 0.12\\

\multicolumn{5}{c}{\uperp\ component} \\
$\Zmm$        & 14.2 $\pm$ 0.04 & 13.6 $\pm$ 0.59 & 1.78 $\pm$ 0.01 & 0.97 $\pm$ 0.09\\
$\Zee$        & 14.3 $\pm$ 0.05 & 13.6 $\pm$ 0.59 & 1.80 $\pm$ 0.02 & 0.96 $\pm$ 0.09\\
\hline
\end{tabular}
\end{table}

\subsection{Performance of \texorpdfstring{\ptmiss}{pT miss} in single-lepton samples}

Also single-lepton events, which contain genuine \ptmiss, are utilized to study the performance of the \ptmiss algorithms. In events with a $\PW$ boson,
the magnitude of the \ptmiss is approximately equal to the \pt of the lepton,
and its resolution is dominated by the hadronic recoil.

\begin{figure}[!htp]
  \centering
   \includegraphics[width=0.42\textwidth]{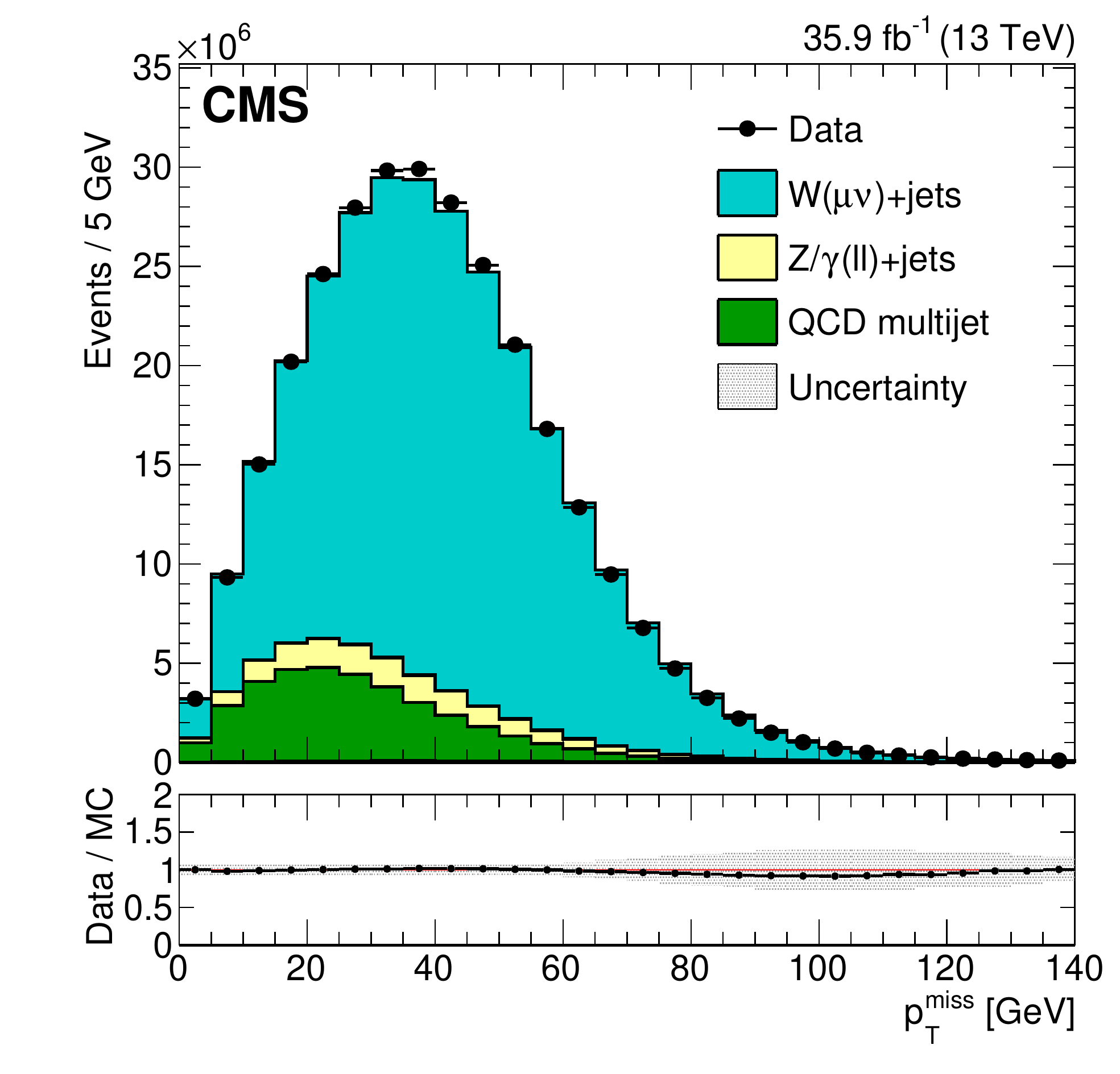}
   \includegraphics[width=0.42\textwidth]{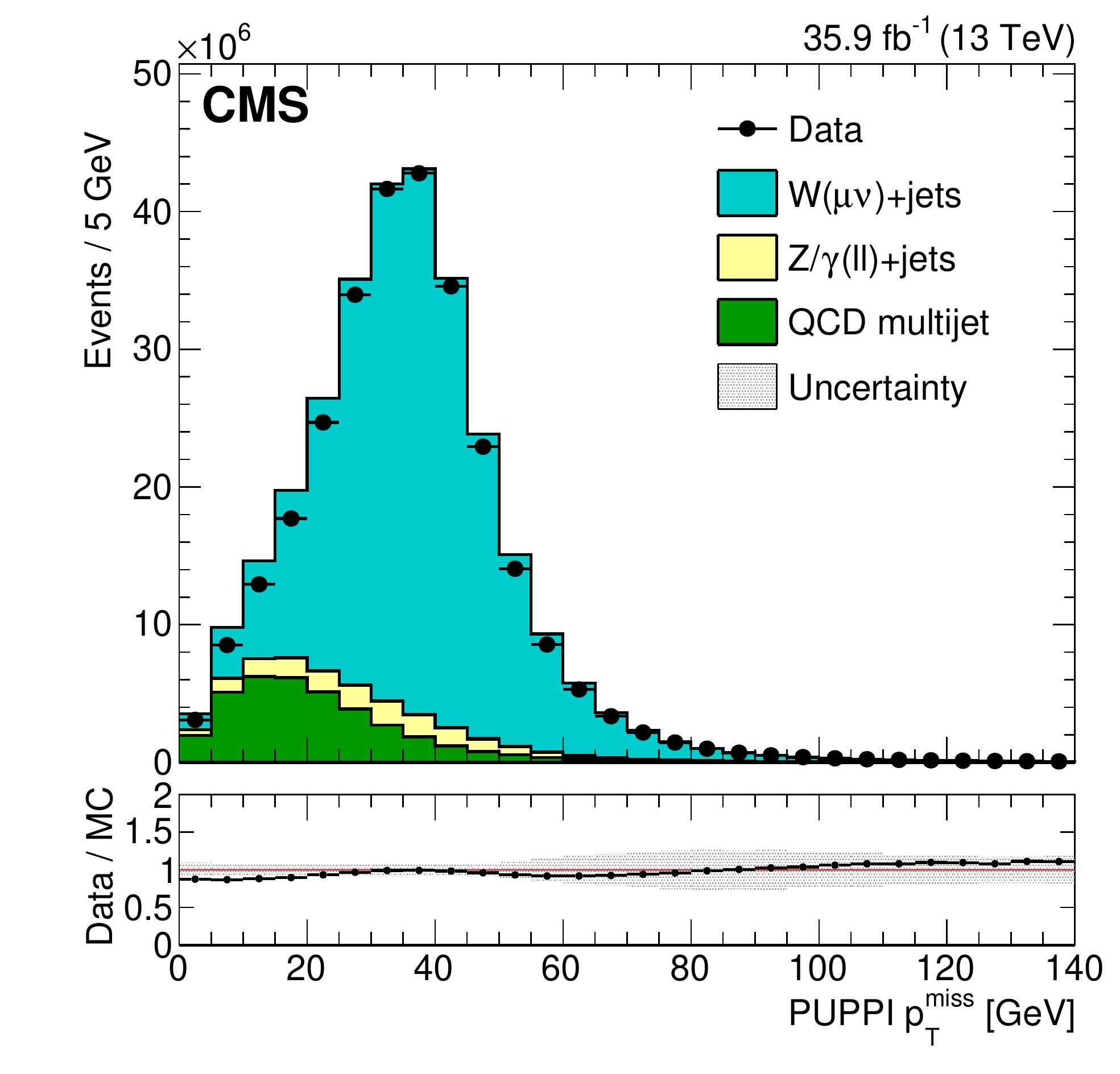}\\
   \includegraphics[width=0.42\textwidth]{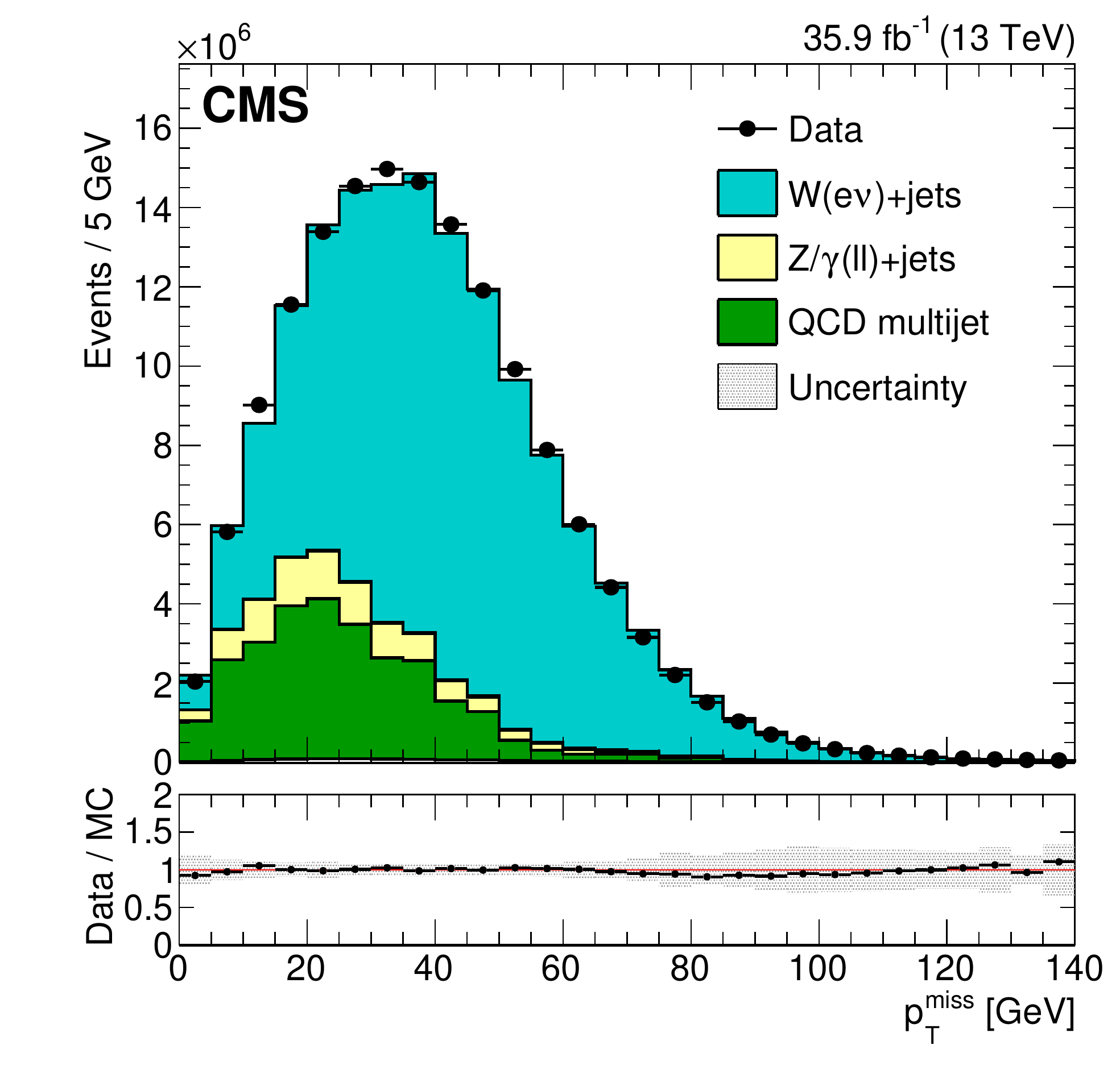}
   \includegraphics[width=0.42\textwidth]{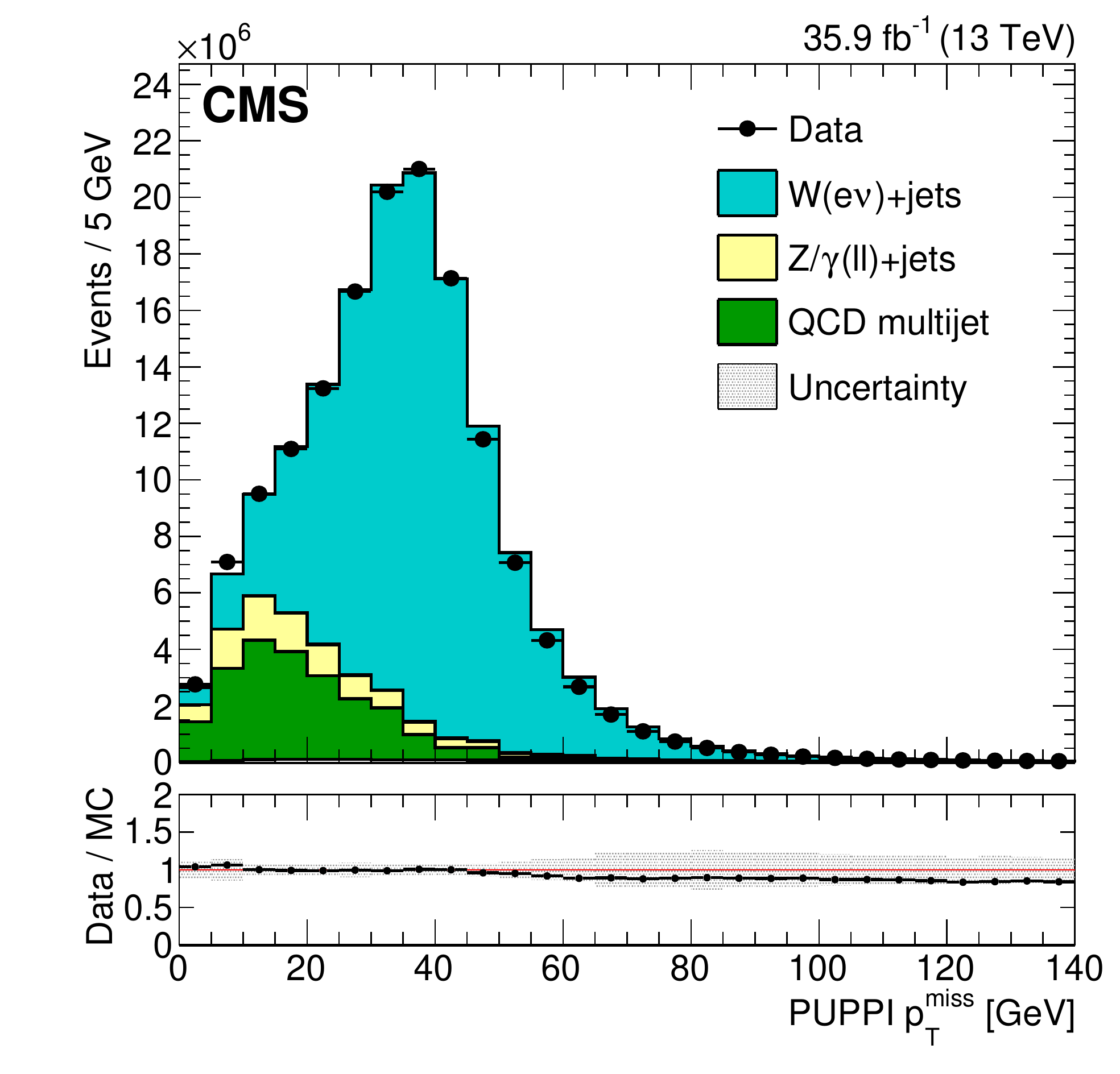}
   \caption{The PF (left) and \puppi (right) \ptmiss distributions are shown for single-muon (upper) and single-electron (lower) events. The last bin includes all events with $\ptmiss>135\GeV$. In all the distributions, the lower panel shows the ratio of data to simulation. The systematic uncertainties due to the JES, the JER, and variations in the $E_{U}$ are added in quadrature and represented by the shaded band.}
  \label{fig:wmet}
\end{figure}

In Fig.~\ref{fig:wmet}, the PF and \puppi \ptmiss distributions are compared in single-muon and -electron samples, where the normalization of the
QCD multijet background is corrected using the method discussed in Section~\ref{sec:onelepselection}. A larger discrimination between events with and without genuine \ptmiss\ is observed for the \puppi \ptmiss algorithm.

\begin{figure}[!htp]
  \centering
   \includegraphics[width=0.42\textwidth]{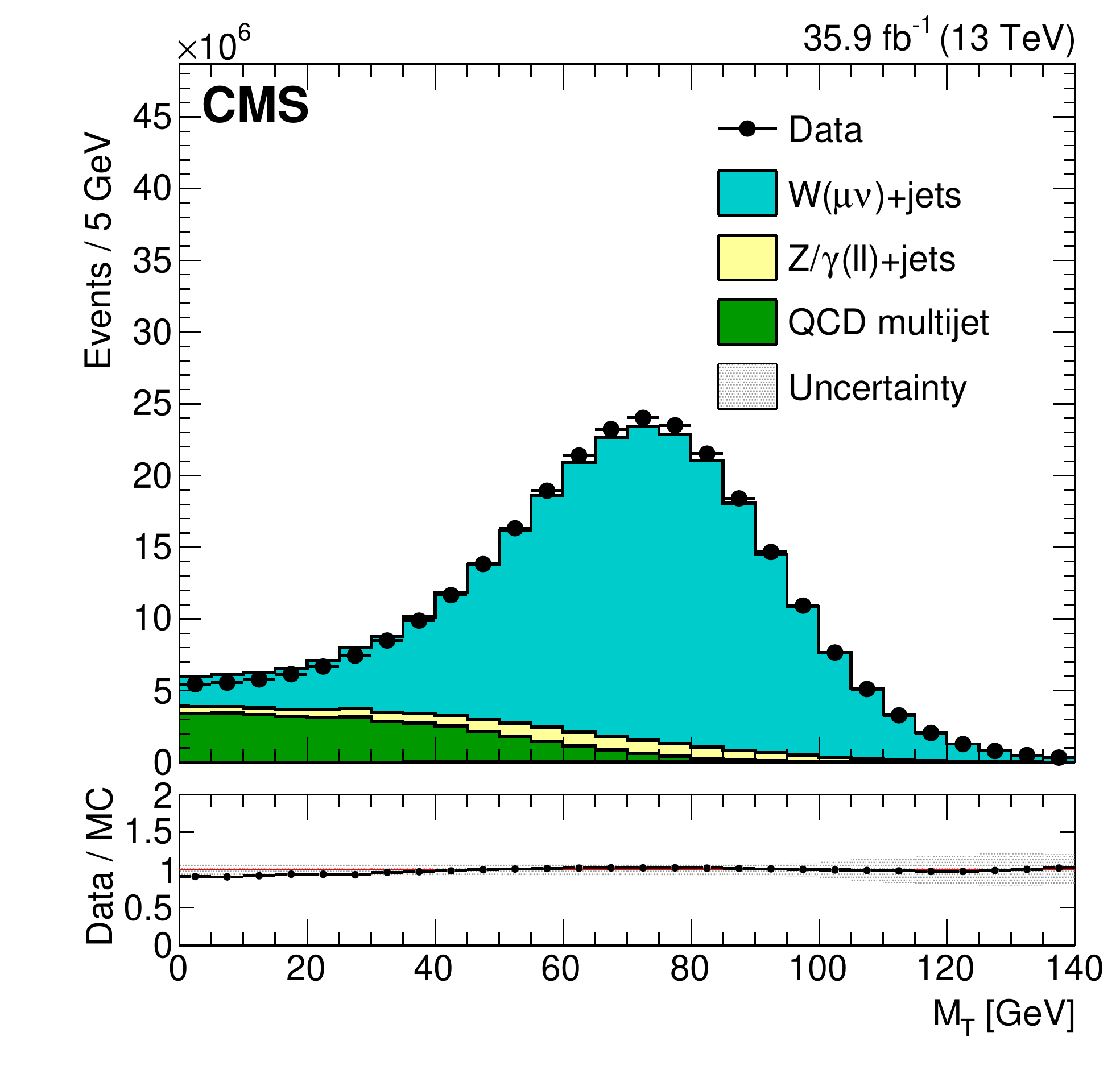}
   \includegraphics[width=0.42\textwidth]{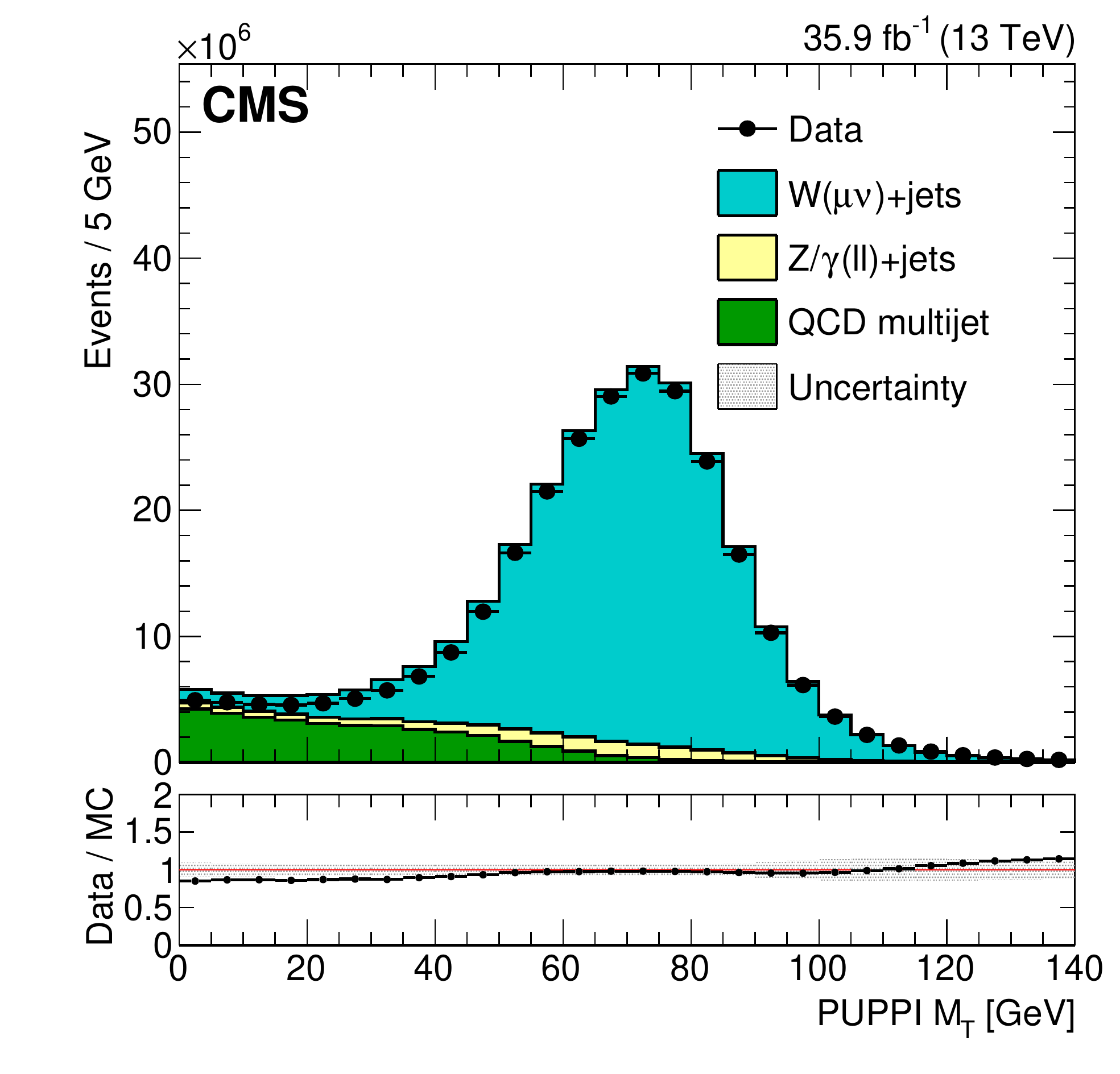}\\
   \includegraphics[width=0.42\textwidth]{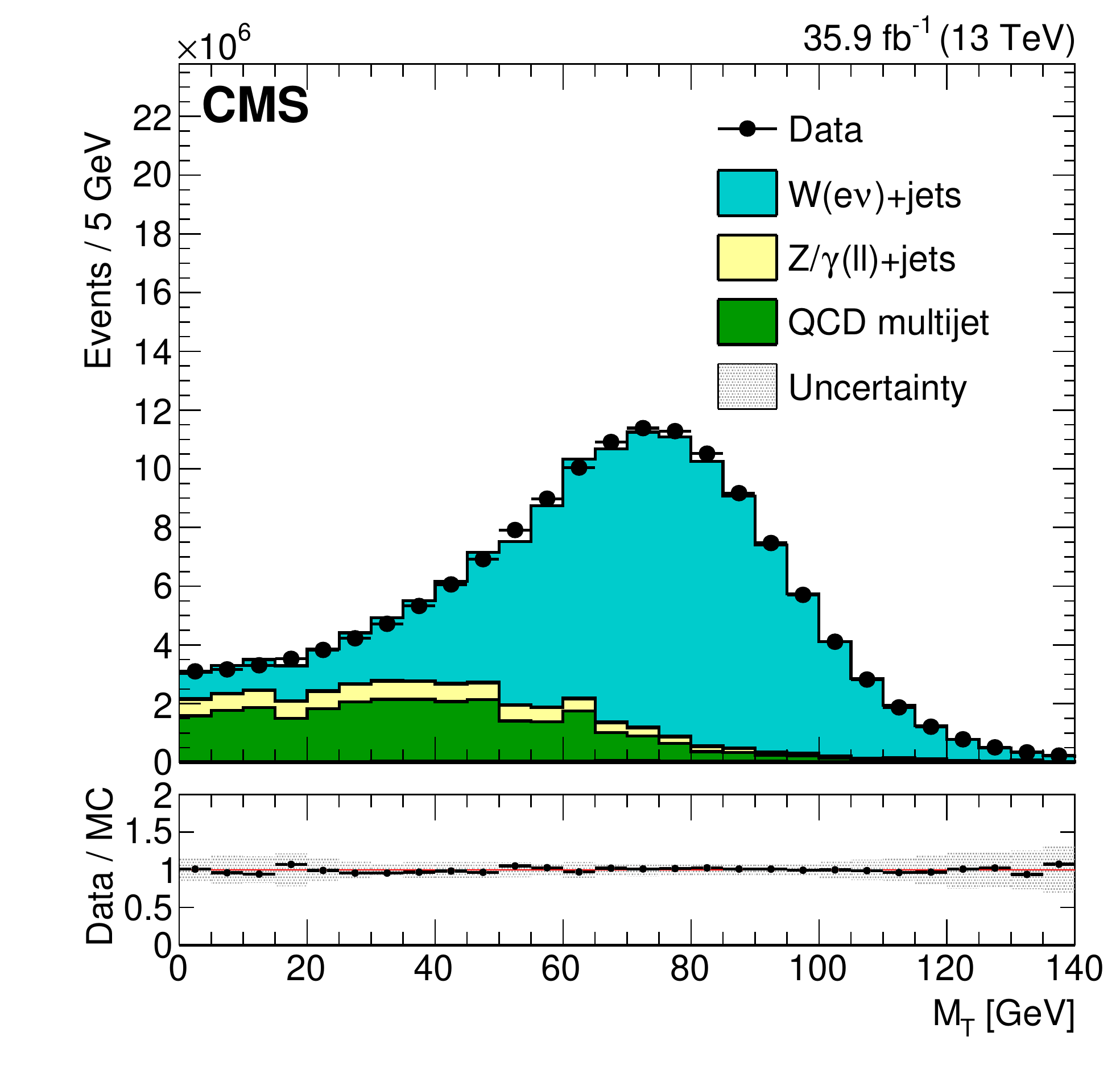}
   \includegraphics[width=0.42\textwidth]{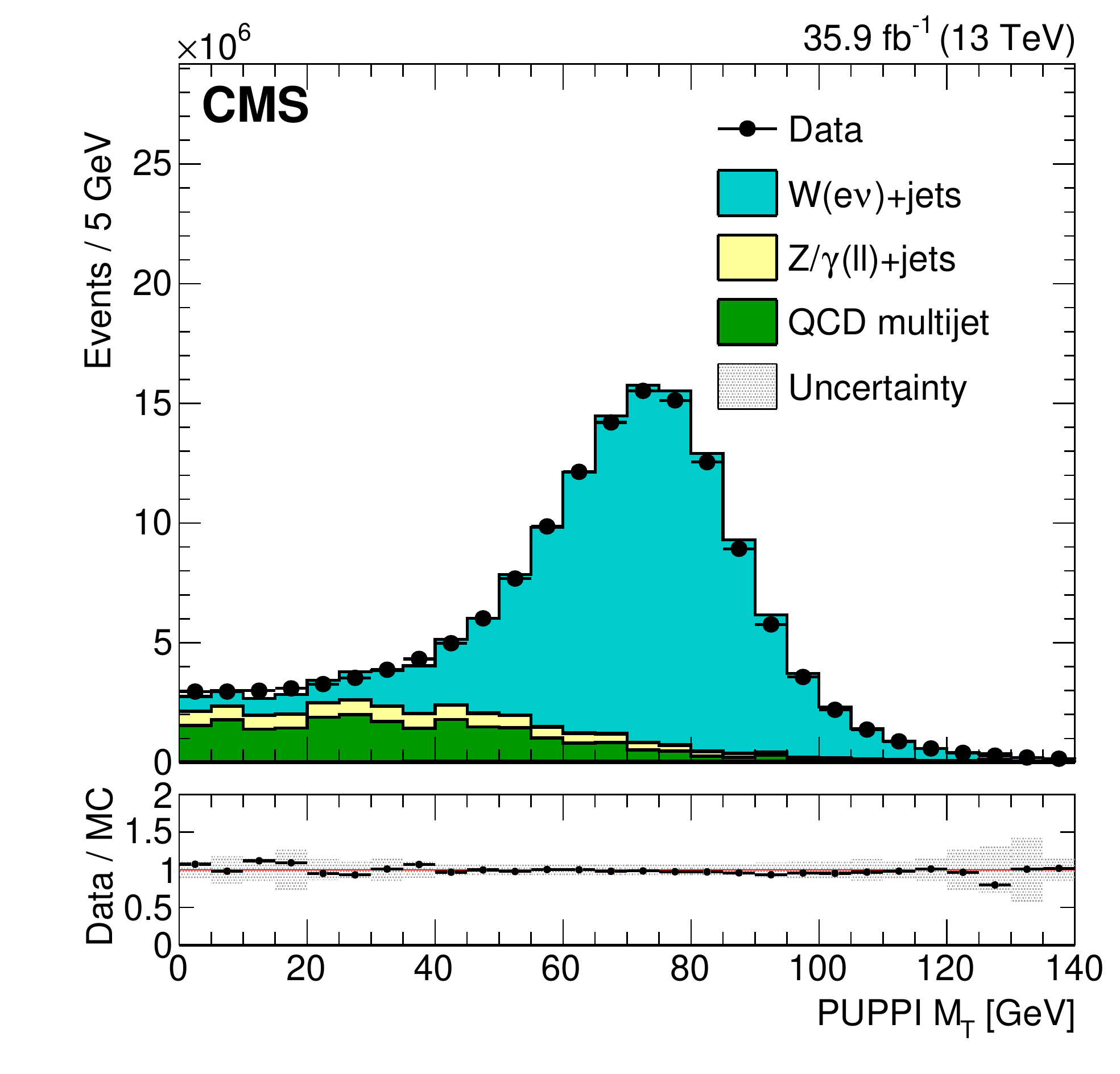}
   \caption{The PF (left) and \puppi (right) \mt~distribution are shown for single-muon (upper) and single-electron (lower) events. The last bin includes all events with $\mt>135\GeV$
In all the distributions, the lower panel shows the ratio of data to simulation. The systematic uncertainties due to the JES, the JER, and variations in the $E_{U}$ are added in quadrature and represented by the shaded band.}
  \label{fig:wmt}
\end{figure}

The transverse mass (\mt) of the lepton-\ptvecmiss system is compared between the algorithms, as shown in  Fig.~\ref{fig:wmt}. The \mt~of the system is computed as:
\begin{linenomath}\begin{equation}\label{eq:mt}
\mt = \sqrt{2\ptmiss \pt^{\text{lepton}} (1 - \cos\Delta\phi)},
\end{equation}\end{linenomath}
where $\pt^{\text{lepton}}$ is the \pt of the lepton, and $\Delta\phi$ is the angle between $\ptvec^{\text{lepton}}$ and $\ptvecmiss$. As in the \ptmiss case, the \puppi algorithm has a better discrimination  between events with and without genuine \ptmiss. In addition, the spread of the Jacobian mass peak
is  smaller when {\mt} is computed using \puppi \ptmiss. The summary of the mean and the spread of the Jacobian mass peak, calculated in simulated $\PW$+jets events, is provided in Table~\ref{tab:wtable}. Utilizing \puppi \ptmiss for the \mt~calculation results in a 10--15\% relative improvement in the resolution of the Jacobian mass peak with respect to PF \ptmiss.

\begin{table}[hbtp]

\centering
\def\arraystretch{1.2}
\topcaption{The summary of the mean and the spread of the Jacobian mass peak in the \mt~distribution in single-lepton events for PF and \puppi \ptmiss algorithms. The results are obtained using simulated $\PW+$jets events.}
\label{tab:wtable}
\begin{tabular}{l  c c  c c}
\hline

Process        & Mean $[\GeVns{}]$ & RMS $[\GeVns{}]$ & Mean $[\GeVns{}]$ & RMS $[\GeVns{}]$ \\ \hline
&  \multicolumn{2}{c}{PF algorithm} & \multicolumn{2}{c}{\puppi algorithm} \\
& \multicolumn{4}{c}{$0<$ \nvtx\ $\leq20$} \\
$\Wmn$          & 76.26 $\pm$ 0.01 & 15.01 $\pm$ 0.01 & 73.44 $\pm$ 0.01 & 13.01 $\pm$ 0.01 \\
$\Wenu$         &  77.46 $\pm$ 0.01 & 15.37 $\pm$ 0.01 & 74.61 $\pm$ 0.01 & 13.18 $\pm$ 0.01\\

& \multicolumn{4}{c}{$20<$ \nvtx\ $\leq30$} \\
$\Wmn$          & 78.58 $\pm$ 0.01 & 16.45 $\pm$ 0.01 & 74.21 $\pm$ 0.01 & 13.65 $\pm$ 0.01\\
$\Wenu$         & 79.96 $\pm$ 0.01 & 16.74 $\pm$ 0.01 & 75.45 $\pm$ 0.01 & 13.87 $\pm$ 0.01\\

& \multicolumn{4}{c}{ \nvtx\ $\geq30$} \\
$\Wmn$          & 80.75 $\pm$ 0.02 & 17.47 $\pm$ 0.01 & 75.29 $\pm$ 0.01 & 14.43 $\pm$ 0.01\\
$\Wenu$         & 82.26 $\pm$ 0.03 & 17.73 $\pm$ 0.02 & 76.68 $\pm$ 0.02 & 14.70 $\pm$ 0.02\\
\hline
\end{tabular}
\end{table}

\section{The \texorpdfstring{\ptmiss}{pT miss} significance}
\label{sec:metsignificance}

The ability to distinguish between events with genuine \ptmiss and those
with spurious \ptmiss is important for analyses targeting
signatures with weakly interacting particles.
The \ptmiss significance variable, denoted by \metsig, quantifies
the degree of compatibility of  \ptmiss with zero
on an event-by-event basis, and it is computed using all clustered objects and
the $E_{U}$ in each event.
A factorized approach leads to the construction of a significance variable that is
applicable to a variety of event topologies.
The variable is described in detail
in Refs.~\cite{2010,Khachatryan:2014gga}. Here we give an overview of
updates and performance studies conducted using the 13 TeV data set.

The significance is defined as the log-likelihood ratio
\begin{linenomath}\begin{equation}
\metsig \equiv 2\ln\left(\frac
{\like(\vet=\sum\vet_{i})}
{\like(\vet=0)}
\right),
\label{eq:metsig:defn}
\end{equation}\end{linenomath}
where the \vet\ is the true \ptmiss\ and $\sum\vet_i$ is the observed \ptmiss.
In the numerator, we evaluate the likelihood that the true value of \ptmiss\ equals the observed value, while the denominator corresponds to the null hypothesis, i.e., that the true \ptmiss\ is zero.
To a very good approximation the likelihood $\like(\vet)$ has the form of a Gaussian distribution. The significance can be therefore written as:
\begin{linenomath}
\begin{equation}
\metsig = \Big(\sum\vet_i \Big)\!^{\text{T}} {\mathbf V^{-1}} \Big( \sum\vet_i \Big),
\label{e:metsig-gaussian}
\end{equation}
\end{linenomath}
where $\mathbf{V}$ is the 2$\times$2 \ptmiss\ covariance matrix.
In this formulation, \metsig\ is conveniently a $\chi^2$ variable with two degrees of freedom (one degree of freedom each for the $x$-- and $y$--axis components of \ptmiss) for events with zero true \ptmiss.

The covariance matrix $\mathbf{V}$ in Eq.~(\ref{e:metsig-gaussian}) models the \ptmiss\ resolution in each event.
It is constructed by propagating the individual resolutions of the objects entering the \ptmiss\ sum.  In most cases, the \ptmiss\ resolution captured in $\mathbf{V}$ is primarily determined by the hadronic components of the event, which includes jets with $\pt > 15\GeV$ and the $E_{U}$. Jets enter the total covariance $\mathbf{V}$ with an individual covariance of the form:
\begin{linenomath}
\begin{equation}
\mathbf{U} = \left( \begin{matrix}
   \sigma_{\pt}^2 & 0 \\
   0 & \pt^2\,\sigma_{\phi}^2
   \end{matrix} \right),
\label{eq:cov}
\end{equation}
\end{linenomath}
where the quantities $\sigma_{\pt}$ and $\sigma_{\phi}$ are measured and then recalculated based on a combination of simulation and data control samples, as explained in Ref.~\cite{Khachatryan:2014gga}. The momenta of the PF candidates $i$ that is not included in a jet are summed vectorially, and the resulting momentum is assigned to a single pseudo-object i.e., $\vec{\pt} = \sum_i{\vec{\pt}_i}$.  The resolution of this pseudo-object is parameterized by the scalar \pt sum of its constituents:
\begin{linenomath}
\begin{equation}
  \sigma_{\text{uc}}^2 = \sigma_0^2 + \sigma_\text{s}^2\sum_{i=1}^{n}{\abs{\vec{p}_{\text{T}_i}}},
  \label{e:pseudo}
\end{equation}
\end{linenomath}
where the values of $\sigma_0^2$ and $\sigma_\text{s}^2$ are determined using control samples in data, as explained in Ref.~\cite{Khachatryan:2014gga}.  The resolution of this object is assumed to be isotropic in the transverse plane of the detector.
The finite (small) resolution of electrons and muons is negligible, compared to the hadronic component of the event, and hence their contribution to $\mathbf{V}$ is neglected

\subsection{Unclustered energy studies}
The unclustered PF candidates are combined into a pseudo-object.
Its resolution should be isotropic in the transverse plane, and  proportional to the magnitude of the \pt of the pseudo-object.
This approach, called the ``standard'' method of \metsig\ in what follows,
is motivated by its simplicity, and  shows good agreement
between data and simulation.
The diagonal elements of the contribution of the $E_{U}$ to the covariance matrix are given by Eq.~(\ref{e:pseudo}).

During the data--taking run, an alternative method to obtain the covariance matrix was explored, the so-called ``jackknife technique''~\cite{jk1,jk2}. The jackknife technique allows the estimation of a covariance matrix that is not necessarily isotropic, and also includes offdiagonal elements. The covariance matrix is calculated using the ``delete-1 method'', in which a single PF candidate is removed.
This approach leads to $N-1$ samples per event, with N the total numbers of constituents contributing to the $E_{U}$. The covariance matrix takes the form:
\begin{linenomath}
\begin{equation}
  \hat{V}_{ij} = \frac{N-1}{N}\sum_{k=1}^{N}(p_i^k-\overline{p}_i)(p_j^k-\overline{p}_j),
\end{equation}
\end{linenomath}
where $k$ is the removed candidate, $p_i^k$ and $p_j^k$ are x and y components of the $E_{U}$ calculated after removing the $k$-th candidate, whereas the indexes i and j both span x and y. The $\overline{p}_i$ and $\overline{p}_j$ are mean values of x and y components of the $E_{U}$ over all samples, defined as:
\begin{linenomath}
\begin{equation}
  \overline{p}_{i,j} = \frac{1}{N} \sum_{k=1}^{N}p_{i,j}^k.
\end{equation}
\end{linenomath}
Again, the resolution is scaled by the parameters tuned in data and simulated samples, referred to as $a_{x}$ and $a_{y}$.
The parameters are determined following a similar approach as in the standard method of \metsig.
The resolutions of the components of the $E_{U}$ are then defined as
\begin{linenomath}
\begin{equation}
\begin{split}
  \sigma^2_{x} & = a^2_{x} \, \hat{V}_{xx},\\
  \sigma^2_{xy} & = a_{x} \, a_{y} \, \hat{V}_{xy},\\
  \sigma^2_{y} & = a^2_{y} \, \hat{V}_{yy}.
\end{split}
\end{equation}
\end{linenomath}

\subsection{Performance evaluation}
\begin{figure}[!htp]
  \centering
  \includegraphics[width=0.42\textwidth]{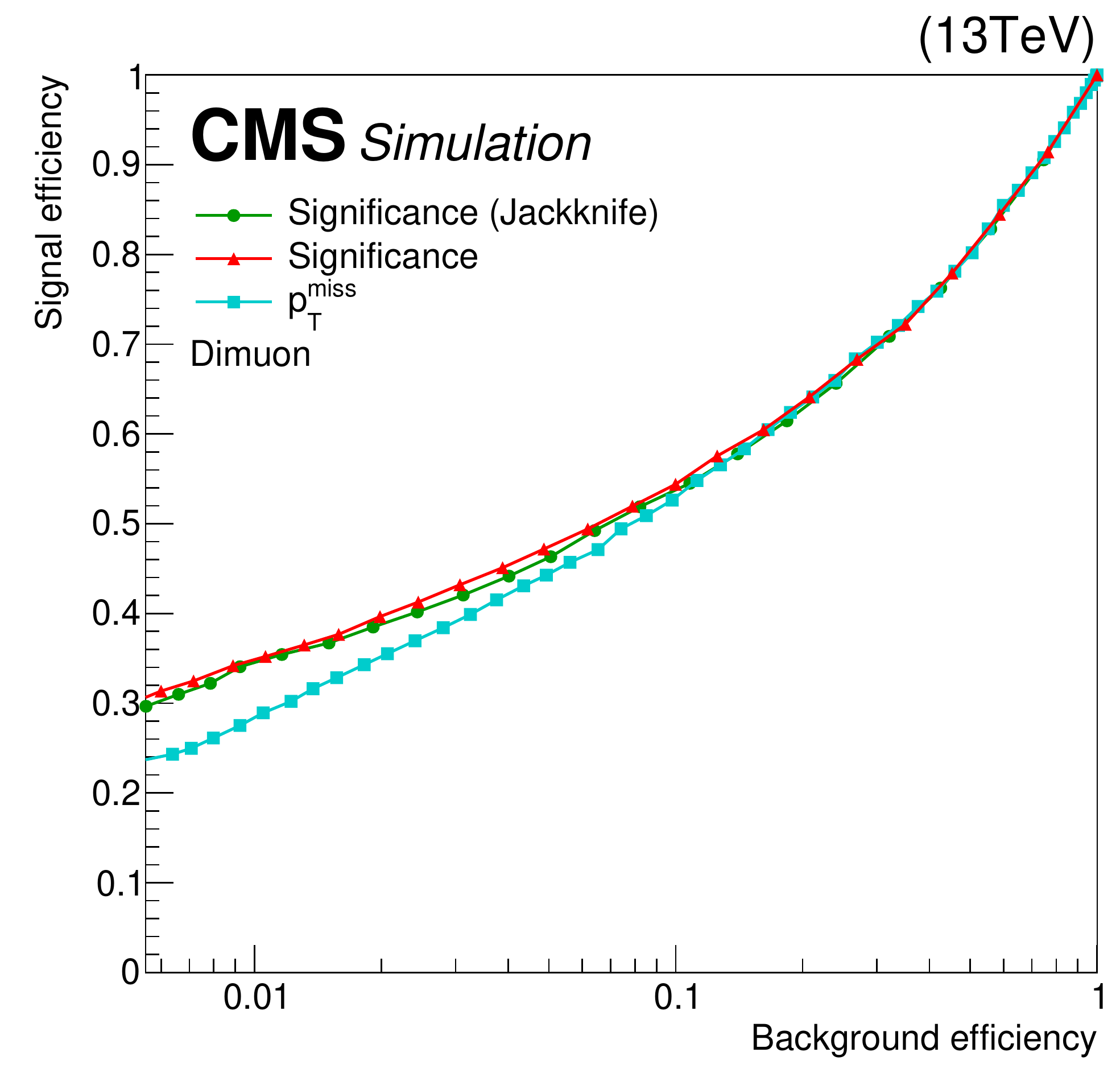}
  \includegraphics[width=0.42\textwidth]{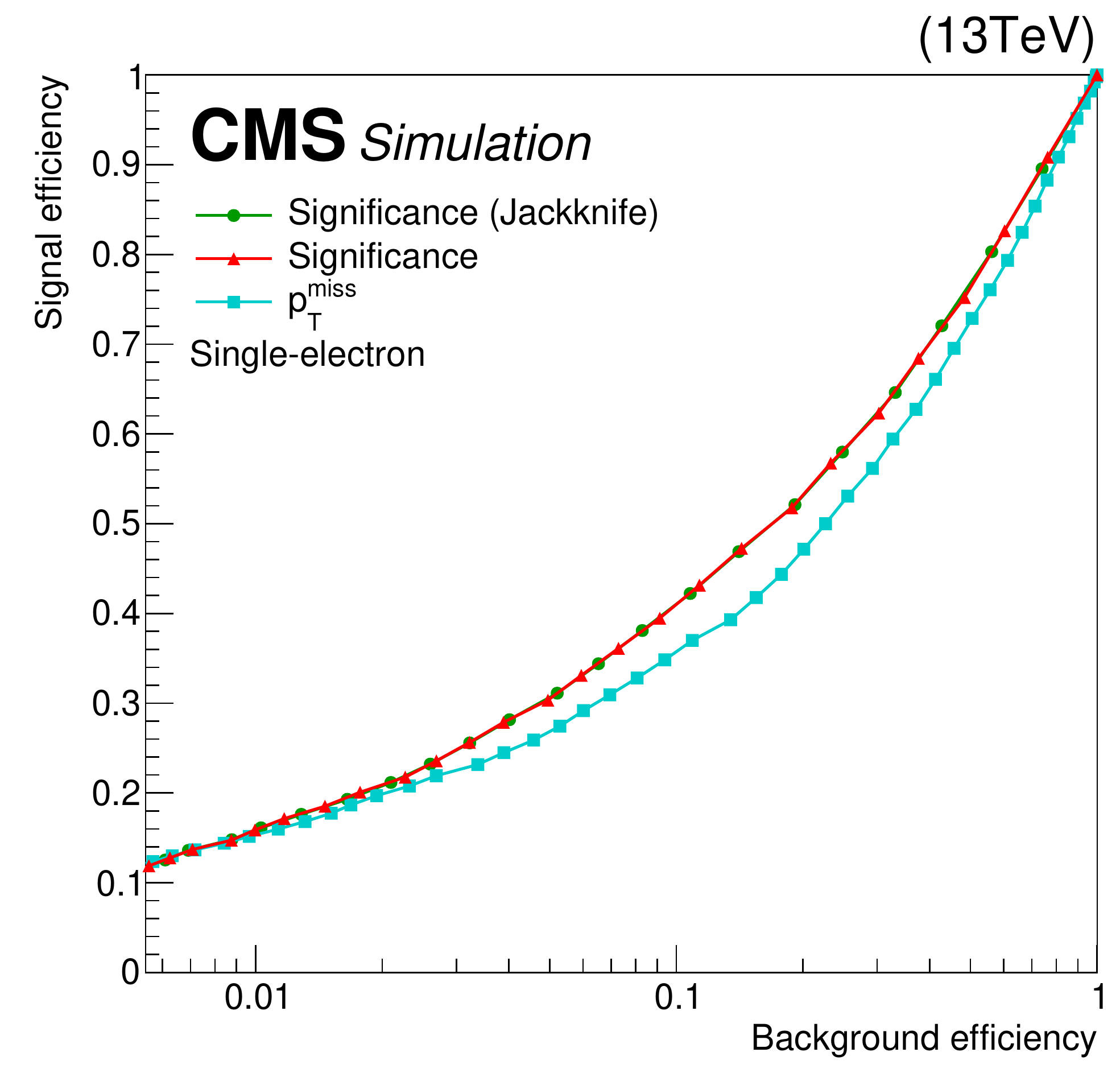}
  \caption{ROC curves comparing the signal (events with genuine \ptmiss) versus background (events with no genuine \ptmiss) efficiency for the standard version of \metsig\ (red line), the jackknife version of \metsig\ (yellow line), and \ptmiss (cyan line) using simulated dimuon events (left) and single-electron events (right). Similar performance is observed between the two versions of \metsig, which perform better than \ptmiss especially in regions with small background efficiency.}
  \label{fig:ROCsig}
\end{figure}

The discrimination power between events with genuine \ptmiss\ (signal) and those without (background), of the two versions of \metsig, the standard and the jackknife, and the \ptmiss\ algorithms, is compared in terms of receiver operating characteristic (ROC) curves. The results are shown in Fig.~\ref{fig:ROCsig} using simulated dimuon events (a sample dominated by events with no genuine \ptmiss) and single-electron events (a sample dominated by events with genuine \ptmiss). No significant difference between the two \metsig\ versions is observed. Both versions of \metsig\ offer better signal-to-background separation than \ptmiss. For example, choosing a working point with 1\% background efficiency the \metsig\ variables offer 5\% higher signal efficiency than
\ptmiss. For the remainder of the section,
we focus only on the standard version of \metsig.

\begin{figure}[!htp]
  \centering
  \includegraphics[width=0.42\textwidth]{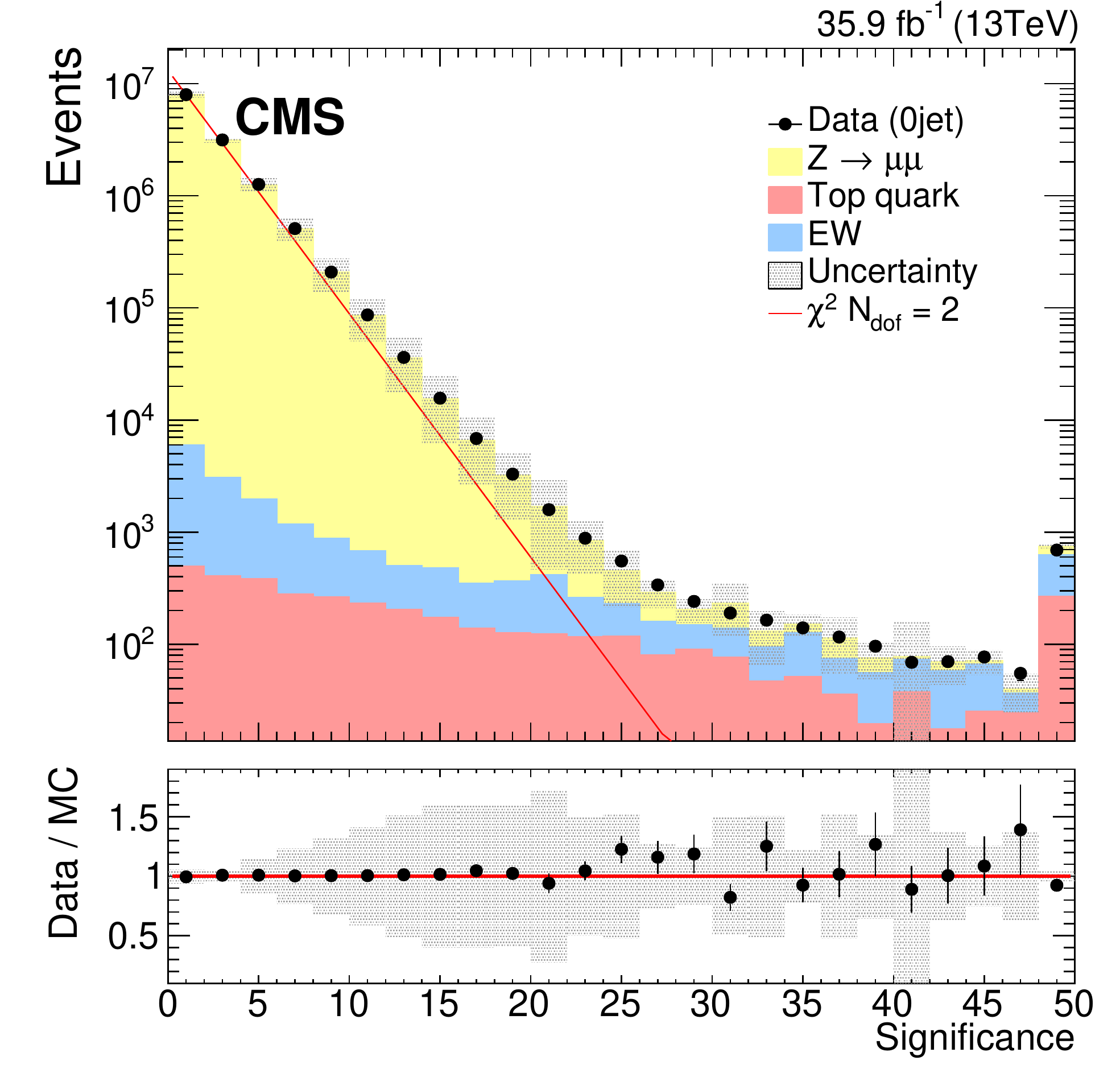}
  \includegraphics[width=0.42\textwidth]{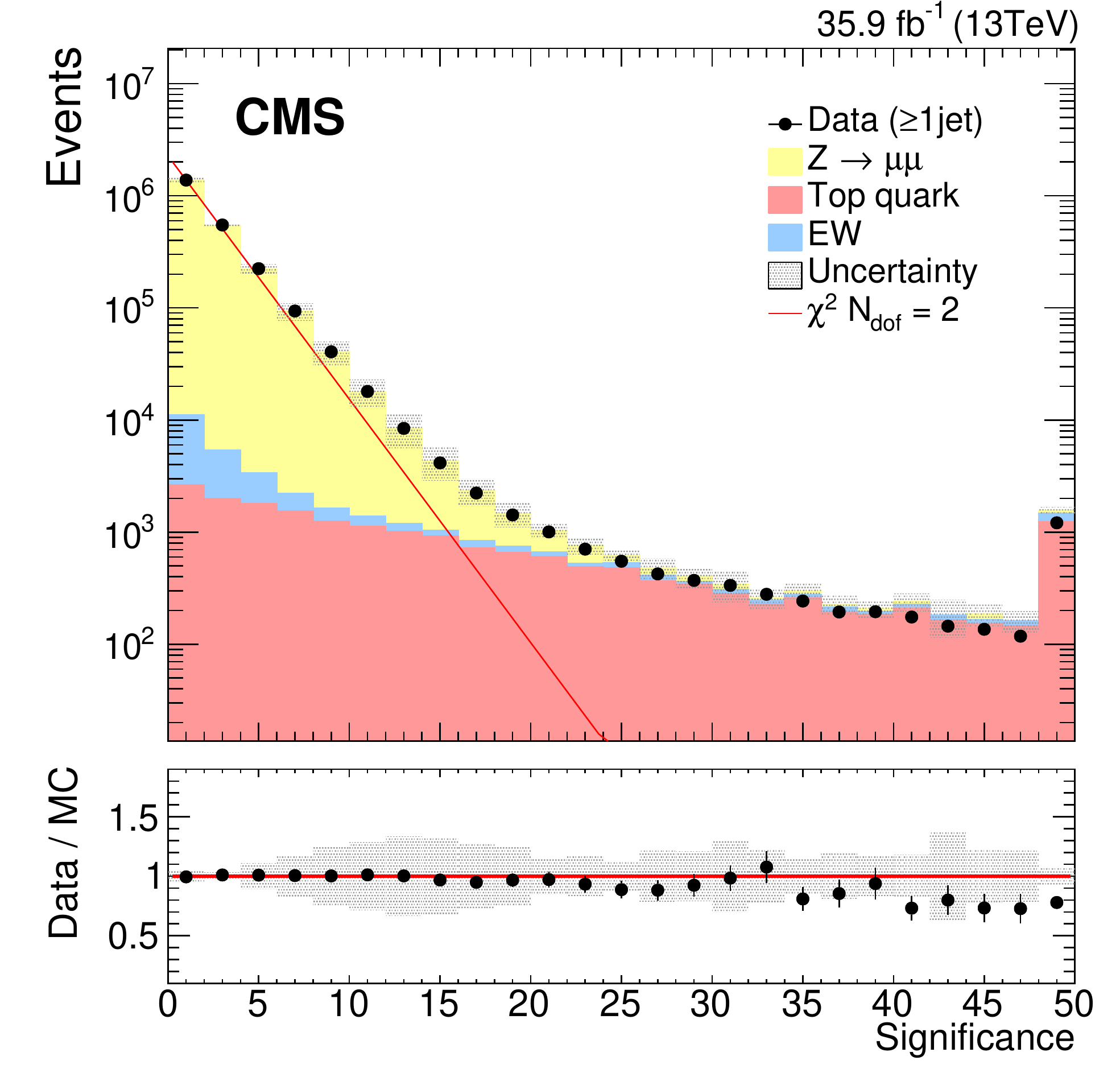} \\
  \includegraphics[width=0.42\textwidth]{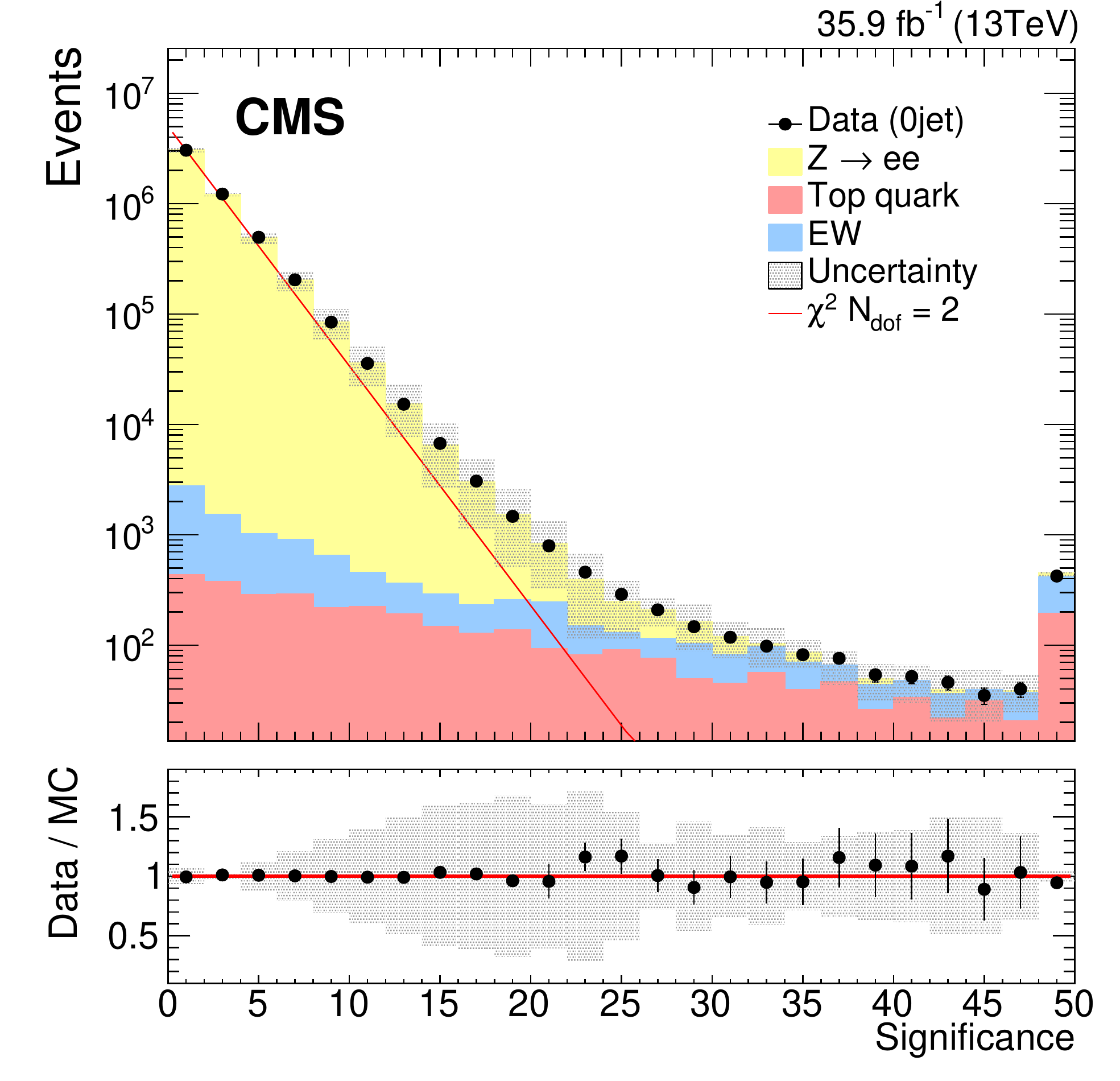}
  \includegraphics[width=0.42\textwidth]{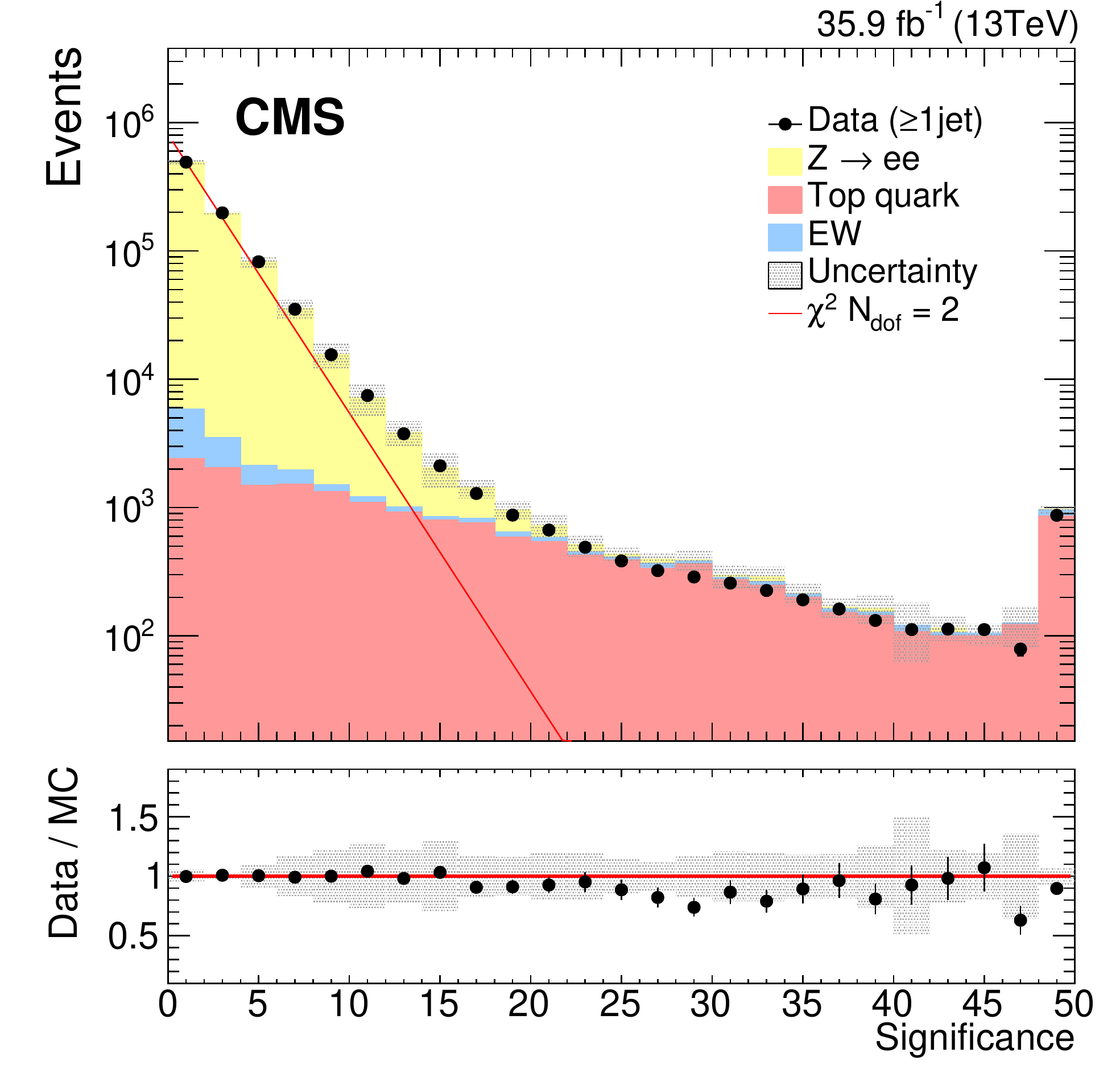}
  \caption{Distributions of \metsig\ in data and simulation in dimuon (upper) and dielectron (lower) samples, for events with zero jet (left) and $\geq$ 1 jet (right). The last bin includes all events with $\metsig>48$. The red straight line corresponds to a $\chi^2$ distribution with two degrees of freedom.  The bands in the bottom panel display systematic uncertainties due to effects from the JES, the JER, and variations in the $E_{U}$ in simulation. Good agreement between data and simulation is observed.}
  \label{fig:metsigInDiLepData}
\end{figure}

\begin{figure}[!htp]
  \centering
  \includegraphics[width=0.42\textwidth]{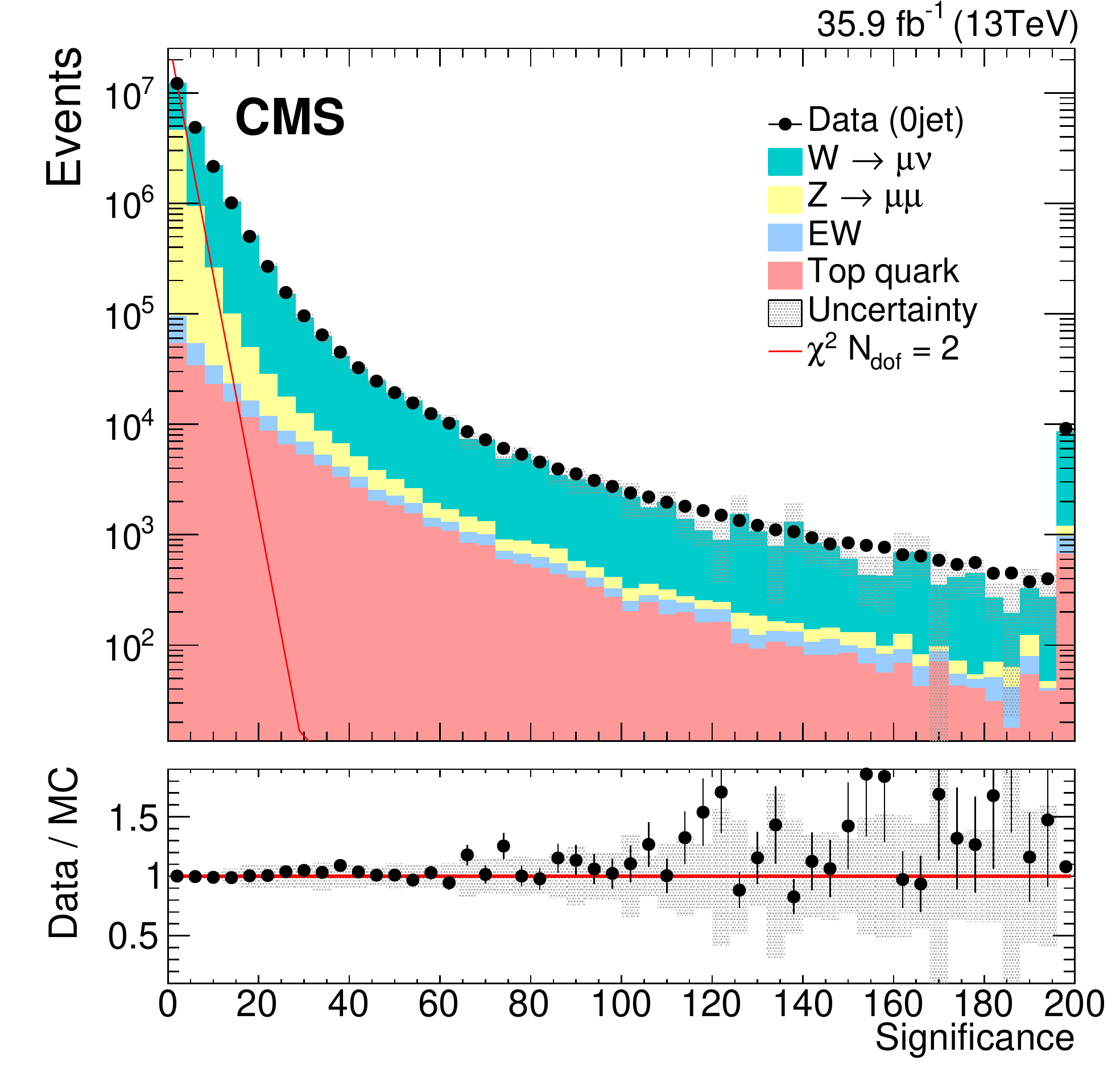}
  \includegraphics[width=0.42\textwidth]{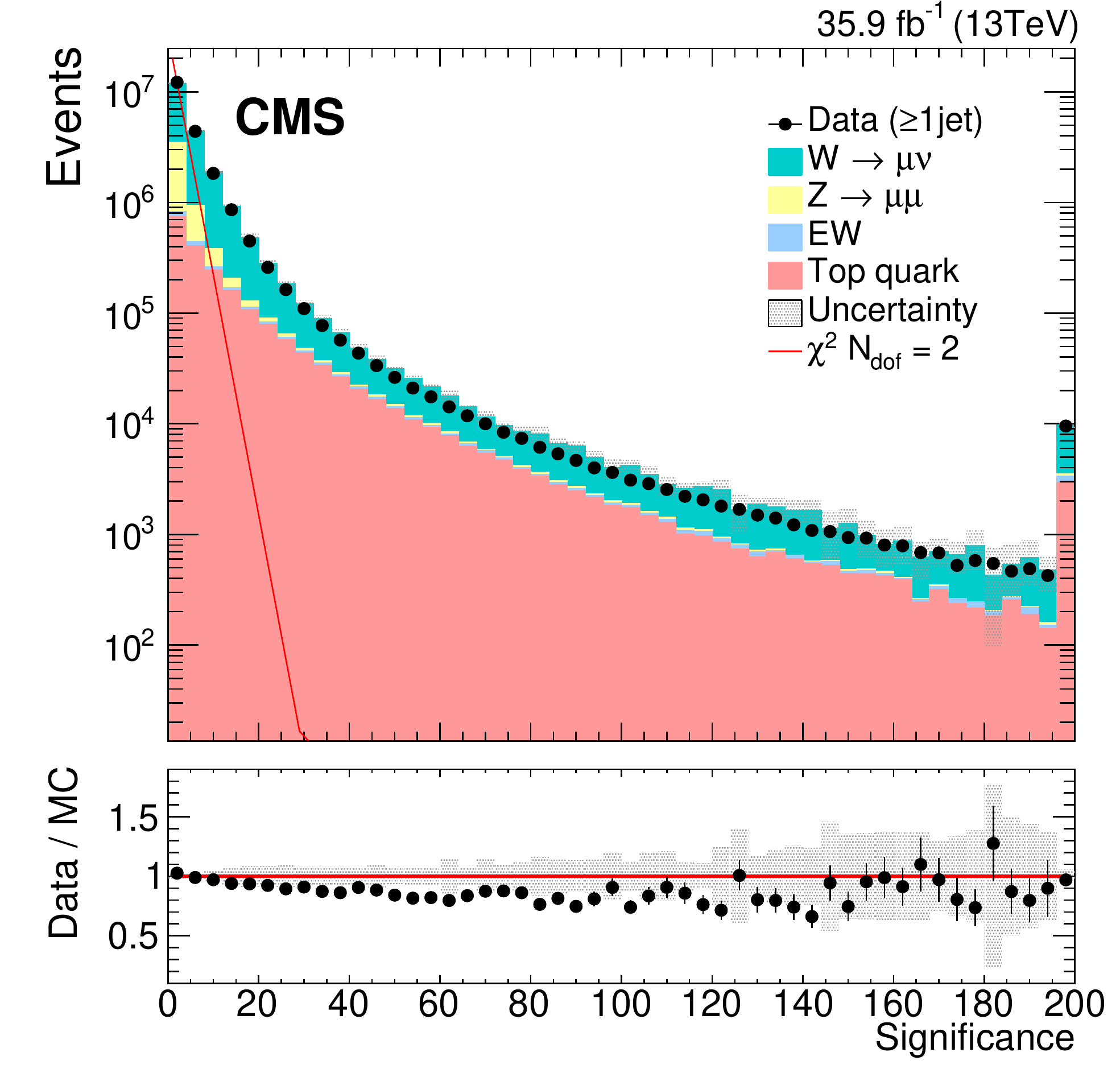} \\
  \includegraphics[width=0.42\textwidth]{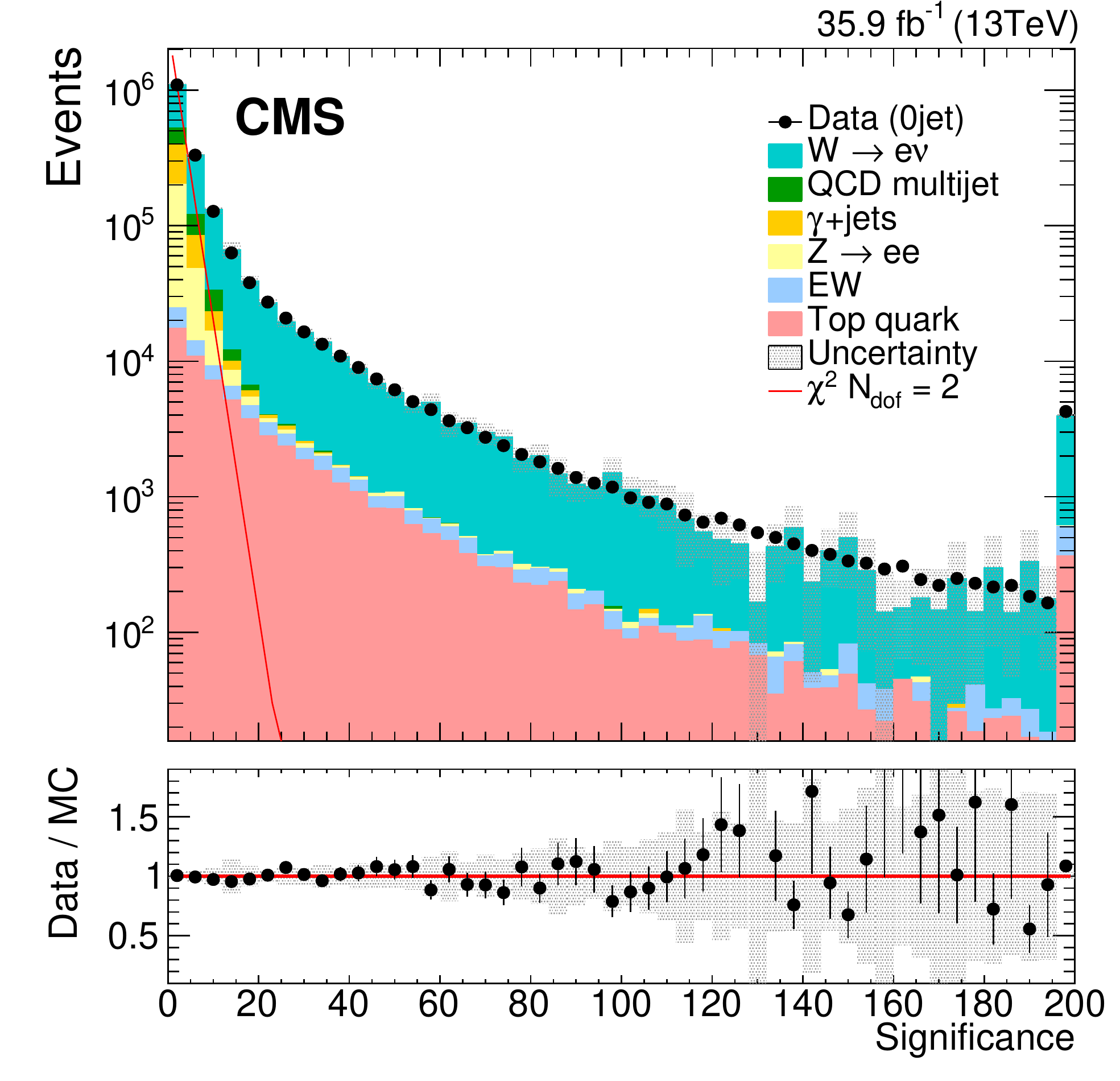}
  \includegraphics[width=0.42\textwidth]{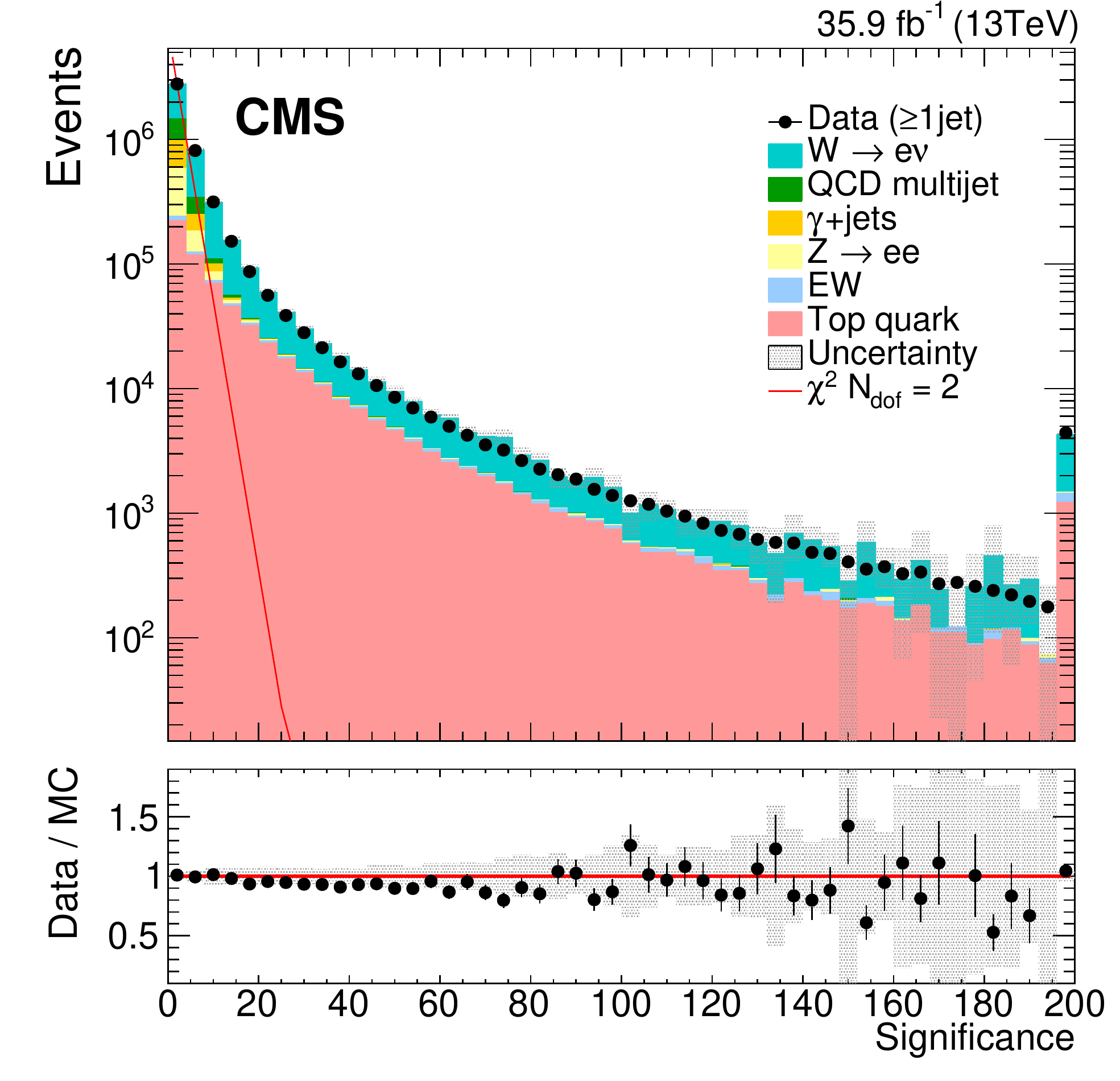}
  \caption{Distributions of \metsig\ in data and simulation in single-muon (upper) and single-electron (lower) samples, for events with zero jet (left) and $\geq$ 1 jet (right).
    The last bin includes all events with $\metsig>48$. The red straight line corresponds to a $\chi^2$ distribution with two degrees of freedom. The bands in the bottom panel display systematic uncertainties due to
    effects from the JES, the JER, and variations in the $E_{U}$ in simulation. Good agreement between data and simulation is observed.}
  \label{fig:metsigInData}
\end{figure}

The performance of \metsig\ is evaluated in data using dilepton and single-lepton events. The results are displayed in Figs. \ref{fig:metsigInDiLepData} and \ref{fig:metsigInData}, respectively, for different jet multiplicities. In Fig.~\ref{fig:metsigInDiLepData}, where events with no genuine \ptmiss dominate, the core of the \metsig\ spectrum  follows an ideal $\chi^2$ distribution. For large values of \metsig\ the spectrum begins to deviate from a perfect $\chi^2$ distribution as the processes with genuine \ptmiss become important.
This deviation also has contributions from the nonGaussian tails of
the jet \pt resolution function, which are not considered in Eq.~(\ref{e:metsig-gaussian}). A detailed discussion of the treatment of nonGaussian resolutions can be found in~\cite{Khachatryan:2014gga}.

\begin{figure}[!htp]
  \centering
  \includegraphics[width=0.42\textwidth]{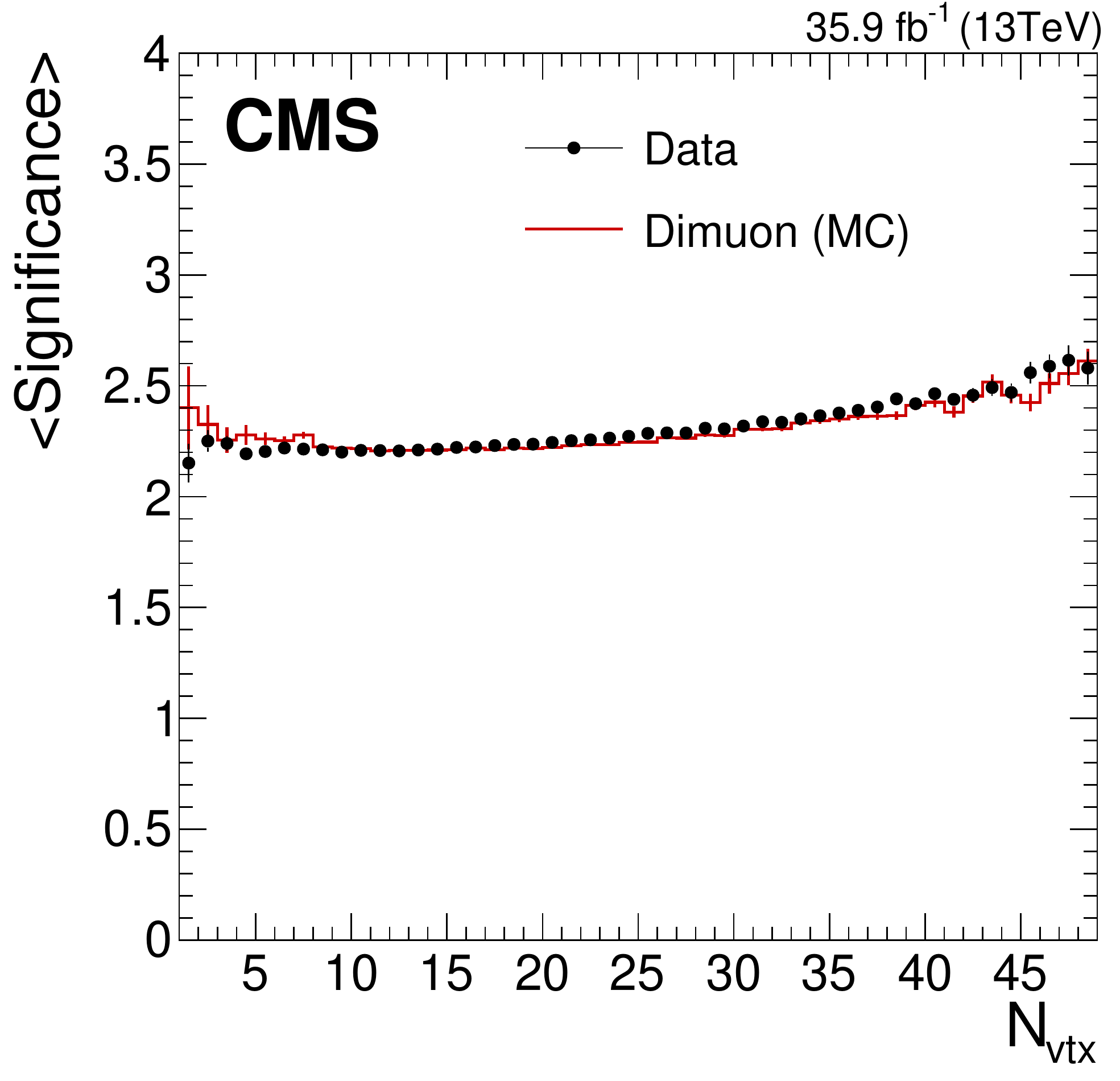}
  \includegraphics[width=0.42\textwidth]{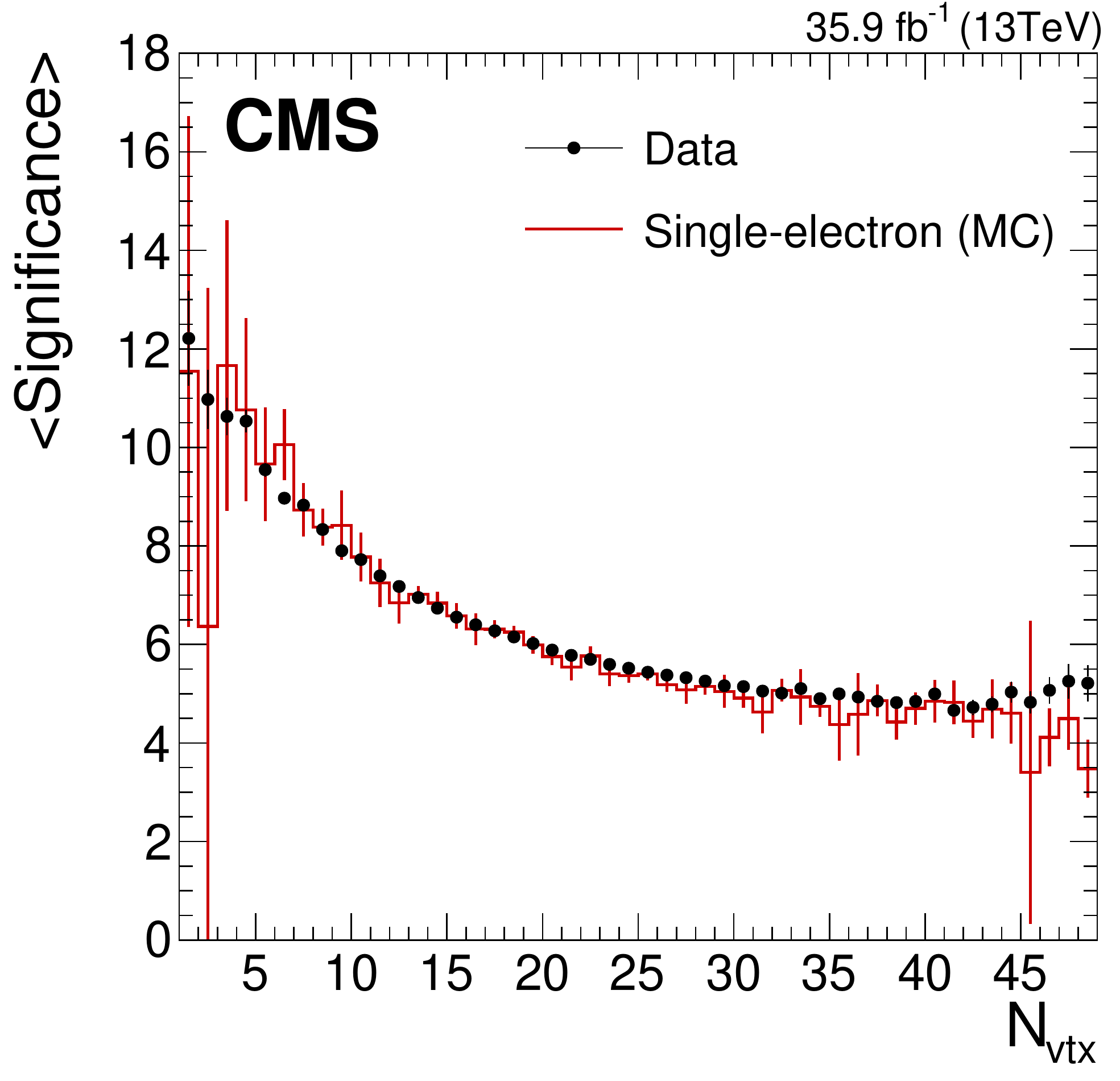}
  \caption{Dependence of the average \metsig\ on pileup, for dimuon (left) and single-electron (right) events.
    Weak dependence is observed for processes with no genuine \ptmiss, whereas in events with genuine \ptmiss the behavior of \metsig\ depends strongly on primary vertex multiplicity. }
  \label{fig:PUsig}
\end{figure}

The stability of \metsig\ against pileup is studied using dimuon and single-electron events. Figure~\ref{fig:PUsig} displays the average \metsig\ as a function of \nvtx.
In the dimuon sample, dominated by events with no genuine \ptmiss, the value of \metsig\ is robust against pileup, with an average value of $\sim2$, as expected for a $\chi^{2}$ variable with two degrees of freedom. This behavior can be explained qualitatively with the following arguments. In the case of events with no genuine \ptmiss, the contribution of pileup affects in a similar manner both \ptmiss and the variance of \ptmiss, since both are dominated by the hadronic resolution.
This results in an essentially constant value of \metsig\, which
does not depend on the number of pileup interactions.
However, in events with genuine \ptmiss, as in the single-electron sample, pileup has a small impact on \ptmiss,
whereas the impact on the resolution in \ptmiss
is similar to the case of no genuine \ptmiss,
leading to a decrease of \metsig\ as pileup increases.
This results in a degradation in the performance of \metsig\ when \nvtx\ is large.

\section{Summary}
\label{sec:summary}

The performance of missing transverse momentum (\ptmiss) reconstruction algorithms in events with or without genuine \ptmiss\ is presented. The results are based on a sample of proton-proton collisions recorded by the CMS experiment at $\sqrt{s} = 13$ TeV in 2016, corresponding to
an integrated luminosity of 35.9\fbinv.

The performance of algorithms
used to identify and remove events with anomalous \ptmiss\ is also studied
in events with one or more jets. The scale and resolution of \ptmiss\ is determined using events with an identified leptonically decaying \PZ boson or an isolated photon.
The measured scale and resolution in data are  in agreement with the expectations from simulation.
Also presented is the performance of an advanced \ptmiss reconstruction algorithm, the ``pileup per particle identification'' \ptmiss, specifically developed to cope with the large  pileup collisions expected at the high--luminosity LHC.
This algorithm shows a significantly reduced dependence of the \ptmiss resolution on the number of pileup collisions ($\gtrsim 10$), particularly important for the upcoming LHC data--taking periods. Finally, the performance of an algorithm (\metsig) used to estimate the compatibility of the reconstructed \ptmiss\ with the hypothesis that it originates from resolution effects, was studied. The \metsig\ shows improved performance in discriminating between events with and without genuine \ptmiss\ compared to the traditional \ptmiss\ reconstruction algorithms.

\section*{Acknowledgments}
\hyphenation{Bundes-ministerium Forschungs-gemeinschaft Forschungs-zentren Rachada-pisek} We congratulate our colleagues in the CERN accelerator departments for the excellent performance of the LHC and thank the technical and administrative staffs at CERN and at other CMS institutes for their contributions to the success of the CMS effort. In addition, we gratefully acknowledge the computing centers and personnel of the Worldwide LHC Computing Grid for delivering so effectively the computing infrastructure essential to our analyses. Finally, we acknowledge the enduring support for the construction and operation of the LHC and the CMS detector provided by the following funding agencies: the Austrian Federal Ministry of Science, Research and Economy and the Austrian Science Fund; the Belgian Fonds de la Recherche Scientifique, and Fonds voor Wetenschappelijk Onderzoek; the Brazilian Funding Agencies (CNPq, CAPES, FAPERJ, and FAPESP); the Bulgarian Ministry of Education and Science; CERN; the Chinese Academy of Sciences, Ministry of Science and Technology, and National Natural Science Foundation of China; the Colombian Funding Agency (COLCIENCIAS); the Croatian Ministry of Science, Education and Sport, and the Croatian Science Foundation; the Research Promotion Foundation, Cyprus; the Secretariat for Higher Education, Science, Technology and Innovation, Ecuador; the Ministry of Education and Research, Estonian Research Council via IUT23-4 and IUT23-6 and European Regional Development Fund, Estonia; the Academy of Finland, Finnish Ministry of Education and Culture, and Helsinki Institute of Physics; the Institut National de Physique Nucl\'eaire et de Physique des Particules~/~CNRS, and Commissariat \`a l'\'Energie Atomique et aux \'Energies Alternatives~/~CEA, France; the Bundesministerium f\"ur Bildung und Forschung, Deutsche Forschungsgemeinschaft, and Helmholtz-Gemeinschaft Deutscher Forschungszentren, Germany; the General Secretariat for Research and Technology, Greece; the National Research, Development and Innovation Fund, Hungary; the Department of Atomic Energy and the Department of Science and Technology, India; the Institute for Studies in Theoretical Physics and Mathematics, Iran; the Science Foundation, Ireland; the Istituto Nazionale di Fisica Nucleare, Italy; the Ministry of Science, ICT and Future Planning, and National Research Foundation (NRF), Republic of Korea; the Lithuanian Academy of Sciences; the Ministry of Education, and University of Malaya (Malaysia); the Mexican Funding Agencies (BUAP, CINVESTAV, CONACYT, LNS, SEP, and UASLP-FAI); the Ministry of Business, Innovation and Employment, New Zealand; the Pakistan Atomic Energy Commission; the Ministry of Science and Higher Education and the National Science Centre, Poland; the Funda\c{c}\~ao para a Ci\^encia e a Tecnologia, Portugal; JINR, Dubna; the Ministry of Education and Science of the Russian Federation, the Federal Agency of Atomic Energy of the Russian Federation, Russian Academy of Sciences and the Russian Foundation for Basic Research; the Ministry of Education, Science and Technological Development of Serbia; the Secretar\'{\i}a de Estado de Investigaci\'on, Desarrollo e Innovaci\'on, Programa Consolider-Ingenio 2010, Plan Estatal de Investigaci\'on Cient\'{\i}fica y T\'ecnica y de Innovaci\'on 2013-2016, Plan de Ciencia, Tecnolog\'{i}a e Innovaci\'on 2013-2017 del Principado de Asturias and Fondo Europeo de Desarrollo Regional, Spain; the Swiss Funding Agencies (ETH Board, ETH Zurich, PSI, SNF, UniZH, Canton Zurich, and SER); the Ministry of Science and Technology, Taipei; the Thailand Center of Excellence in Physics, the Institute for the Promotion of Teaching Science and Technology of Thailand, Special Task Force for Activating Research and the National Science and Technology Development Agency of Thailand; the Scientific and Technical Research Council of Turkey, and Turkish Atomic Energy Authority; the National Academy of Sciences of Ukraine, and State Fund for Fundamental Researches, Ukraine; the Science and Technology Facilities Council, UK; the US Department of Energy, and the US National Science Foundation.

Individuals have received support from the Marie-Curie programme and the European Research Council and Horizon 2020 Grant, contract No. 675440 (European Union); the Leventis Foundation; the A. P. Sloan Foundation; the Alexander von Humboldt Foundation; the Belgian Federal Science Policy Office; the Fonds pour la Formation \`a la Recherche dans l'Industrie et dans l'Agriculture (FRIA-Belgium); the Agentschap voor Innovatie door Wetenschap en Technologie (IWT-Belgium); the F.R.S.-FNRS and FWO (Belgium) under the ``Excellence of Science - EOS'' - be.h project n. 30820817; the Ministry of Education, Youth and Sports (MEYS) of the Czech Republic; the Lend\"ulet (``Momentum'') Programme and the J\'anos Bolyai Research Scholarship of the Hungarian Academy of Sciences, the New National Excellence Program \'UNKP, the NKFIA research grants 123842, 123959, 124845, 124850 and 125105 (Hungary); the Council of Scientific and Industrial Research, India; the HOMING PLUS programme of the Foundation for Polish Science, cofinanced from European Union, Regional Development Fund, the Mobility Plus programme of the Ministry of Science and Higher Education, the National Science Center (Poland), contracts Harmonia 2014/14/M/ST2/00428, Opus 2014/13/B/ST2/02543, 2014/15/B/ST2/03998, and 2015/19/B/ST2/02861, Sonata-bis 2012/07/E/ST2/01406; the National Priorities Research Program by Qatar National Research Fund; the Programa de Excelencia Mar\'{i}a de Maeztu and the Programa Severo Ochoa del Principado de Asturias; the Thalis and Aristeia programmes cofinanced by EU-ESF and the Greek NSRF; the Rachadapisek Sompot Fund for Postdoctoral Fellowship, Chulalongkorn University and the Chulalongkorn Academic into Its 2nd Century Project Advancement Project (Thailand); the Welch Foundation, contract C-1845; and the Weston Havens Foundation (USA).
\clearpage
\bibliography{auto_generated}
\cleardoublepage \appendix\section{The CMS Collaboration \label{app:collab}}\begin{sloppypar}\hyphenpenalty=5000\widowpenalty=500\clubpenalty=5000\vskip\cmsinstskip
\textbf{Yerevan Physics Institute, Yerevan, Armenia}\\*[0pt]
A.M.~Sirunyan, A.~Tumasyan
\vskip\cmsinstskip
\textbf{Institut f\"{u}r Hochenergiephysik, Wien, Austria}\\*[0pt]
W.~Adam, F.~Ambrogi, E.~Asilar, T.~Bergauer, J.~Brandstetter, M.~Dragicevic, J.~Er\"{o}, A.~Escalante~Del~Valle, M.~Flechl, R.~Fr\"{u}hwirth\cmsAuthorMark{1}, V.M.~Ghete, J.~Hrubec, M.~Jeitler\cmsAuthorMark{1}, N.~Krammer, I.~Kr\"{a}tschmer, D.~Liko, T.~Madlener, I.~Mikulec, N.~Rad, H.~Rohringer, J.~Schieck\cmsAuthorMark{1}, R.~Sch\"{o}fbeck, M.~Spanring, D.~Spitzbart, A.~Taurok, W.~Waltenberger, J.~Wittmann, C.-E.~Wulz\cmsAuthorMark{1}, M.~Zarucki
% ...
\vskip\cmsinstskip
% ...
1:  Also at Vienna University of Technology, Vienna, Austria\\
% ...
\end{sloppypar}
\end{document}